\documentclass[aps,nofootinbib,amsmath,prd,onecolumn,notitlepage,showpacs,superscriptaddress,11pt]{revtex4-1}

\usepackage[american]{babel}
\usepackage{amsfonts}
\usepackage{amsmath}
\usepackage{amssymb}
\usepackage{slashed}
\usepackage{upgreek}

\usepackage{xcolor}
\usepackage{bm}
\usepackage{float}
\usepackage[utf8]{inputenc}
\usepackage{nicefrac}
\usepackage{hyperref}
\usepackage{url}
\usepackage{tikz-cd}
\usepackage[normalem]{ulem}
\usepackage{mathtools}
\usepackage{graphicx,txfonts}

\definecolor{darkblue}{rgb}{0.1,0.1,0.7}
\hypersetup{colorlinks,
           linkcolor={darkblue},
           citecolor={darkblue},
           urlcolor={darkblue}
}

\newcommand{\lvec}[2]{\raise #1\hbox{$^\leftarrow$} \hspace{-9pt} #2}
\newcommand{\rvec}[2]{\raise #1\hbox{$^\rightarrow$} \hspace{-9pt} #2}
\newcommand{\lrvec}[2]{\raise #1\hbox{$^\leftrightarrow$} \hspace{-9pt} #2}

\allowdisplaybreaks

\usepackage[normalem]{ulem}

\allowdisplaybreaks

 \DeclareMathOperator{\Tr}{Tr}

 \DeclarePairedDelimiter\abs{\lvert}{\rvert}%
\DeclarePairedDelimiter\norm{\lVert}{\rVert}%

 \makeatletter
\let\oldabs\abs
\def\abs{\@ifstar{\oldabs}{\oldabs*}}
\let\oldnorm\norm
\def\norm{\@ifstar{\oldnorm}{\oldnorm*}}
\makeatother

\begin{document}

\title{Weyl cohomology and the conformal anomaly in the presence of torsion}

\author{Gregorio Paci}
\email{gregorio.paci@phd.unipi.it (Corresponding Author)}
\affiliation{
Dipartimento di Fisica, Universit\`a di Pisa, Largo Bruno Pontecorvo 3, 56127 Pisa, Italy}
\affiliation{INFN - Sezione di Pisa, Largo Bruno Pontecorvo 3, 56127 Pisa, Italy}

\author{Omar Zanusso}
\email{omar.zanusso@unipi.it}
\affiliation{
Dipartimento di Fisica, Universit\`a di Pisa, Largo Bruno Pontecorvo 3, 56127 Pisa, Italy}
\affiliation{INFN - Sezione di Pisa, Largo Bruno Pontecorvo 3, 56127 Pisa, Italy}

\begin{abstract}
%
Using cohomological methods, we identify both trivial and nontrivial contributions to the conformal anomaly in the presence of vectorial torsion in $d=2,4$ dimensions. In both cases, our analysis considers two scenarios: one in which the torsion vector transforms in an affine way, i.e., it is a gauge potential for Weyl transformations, and the other in which it is invariant under the Weyl group. An important outcome for the former case in both $d=2,4$ is the presence of anomalies of a ``mixed'' nature in relation to the classification of Deser and Schwimmer.  
For invariant torsion in $d=4$, we also find a new type of anomaly which we dub $\Psi$-anomaly.
Taking these results into account, we integrate the different anomalies to obtain renormalized anomalous effective actions. Thereafter, we recast such actions in the covariant nonlocal and local forms, the latter being easier to work with. Along the way, we pause to comment on the physical usefulness of these effective actions, in particular to obtain renormalized energy-momentum tensors and thermodynamics of $2d$ black holes.
\end{abstract}

\maketitle

\section{Introduction}\label{sect:intro}

Below the Planck scale, an important framework in which to account for elementary particles and fields in the presence of gravity is quantum field theory in curved spacetimes (QFTCS) \cite{Parker:2009uva,Birrell:1982ix}. 
QFTCS has led to remarkable advances in topics such as inflationary cosmology and the thermodynamic theory of black holes. Still, within this context, different computational strategies are possible. One possibility is to resort to a mode expansion on some specific and suitable manifold. Another widely spread technique combines the background field method and the effective action. While the first has the advantage of being more physically transparent in many ways, the second is much more general, allowing us to work with a generic background metric and to study interacting fields in a relatively straightforward way. Naturally, the \emph{full} and \emph{exact} effective action is usually unknown. In this regard, an intriguing scheme to obtain at least part of such action is offered by the integration of the conformal anomaly. 

The conformal anomaly arises because quantum effects break conformal invariance. For example, in dimensional regularization, it is linked to the fact that the divergent part of the effective action is not conformally invariant in $d \neq 4$. Since we have to return to $d=4$ \emph{after} computing a physical quantity, it is possible to find a remnant of such noninvariance \cite{Birrell:1982ix}. Since it will be important in the following, let us recall that the contributions to the conformal anomaly can be geometrically classified in the approach of Deser and Schwimmer \cite{Deser:1993yx}. According to this classification, we have three kinds of anomalies: the $b$-anomalies, which are the conformal invariants, the $a$-anomaly, which is the Euler density, and the $a'$-anomalies, which can be changed by a local counterterm.

For us, the key point concerning the conformal anomaly is as follows. By integrating the differential functional equation defining it \cite{Riegert:1984kt}, we can obtain the part of the renormalized effective action responsible for the anomaly itself, i.e., the anomalous effective action $\Gamma_A$. In general, $\Gamma_A$ is a nonlocal action since it involves the Green's function of some differential operators. However, it can be recast in a more practical local form at the price of introducing auxiliary scalar fields. This way, we can obtain the vacuum expectation value and the higher point correlation functions of the energy-momentum tensor (emt) by a standard variation w.r.t.\ the metric. 

From a physical point of view, $\Gamma_A$ can offer interesting insights into the backreaction problem of quantum matter fields on the background geometry. However, it is important to notice that the anomalous action is defined up to a conformally invariant action $\Gamma_c$, which acts as an integration constant. For this reason, these kinds of actions are especially useful when the physics is mainly driven by the anomaly. This is the case of black hole physics, since the Hawking effect is intimately related to the conformal anomaly \cite{Balbinot:1999ri,Fursaev:1994te,Boos:2019vcz,Christensen:1977jc}. As a matter of fact, the approximately flat near horizon geometry plays a crucial role in this context since it determines the Hawking temperature. When only the metric is involved, another relevant context is cosmology, where conformally flat backgrounds are employed. An important example is Starobinsky inflation \cite{Starobinsky:1980te}. An application in which torsion is involved can be found in \cite{Buchbinder:1985ym}, where a nonsingular cosmological model was found.

It is suggested by QFTCS that, for consistency of the quantum theory, we need to extend General Relativity (GR). Indeed, it is necessary to include higher derivative operators to address the divergences in the effective action. Moreover, although GR works well on astrophysical scales, it exhibits features at both the classical and quantum levels that indicate the need for modifications or extensions of the theory. Inflation can be considered as one such modification. Other often mentioned issues are the nonrenormalizability at the perturbative level \cite{tHooft:1974toh,Christensen:1979iy,Goroff:1985th} and even the presence of spacetime singularities \cite{Penrose:1964wq,Hawking:1970zqf}.

When seeking a theoretical extension of GR, it is natural to consider a richer geometrical framework. One possibility that has been extensively studied is the inclusion of spacetime torsion. The first to include torsion in the game was Cartan shortly after GR and after that many different instances were considered. An agnostic approach to even more general geometries is offered by metric affine theories of gravity (MAG) \cite{Hehl:1994ue,Gronwald:1995em,Baldazzi:2021kaf,Sauro:2022chz,Sauro:2022hoh}. In an index-free notation, the torsion tensor, which is a vector-valued $2$-form $T(X,Y)$ acting on two vectors $X$ and $Y$, can be defined as
\begin{align}
T(X,Y)
=
\tilde{\nabla}_X Y 
-
\tilde{\nabla}_Y X
-
\left[X,Y\right] 
\, .
\end{align}
Using a holonomic basis, its components read $T^\lambda{}_{\mu\nu}= \tilde{\Gamma}^\lambda{}_{\nu\mu}-\tilde{\Gamma}^\lambda{}_{\mu\nu}$, where $\tilde{\Gamma}^\lambda{}_{\nu\mu}$ are the components of a metric-compatible but nonsymmetric affine connection. In this paper, we exploit the well-known decomposition 
\begin{align}\label{eq:connection_decomposition}
\tilde{\Gamma}^\mu{}_{\nu\rho} 
=
\Gamma^\mu{}_{\nu\rho} 
+
\frac{1}{2}\Bigl(T_\rho{}^\mu{}_\nu+T_\nu{}^\mu{}_\rho-T^\mu{}_{\nu\rho}\Bigr)
\equiv
\Gamma^\mu{}_{\nu\rho} 
+
K^\mu{}_{\nu\rho}
\, ,
\end{align}
in which we introduced the contortion tensor $K^\mu{}_{\nu\rho}$, and denoted with $\Gamma^\mu{}_{\nu\rho}$ the Christoffel symbols, i.e., the components of the unique metric-compatible symmetric connection $\nabla$. We adopt the convention that the last index is the directional one. Thanks to Eq.~\eqref{eq:connection_decomposition}, it is always possible to write any metric-compatible MAG's action in terms of the Levi-Civita connection, its curvature, and the torsion because
\begin{align}\label{eq:curvature_decomposition}
 \tilde{R}^\mu{}_{\nu\rho\theta} = {R}^\mu{}_{\nu\rho\theta}
  + {\nabla}_\rho K^\mu{}_{\nu\theta}
  -{\nabla}_\theta K^\mu{}_{\nu\rho} + K^\mu{}_{\eta\rho} K^\eta{}_{\nu\theta}
  -K^\mu{}_{\eta\theta} K^\eta{}_{\nu\rho}
\,.
\end{align}
Therefore, we can treat the torsion as a new field on a Riemannian background. In the following, we take this field-theoretical perspective on theories involving torsion and refer to the Levi-Civita connection and the Riemann curvature always as $\nabla$ and $R$, respectively, as we have done in Eq.~\eqref{eq:curvature_decomposition}.

It is convenient to decompose the torsion in general dimension $d$ into irreducible tensors as
\begin{align}\label{eq:T-Q-decomposition}
T^\mu{}_{\nu\rho} 
=
\frac{1}{d-1} \left(\delta^\mu{}_\rho \tau_\nu - \delta^\mu{}_\nu \tau_\rho \right)
+ H^\mu{}_{\nu\rho}
+\kappa^\mu{}_{\nu\rho}
\, ,
\end{align}
where $\tau_\mu \equiv T^\rho{}_{\mu\rho} $ is the torsion vector,  $H_{\mu\nu\rho} \equiv \frac{1}{3} T_{[\mu\nu\rho]}$ is the completely antisymmetric part, and $\kappa^\mu{}_{\nu\rho}$ is a tracefree hook antisymmetric tensor \cite{Paci:2023twc,Shapiro:2001rz}. We stress that in the following we will not consider the quantum properties of the torsion itself, which were instead considered in \cite{Martini:2023apm,Martini:2023rnv} from a path integral perspective. We will focus on the torsion vector $\tau$ considered as a classical background field along with the metric. 
This is because, as thoroughly analyzed before in Ref.~\cite{Shapiro:2001rz} (see also Ref.~\cite{Buchbinder:1985ux}), the torsion vector is the only component that admits a nontrivial transformation under the Weyl group.

 In particular, by borrowing the names used in Ref.~\cite{Shapiro:2001rz}, we shall consider the case of weak and strong conformal symmetry. The former means that the torsion tensor does not transform at all, while the latter allows for an affine transformation of only $\tau$ of the kind $\tau_\mu = \tau'_\mu + b \,  \partial_\mu \sigma$. This second possibility is deeply linked to Weyl geometry \cite{Sauro:2022hoh} and the resulting anomaly in $d=2$ has been studied also in \cite{Zanusso:2023vkn} in combination with the local renormalization group \cite{Osborn:1991gm}. The flexibility of assigning different Weyl transformations to the torsion can be understood geometrically as follows. In a torsionful geometry, conformal transformations may be extended in two primary ways \cite{Sauro:2022chz}. The first strategy is to assume that the full connection $\tilde{\Gamma}^\mu{}_{\nu\rho}$ transforms identically to the Christoffel connection ${\Gamma}^\mu{}_{\nu\rho} $. Under this assumption, the contortion tensor $K^\mu{}_{\nu\rho} $ remains unchanged, which also implies that the torsion is invariant. Alternatively, the second strategy is to require that the full connection is invariant under Weyl transformations, leveraging on the fact that, in metric-affine models, the metric $g_{\mu\nu}$ and the connection $\tilde{\Gamma}^\mu{}_{\nu\rho}$ are independent variables, and thus their conformal properties need not be related. This second case, naturally leads to the inclusion of a Weyl potential that transforms affinely in $\sigma$ \cite{Sauro:2022chz} (see also the beginning of the next section), which can be associated with the torsion vector in a mechanism of breaking of projective invariance of the lightcone \cite{Sauro:2022hoh}. We do not delve into these additional geometric structures here, as they are not particularly important for the rest of the analysis.
  
It is noteworthy to mention that Ref.~\cite{Shapiro:2001rz} discusses also a third type of conformal transformation, initially introduced in Ref.~\cite{Helayel-Neto:1999acz} and referred to as ``compensating'' conformal symmetry. While intriguing, the compensating conformal symmetry for the torsion is often structurally the same as the strong symmetry and can be derived from the latter by choosing a specific value of the parameter $b$, at least in the case of the nonminimally coupled scalar field examined in Refs.~\cite{Shapiro:2001rz,Helayel-Neto:1999acz}. Consequently, this does not impact our cohomological analysis, which keeps $b$ arbitrary. For comparison, see also the examples which we examine after the cohomological study of the strong case, where we explicitly compute the anomaly in the presence of a scalar field.

Since there is no model independent way to look at torsion, it is important to obtain the anomaly independently from the starting action generating it. This can be carried out by inspecting the de Rham cohomology for the Weyl group. This procedure is equivalent to employ the Wess-Zumino consistency conditions \cite{Wess:1971yu}, as can be seen from the finite version of the cohomological method \cite{Mazur:2001aa}. The latter approach has been put forward in \cite{Bonora:1983ff,Bonora:1985cq}, and has the advantage of being technically much simpler than the Wess-Zumino consistency conditions.

By applying this technique, we derive the most general consistent conformal anomaly in $d=2,4$ in the presence of background vectorial torsion (in addition to the metric) for both transformations of $\tau$. We also study in detail the associated anomalous actions. Our findings extend previous results \cite{Camargo:2022gcw,Buchbinder:1985ym} not only because we do not have to choose any particular theory, but also because we obtain the $\tau$-dependent anomaly. While, differently from \cite{Zanusso:2023vkn}, we use cohomological methods for different transformations for $\tau$, and consider also the $d=4$ case. However, we will not consider the couplings to be spacetime dependent. 

The paper is organized as follows. In Sect.~\ref{sect_conformal_tensors}, we review and clarify the cohomological analysis of \cite{Bonora:1985cq} and the procedure of \cite{Riegert:1984kt} for the anomalous action to encompass torsion. In the same section, some combinations of curvature tensors and $\tau$ with nice conformal transformations are introduced. Sect.~\ref{sect_2d_cohomological_analisys} presents the cohomological analysis in $d=2$ for strong and weak conformal symmetry. We check the analysis with some examples in which to compute the anomaly explicitly. This can be done with the heat kernel method \cite{Vassilevich:2003xt,DeWitt:1988fm,Barvinsky:1985an} since in general even dimension $d$ the anomaly and the Seeley-DeWitt coefficient $\hat{a}_{d/2}$ are proportional \cite{Zanusso:2023vkn,Birrell:1982ix,Parker:2009uva}. In addition, we also obtain the anomalous actions and write them in the Wess-Zumino, covariant nonlocal, and covariant local forms (see below). We discuss an unphysical ambiguity of these actions linked to the presence of an $a'$-anomaly mixed with an $a$-anomaly, and comment on the physical meaning of this result by examining the Wald entropy \cite{Wald:1993nt,Myers:1994sg}. Interestingly, we show it is insensitive to the mentioned ambiguity. We also evaluate this quantity for invariant torsion on a Rindler spacetime with a spherically symmetric vector on top of it. Thereafter, we perform in Sect.~\ref{sect_4d_cohomological_analisys} an analogous, although much more involved, cohomological treatment for $d=4$.  Also in this case we compute the anomaly in some examples. Moreover, for invariant torsion, we find a new kind of anomaly which does not fit in the original classification of Deser and Schwimmer, while, for transforming torsion, we find a mixed anomaly. The latter contribution is a $b$- mixed with an $a'$-anomaly, which, again, jeopardizes the usefulness of $\Gamma_A$ to some extent. This ambiguity seems to be more severe than in the $d=2$ case, as we discuss in Sect.~\ref{sect:Conclusions}. In this Section, we also give our conclusions and speculate on interesting concrete applications of our results.

\section{The general formalism and modified conformal tensors}\label{sect_conformal_tensors}

%
As we mentioned in the Introduction, we will consider both ``strong'' and ``weak'' conformal symmetry, borrowing the names from Ref.~\cite{Shapiro:2001rz}. In the subsequent analysis, we will always start with the \emph{strong} version of conformal transformations first, and the weak version later. In concrete terms, the strong transformation is defined as follows: we take the usual Weyl transformation for the metric tensor
\begin{align}\label{eq:transformation_metric}
g_{\mu\nu}=e^{2\sigma}g'_{\mu\nu} 
\, ,
\end{align}
where we could think of $g'$ as a fiducial metric with fixed determinant, while for the torsion vector we choose instead an affine transformation  
\begin{align}\label{eq:affine_transformation_torsion}
\tau_\mu = \tau'_\mu + b \,  \partial_\mu \sigma
\,  .
\end{align}
In the previous equation, $b$ is a nonzero parameter that can be chosen at will by normalizing $\tau$, and $\tau'$ could be though of a transverse vector. A somewhat natural choice for the parameter $b$ would be $(d-1)$, in which case the vector $S_\mu=-\tau_\mu/(d-1)$ could be interpretated as a gauge potential for Weyl symmetry, i.e., a Weyl potential \cite{Sauro:2022hoh}.\footnote{The gauge Weyl potential enjoys also a Weyl-covariant derivative, which includes a disformation term that is constructed with $S_\mu$, besides the usual Abelian gauge term. More details can be found in Refs.~\cite{Iorio:1996ad,Sauro:2022hoh} and references therein. In the present paper we purposedly try to avoid these additional geometric structures to make more general statements.} However, we will leave $b$ as a constant free parameter since we want to stick with the interpretation of $\tau$ as the torsion vector.

The use of a general $b$ also allows us to simplify the connection between the two kinds of transformation for $\tau$.
In fact, for \emph{weak} conformal symmetry, the metric transforms again as in \eqref{eq:transformation_metric}, but the torsion vector is left invariant
\begin{align}\label{eq:invariant_transformation_torsion}
\tau_\mu = \tau'_\mu 
\, .
\end{align}
Therefore, the weak case could be formally obtained as the $b \to 0$ limit of the strong one.
It is important to notice that in both cases $\tau_\mu$ is naturally defined with a covariant (down) index, so, for example, when varying $\tau^\mu$ should be regarded as a function of both the metric and the torsion as $\tau^\mu = g^{\mu\nu} \tau_\nu$.

\subsection{The general formalism}\label{subsect_general_formalism}

Take now $\sigma$ to be a Grassmannian function~\cite{Bonora:1983ff,Bonora:1985cq}.
For $\tau$ transforming affinely (that is, for strong transformations), the generator of Weyl transformations acquires the form
\begin{align}\label{eq:coboundary_operator}
\delta_\sigma
=
\delta^{g}_\sigma
+
\delta^{\tau}_\sigma
=
2\int {\rm d}^dx \, \sigma \, g_{\mu\nu} \frac{\delta}{\delta g_{\mu\nu}}
+b
\int {\rm d}^dx \,  \partial_\mu \sigma \,  \frac{\delta}{\delta \tau_\mu}
\, .
\end{align}
It is simple to verify that, for anticommuting $\sigma$, Eq.~\eqref{eq:coboundary_operator} gives a proper coboundary operator, i.e., it is \emph{nilpotent} $\delta^2_\sigma=0$ like the exterior derivative in differential geometry. Similarly, for invariant torsion (weak transformations), the generator is defined as
\begin{align}\label{eq:coboundary_operator_inv_torsion}
\delta_\sigma
=
\delta^{g}_\sigma
=
2\int {\rm d}^dx \, \sigma \, g_{\mu\nu} \frac{\delta}{\delta g_{\mu\nu}}
\, ,
\end{align}
which is again nilpotent. 
Thus, in both cases, we can construct analogues of de Rham's cohomology for $\delta_\sigma$ and of closed and exact forms.

Classically, it is clear that $\delta_\sigma S =0$ for a Weyl invariant theory, for both weak and strong transformations. At the quantum level, Weyl invariance is broken and the same equation displays an anomaly
\begin{align}\label{eq:delta-sigma}
\delta_\sigma \Gamma^{(1)}
=
\omega[\sigma; g, \tau] 
\, .
\end{align}
Thanks to the nihilpotency of $\delta_\sigma$, the previous equation enforces a consistency condition on the anomaly~\cite{Bonora:1983ff,Bonora:1985cq}
\begin{align}\label{eq:1-cocycle}
\delta_\sigma \omega[\sigma; g, \tau]
=
0 \, .
\end{align}
A general solution of \eqref{eq:1-cocycle} can be written as 
\begin{align}
\omega[\sigma; g, \tau]
=
\delta_\sigma F[g, \tau]
+
\omega_{NT}[\sigma; g, \tau]
\, ,
\end{align}
where $F[g, \tau]$ is a local functional, while the functional $\omega_{NT}$ is such that $\delta_\sigma \omega_{NT}=0$, but a local action $\bar{F}[g, \tau]$ for which $\delta_\sigma \bar{F}= \omega_{NT}$ \emph{does not exist}. In words, $\omega_{NT}$ cannot be expressed as the variation of a local functional.

We see that, geometrically, the consistent tensor structures in the anomaly are the closed $1$-forms w.r.t.\ the coboundary operator $\delta_\sigma$, while local functionals correspond to $0$-forms. The contributions to the trace anomaly can be trivial ($a'$-anomaly) or nontrivial ($a$- and $b$-anomalies). The difference between the two types lies in the fact that only the former are those that can be expressed as the variation of local functionals, i.e., they are exact $1$-forms. On the other hand, the nontrivial anomalies are closed but not exact $1$-forms. In other words, they belong to the (first) de Rham cohomology group constructed with $\delta_\sigma$.

In the construction of the de Rham cohomology \cite{Fecko:2006zy} one uses the cochain complex of p-forms on a differential manifold $M$ built with the exterior derivative $d$ 
\begin{align}
0 \overset{d}{\rightarrow} \Lambda^0(M) \overset{d}{ \rightarrow}  \Lambda^1(M) \overset{d}{\rightarrow} \Lambda^2(M) \dots
\, ,
\end{align}
where $ \Lambda^0(M)$ are the smooth functions ($0$-forms), $ \Lambda^1(M)$ the $1$-forms, etc.
In complete analogy, we can construct a cochain complex substituting $d$ with $\delta_\sigma$
\begin{align}
0 \overset{\delta_\sigma}{\rightarrow} \Omega^0 \overset{\delta_\sigma}{ \rightarrow}  \Omega^1 \overset{\delta_\sigma}{\rightarrow} \Omega^2 \dots
\, ,
\end{align}
where $\Omega^0$ are the local functionals independent of $\sigma$, $\Omega^1$ is the space to which the anomaly belongs (functionals linear in $\sigma$), while the consistency conditions belong to $\Omega^2$ (functionals quadratic in $\sigma$).
This formally justify calling $\omega$ in Eq.~\eqref{eq:1-cocycle} a $1$-cochain, which is a standard terminology that we will also adopt, and in the same way we will refer to the $2$-forms as $2$-cochains.

Naturally, $\omega$ can be expanded as the linear combination $\omega=\sum_i c_i\omega_i$ of $1$-cochains $\omega_i$. These are built using derivatives, metric, inverse metric, and torsion vector. They are also linear in $\sigma$. It is clear that the properties the $\omega_i$ have to satisfy are
\begin{itemize}
\item being diffeomorphism invariant,
\item being of dimension four in $d=4$, i.e., they counts as $4$ derivatives (or two in $d=2$),
\item and being linear in $\sigma$, i.e., they have ``ghost'' number $1$.
\end{itemize}
For simplicity, we also require parity invariance, so that a parity odd Hirzebruch-Pontryagin density, defined as $\epsilon^{\mu\nu\alpha\beta}{R}_{\mu\nu\rho\lambda}R^{\rho\lambda}{}_{\alpha\beta}$, is not allowed from the onset. However, since
\begin{align}
\delta_\sigma \int {\rm d}^4x \sqrt{g} \sigma \epsilon^{\mu\nu\alpha\beta}{R}_{\mu\nu\rho\lambda}R^{\rho\lambda}{}_{\alpha\beta}
=
4 \sigma (\partial_\rho \sigma) \epsilon_{\beta\alpha\mu\nu} \nabla^{\beta}{R}^{\rho\alpha\mu\nu}
=
 0 
 \, ,
\end{align}
the Pontryagin density is cohomologically allowed because of Bianchi identities. In addition, notice that ($\sqrt{g}$ times) the Pontryagin density is conformally invariant so that it could be easily included in the anomalous action, see Eq.~\eqref{eq:integration_b_anomalies} below.

To construct a minimal basis for the $1$-cochains, we will require that there are no derivatives acting on $\sigma$, which is a useful property for identifying the trivial anomalies and it can always be achieved with simple integration by parts. Therefore, the bases that we will construct have the schematic form
\begin{align}
\omega_i =  \int {\rm d}^dx \sqrt{g} \sigma F_i[g,\tau]
\, ,
\end{align}
where $F_i[g,\tau]$ are functionals of the appropriate dimension constructed with $g_{\mu\nu}$ and $\tau_\mu$. Notice that we can choose a basis of $1$-cochains of this form in each dimension $d$. Using \eqref{eq:delta-sigma} for a linear combination 
$\omega[\sigma; g, \tau]= \sum_i c_i \omega_i$,
we constrain the coefficients $c_i$ in $\omega$, selecting admissible tensor structures of the anomaly. 

Along with $1$-cochains, we are interested also in $0$- and $2$-cochains. Operationally, the main difference of $0$-cochains w.r.t.\ $1$-cochains is that they have ghost number $0$, i.e., they are local functional containing marginal operators. In this language, trivial anomalies $\xi$ can be expressed as $\xi=\delta_\sigma  \sum l_i \xi_i$ where $\xi_i$ constitute a basis for $0$-cochains. Similarly, $2$-cochains have ghost number $2$, i.e., they are $2$-forms. Another concept that we will refer to is that of cocycles, which are defined as closed cochains. For example, nontrivial Weyl anomalies are $1$-cocycles, Weyl invariant functionals are $0$-cocycles and, finally, trivial anomalies are called coboundaries. Therefore, the problem of finding the first de Rham cohomology group (nontrivial anomalies) can be restated as that of finding all $1$-cocycles which are not coboundaries. However, we anticipate here that the inclusion of $\tau_\mu$ leads to a slightly more complicate structure than that of the de Rham cohomology built with the exterior derivative. This will turn out to be important when integrating the anomaly.

It is useful to notice that for $d=4$ the variations $\delta_\sigma  \omega_i[\sigma; g, \tau]$ can be expressed in terms of a linear combinations of
\begin{align}\label{eq:1_basis_2_cochains}
\int {\rm d}^4x \sqrt{g} \left(\sigma \nabla_\mu \sigma \right) v_i^{\mu}
\, ,
\qquad \int {\rm d}^4x \sqrt{g}  \left(\nabla_\mu \sigma \Box\sigma\right)  w_j^{\mu}
\, ,
\end{align}
where the $v_i^{\mu}$ and $w_j^{\mu}$ are all possible vectors counting, respectively, as $3$ and $1$ derivatives that can be built from of curvatures, $\tau$ and covariant derivatives.
Instead, in $d=2$ the only term in the basis is that depending on
the analog of $v_i^{\mu}$, as it is clear also for dimensional reasons. Notice that for us the only tensor dimensionally counting as one derivative is $\tau$ itself and we will always sort the covariant derivatives so that they act as divergences on the torsion vector when possible.  Additionally, it is important to keep in mind the Grassmannian nature of $\sigma$ in Eq.~\eqref{eq:1_basis_2_cochains}. With these considerations, it is simple to see that these are a complete basis for $2$-cochains for $d=2,4$. Indeed, we notice that $\int\sqrt{g} \nabla_\mu \sigma \nabla_\nu \sigma V^{[\mu\nu]} =\int\sqrt{g} \sigma \nabla_\mu \sigma \nabla_\nu V^{[\mu\nu]} $, while $\int\sqrt{g}  \sigma \nabla_\mu \nabla_\nu \sigma W^{(\mu\nu)} = - \int\sqrt{g} \sigma \nabla_\mu \sigma \nabla_\nu W^{(\mu\nu)} $. Similarly, any term with three derivatives arbitrarily distributed on the $\sigma$s can be written as a combination of the two structures that we reported above. In fact, the only tensor with three free indices and counting as one derivative has the schematic form $\tau\otimes g$. The same happens for terms with four derivatives on $\sigma$, which are either zero or of the type \eqref{eq:1_basis_2_cochains}. For $d > 4$ additional terms appear and the analysis becomes more involved. For example, in $d=6$ we would consider tensors with three free indices counting as three derivatives, e.g., $\tau^\mu R^{\nu\rho}$, but in this paper we concentrate only on $d=2,4$.

Let us now turn to a more physical point of view on the conformal anomaly,
introducing the variational energy-momentum tensor $T_{\mu\nu}$ and a current $\mathcal{D}^\mu$ coupled to $\tau_\mu$ \cite{Sauro:2022hoh} (for their precise definition, see Eq.~\eqref{eq:vev_affine_tau} below and replace the effective action with a classical action). In the case of the torsion transforming affinely (strong version), classical Weyl invariance of $S[g,\tau]$ does not imply a vanishing $T^{\mu}{}_{\mu}$ as it does for torsionless theories or for invariant torsion. Instead we have
\begin{align}\label{eq:classical_Tr_emt_affine_tau}
T^{\mu}{}_{\mu}
=
- b \nabla_\mu   \mathcal{D}^{\mu}
 \, ,
\end{align}
which implies scale invariance in the flat space limit $g_{\mu\nu}\to \delta_{\mu\nu}$ and $\tau_\mu\to 0$. We refer to $\mathcal{D}_\mu$ as the virial current.\footnote{
The N\"other identity of scale invariance implies $\int \sqrt{g} T^\mu{}_\mu=0$,
so, for appropriate boundary conditions, there exists a vector, known as the virial current, such that its total divergence is the trace of the emt. Up to a normalization
we thus have that $\mathcal{D}_\mu$ plays the role of the virial current.}
Let us now introduce, similarly to Ref.~\cite{Zanusso:2023vkn}, the following functional derivatives of the effective action
\begin{align}\label{eq:vev_affine_tau}
\langle T^{\mu\nu} \rangle=-\frac{2}{\sqrt{g}} \frac{\delta \Gamma}{\delta g_{\mu\nu}} \, , 
\qquad 
\langle \mathcal{D}^{\mu} \rangle=\frac{1}{\sqrt{g}} \frac{\delta \Gamma}{\delta \tau_{\mu}} \, ,
\end{align}
which, by definition, give the vacuum expectation values of the energy-momentum tensor (emt) and of the virial current from the path-integral.
Therefore, at the quantum level we get
\begin{align}\label{eq:Asigma_affine_tau}
\delta_\sigma \Gamma
=
\int {\rm d}^dx \sqrt{g} \Bigg\{
- \sigma \langle T^{\mu}{}_{\mu} \rangle
+
b \partial_\mu \sigma  \langle \mathcal{D}^{\mu} \rangle
 \Bigg\} 
 =
 A_\sigma
 \, ,
\end{align}
that defines the integrated anomaly $A_\sigma$, that gives the nonintegrated anomaly $\mathcal{A}$ upon variation w.r.t.\ $\sigma$ ($\sqrt{g}\mathcal{A}= \frac{\delta A_\sigma}{\delta\sigma}$, see also Eq.~\eqref{eq:A_affine_tau} below).

Moving on to the case of invariant torsion (weak version), the equivalent of Eqs.~\eqref{eq:classical_Tr_emt_affine_tau} and \eqref{eq:Asigma_affine_tau} can be obtained by taking the limit $b \to 0$, as it is simple to verify. 

In the following, we assume that the anomaly depends only on the underlying geometric structure, i.e., on curvature tensors and torsion. For example, this is equivalent to consider the anomaly as generated by an action quadratic in quantum fields of arbitrary spin
that are integrated-out. For a general interacting theory, additional contributions to the anomaly would arise, for example from the renormalization group (RG) beta functions, meaning that our assumptions hold strictly at fixed points of the RG. Nevertheless, as we will discuss in Sect.~\ref{sect:Conclusions}, this assumption can be relaxed by including additional terms in Eq.~\eqref{eq:1_basis_2_cochains} which depend on the background fields. Of course, the specific structure of this new basis crucially depends on the spin and on the mass dimension of the fields under consideration.

In both cases, it is important to notice that the coefficients in front of the tensors entering the anomaly, being these purely geometrical tensors, are independent of the choice of the quantum state. In particular, by using covariant heat kernel techniques, these coefficients can generally be computed as linear combination of the number of fields of different spins that are integrated away \cite{Vassilevich:2003xt,DeWitt:1988fm,Barvinsky:1985an} and, eventually, rg beta functions for interacting theories.

Invariance under diffeomorphisms (Diff) of the classical action $S[g,\tau]$ implies the following N\"other identity
\begin{align}\label{eq:noether_identity_diff}
\nabla_{\mu}T^{\mu}{}_{\nu}
=
\mathcal{D}^{\mu}\Omega_{\mu\nu}
+
\tau_\nu \nabla_{\mu} \mathcal{D}^{\mu}
\, ,
\end{align}
where $\Omega_{\mu\nu}= \partial_\mu \tau_\nu - \partial_\nu \tau_\mu  $ is the homotetic curvature. Of course, Eq.~\eqref{eq:noether_identity_diff} holds independently from the Weyl transformation of $\tau_\mu$, which is just a vector under Diff.
In the following we will stick with the choice that the anomaly breaks conformal invariance but preserves Diff. This has been tacitly assumed when discussing the cohomology associated only with the Weyl group, which is enough to ensure that we will get a Weyl (and not Diff) anomaly. In fact, by using the anomalous effective action in place of the classical action, Eq.~\eqref{eq:noether_identity_diff} is assumed to hold unchanged for the vevs $\langle T^{\mu}{}_{\nu} \rangle$ and $  \langle\mathcal{D}^{\mu}\rangle$.
Still, we will verify explicitly that this assumption is consistent for our anomalous $\Gamma$ in the computationally simpler $d=2$ cases. 

Let us now describe the general procedure to obtain such anomalous actions even in the presence of torsion, using a technique that is a straightforward generalization of Ref.~\cite{Riegert:1984kt}.
It is convenient to rewrite \eqref{eq:Asigma_affine_tau} as
\begin{align}\label{eq:A_affine_tau}
\frac{\delta}{\delta \sigma} \Gamma[g,\tau]
=
-\sqrt{g}
\Big(
\langle T^{\mu}{}_{\mu}  \rangle 
+
b \nabla_\mu  \langle \mathcal{D}^{\mu} \rangle
\Big)
 =
\sqrt{g} \mathcal{A}
 \, ,
\end{align}
which can be functionally integrated by exploiting the conformal transformations of the objects involved ($b=0$ in the case of invariant torsion).
We recall the (finite) Weyl transformations $g_{\mu\nu}=e^{2\sigma}g'_{\mu\nu}$, and $\tau_\mu=\tau'_\mu + b\partial_\mu\sigma$, and consider the Weyl invariant functional, $ \Gamma[g,\tau]=\int {\rm d}^dx \sqrt{g} F[g,\tau]= \int {\rm d}^dx \sqrt{g'} F[g',\tau']=\Gamma[g',\tau']$. Then, we have that the following functional derivative w.r.t $\sigma$ gives
\begin{equation}\label{eq:integration_b_anomalies}
\begin{split}
\frac{\delta}{\delta \sigma}  \Gamma^F[\sigma; g',\tau']
=
\frac{\delta}{\delta \sigma}\int {\rm d}^dx \sqrt{g'} \sigma F[g',\tau']
=
 \sqrt{g} F[g,\tau] \, ,
\end{split}
\end{equation}
where $\Gamma^F$ is the $1$-cochain built with $F[g,\tau]$.
Therefore, any Weyl invariant contribution to the anomaly $\mathcal{A}$, can be easily integrated by using the functional $\Gamma^F[\sigma; g',\tau']=\int {\rm d}^dx \sqrt{g'} \sigma F[g',\tau']$. A similar situation occurs when we have contributions transforming linearly in $\sigma$ and with a self-adjoint and a conformally covariant \emph{differential} operator $O$ acting on scalars, i.e., $ \sqrt{g} E[g,\tau]=\sqrt{g'} E[g',\tau'] + \sqrt{g'} O(g',\tau')\sigma$. For example, this happens for $a$-anomalies. We have
\begin{align}\label{eq:integration_a_anomalies}
\frac{\delta}{\delta \sigma}  \Gamma^E[\sigma; g',\tau']
=
\frac{\delta}{\delta \sigma}\int {\rm d}^dx \sqrt{g'} \sigma 
\Bigg\{
E[g',\tau']
+
\frac{1}{2} O(g',\tau')\sigma
 \Bigg\}
=
 \sqrt{g} E[g,\tau] 
\, ,
\end{align}
where $\Gamma^E$ has been implicitly defined. Therefore, the contribution $\Gamma^E[\sigma; g',\tau']$ to the action reproduces the relative piece in $\mathcal{A}$ upon variation. Notice that it is easy to write these actions only in terms of non-primed quantities thanks to the conformal properties of $O$ and $F$. The last ingredient are the trivial anomalies, which usually appear in the anomaly as total derivatives. These can be straightforwardly integrated by taking a linear combination of a complete set of marginal operators (or $0$-cochains) and fixing the coefficients accordingly. We will call an action obtained in this way a Wess-Zumino anomalous action.

At this point, we can obtain the nonlocal version of the action. One starts by defining the the Green's function $\mathcal{G}(x,y)$ of the operator $O$ as
\begin{align}
\sqrt{g}_x O_x \, \mathcal{G}(x,y) = \delta^{(4)}(x,y) \, ,
\end{align}
from which we get $ \mathcal{G}(x,y)= \mathcal{G}'(x,y)$, i.e., it is conformally invariant.\footnote{%
Keep in mind that a solution of the Green equation always requires appropriate boundary conditions, which may or may not be compatible with conformal invariance. We assume that they do and check their effect in the explicit examples.
}
In the following, we will use the compact notation $\frac{1}{O}\varphi (x) =\int {\rm d}^4y \sqrt{g} \, \mathcal{G}(x,y)\,\varphi (y) $ for a Green's function acting on a test function. From the conformal transformation of $E[g,\tau]$ and $\mathcal{G}(x,y)$, it is simple to verify that 
\begin{align}
\frac{1}{O}E[g,\tau]
=
\sigma(x) + \text{$(g',\tau')$-dependent terms} \, ,
\end{align}
which can be inserted in equation \eqref{eq:integration_b_anomalies} to get the nonlocal action
\begin{align}
\Gamma^F_{NL}[g,\tau]
=
\int {\rm d}^dx \sqrt{g}  F[g,\tau] \frac{1}{O}E[g,\tau]
\, ,
\end{align}
and it is such that $\frac{\delta}{\delta \sigma}\Gamma^F_{NL}[g,\tau] =  \sqrt{g} F[g,\tau]$, where all the conformal transformations of the involved objects have been used, see Appendix \ref{app:non_local_actions}. Analogously, we also have that the action
\begin{align}
\Gamma^E_{NL}[g,\tau]
=
\frac{1}{2}
\int {\rm d}^dx \sqrt{g} E[g,\tau]  \frac{1}{O}E[g,\tau]
\,  
\end{align}
gives $\frac{\delta}{\delta \sigma}\Gamma^E_{NL}[g,\tau] =  \sqrt{g} E[g,\tau] $. We will, as customary, refer to the contribution to the affective action reproducing $a$- and $b$-anomalies (or a mixing of them) as the nonlocal part of the action.

Working with nonlocal actions it is in general a non trivial task. However, these actions can be localized by using the appropriate number of auxiliary fields. In the following we shall see several examples of this well-known procedure.

\subsection{The modified conformal tensors}\label{subsect_modified_conf_tensors}

Conformally covariant tensors and tensors with special conformal properties are particularly useful when dealing with Weyl transformations. In this section, when the torsion transforms affinely, we modify them by adding some torsion-dependent terms in order to obtain objects with suitable finite conformal transformations, which significantly simplifies the computations of the next sections. We start by introducing the Schouten tensor and its trace
\begin{equation}
\begin{split}
 K_{\mu\nu} = \frac{1}{d-2}\Bigl(R_{\mu\nu}-\frac{1}{2(d-1)} R g_{\mu\nu}\Bigr) \, ,
 \qquad J = g^{\mu\nu} K_{\mu\nu} = \frac{1}{2(d-1)} R
\end{split}
\end{equation}
that are characterized by the very simple Weyl transformations
\begin{equation}
\begin{split}
\delta_\sigma K_{\mu\nu}=-\nabla_\mu \partial_\nu\sigma \, ,
 \qquad \delta_\sigma J = -2J \sigma -\Box \sigma \, .
\end{split}
\end{equation}
If we take $\tau_\mu$ to transform affinely as in \eqref{eq:affine_transformation_torsion},
we can look for combinations of curvature tensors with the torsion vector such that they transform homogeneously. For example, let us consider the scalar combination
\begin{align}\label{eq:TildeJ}
 \tilde{J}
 =
 J
 +
 c \,  \nabla \cdot \tau
 +
 c_1 \, \tau \cdot \tau
 \, .
\end{align}
By imposing that $\tilde{J}$ transforms homogeneously, i.e., that it satisfies $ \tilde{J} =\tilde{J}'- 2 \sigma  \tilde{J}$, we get that the coefficients must be
\begin{align}
c
=
\frac{1}{b} \, ,
 \qquad
c_1
=
-\frac{d-2}{2b^2}   \, .
\end{align}
Accordingly, the quantity $\sqrt{g}\tilde{J}$ is invariant in $d=2$, while $\sqrt{g}\tilde{J}^2$ is invariant in $d=4$. The modified Schouten trace squared can be written in terms of conformal tensors as
\begin{align}\label{eq:TildeJ_Squared}
\tilde{J}^2
=
J^2
-
\frac{d-2}{b^2} J\, \left(\tau \cdot \tau\right)
+
\frac{2}{b} J \, \left( \nabla \cdot \tau \right)
+
\frac{1}{b^2} \left( \nabla \cdot \tau   \right)^2
+
\frac{(d-2)^2}{4b^4}  \left(\tau \cdot \tau\right)^2
-
\frac{d-2}{b^3} \left( \nabla \cdot \tau \right) \left(\tau \cdot \tau\right)
\, ,
\end{align}
that we report here for later convenience.
Similarly, by requiring that for the tensor
\begin{equation}\label{eq:TildeK}
\begin{split}
\tilde{K}^{\mu\nu}
=
{K}^{\mu\nu}
+
b_1 \,  \left(\nabla_\mu \tau_\nu + \nabla_\nu \tau_\mu \right)
+
b_2 \,  \tau_\mu \tau_\nu
+
b_3 \, g^{\mu\nu} \tau \cdot \tau
+
b_4 \, g^{\mu\nu} \nabla \cdot \tau
\end{split}
\end{equation}
the conformal transformation is also homogeneous, $\tilde{K}^{\mu\nu}=\tilde{K}'^{\mu\nu} -4 \sigma  \tilde{K}^{\mu\nu}$, we obtain
\begin{equation}
\begin{split}
b_1
=
\frac{1}{2b} \, ,
 \qquad
b_2
=
\frac{1}{b^2} \, ,
\qquad
b_3
=
-\frac{1}{2b^2}   \, ,
\qquad
b_4
=
0   \, .
\end{split}
\end{equation}
Naturally, we have that
\begin{equation} 
\tilde{K}^{\mu}{}_{\mu}=\tilde{J} \, ,
\end{equation} 
and, for the same choice of constants, that  $ \tilde{{K}'}_{\mu\nu} = \tilde{K}_{\mu\nu}$.\footnote{%
Equations involving the tensors $\tilde{K}_{\mu\nu}$ and $\tilde{J}$ have been explored
in the past in the context of ``Ricci-gauging'' \cite{Iorio:1996ad},
which is a geometric procedure to produce Weyl-invariant actions. They also arise naturally in a generalization of the ambient space approach \cite{Jia:2023gmk}.
}

It should be clear that $\sqrt{g}\tilde{K}^2_{\mu\nu} $ is conformally invariant in $d=4$, where the modified Schouten squared relates to the original tensors as
\begin{align}\label{eq:TildeK_Squared}
\tilde{K}^2_{\mu\nu} 
&=
{K}^2_{\mu\nu} 
-
\frac{1}{b^2} J\, \left(\tau \cdot \tau\right)
+
\frac{2}{b^2} {K}_{\mu\nu} \, \left( \tau^\mu\ \tau^\nu \right)
+
\frac{2}{b} {K}_{\mu\nu} \, \left( \nabla^\mu\ \tau^\nu \right)
+
\frac{d}{4b^4}  \left(\tau \cdot \tau\right)^2\\
&+
\frac{2}{b^3}\left(\tau^\mu \tau^\nu \right)   \nabla_\mu  \tau_\nu  
-
\frac{1}{b^3} \left( \nabla \cdot \tau \right)\left(  \tau \cdot \tau\right)
+
\frac{1}{2b^2} \left(\nabla_\mu \tau_\nu \nabla^\nu\tau^\mu  \right)
+
\frac{1}{2b^2} \left( \nabla_\mu \tau_\nu   \right)^2
\, .
\end{align}
Another useful tensor to consider is the aforementioned homotetic curvature 
\begin{align}\label{eq:Omega}
\Omega_{\mu\nu}
=
 \left(\nabla_\mu \tau_\nu - \nabla_\nu \tau_\mu \right)
 =
 \left(\partial_\mu \tau_\nu - \partial_\nu \tau_\mu \right)  \, .
\end{align}
Notice that the quantity $\sqrt{g}\Omega^2_{\mu\nu}$ is also conformal invariant in $d=4$.

Furthermore, it is also possible to construct a ``boundary'' term that we denote $\tilde{\Box} \tilde J$ (it actually is a boundary term only for $d=4$, see the explicit form below). The most general ansatz for the operator $\tilde{\Box}$ reads
\begin{equation}
\begin{split}
\tilde{\Box}
=
\Box + g_1 \tau^\mu \nabla_\mu + g_2 \left( \nabla_\mu  \tau^\mu \right) + g_3 \left( \tau^\mu  \tau_\mu \right) \, .
\end{split}
\end{equation}
If we impose $\tilde{\Box} \tilde{J} = (\tilde{\Box} \tilde{J})' -4\sigma (\tilde{\Box} \tilde{J})'$, we must set the constants to be
\begin{equation}
\begin{split}
g_1
=
\frac{d-6}{b} \, ,
 \qquad
g_2
=
\frac{2}{b} \, ,
\qquad
g_3
=
-\frac{2(d-4)}{b^2}   \, .
\end{split}
\end{equation}
In terms of the conformal tensors, $ \tilde{\Box} \tilde{J}$ reads
\begin{align}\label{eq:BoxJ}
\tilde{\Box} \tilde{J}
&=
 \Box J
+
\frac{1}{b}  \, \Box \nabla \cdot \tau
-
\frac{d-2}{b^2}  \, \tau^\mu \Box \tau_\mu
-
\frac{d-2}{b^2}  \,  \left(  \nabla_\mu  \tau_\nu   \right)^2 
-
\frac{d-6}{b}  \,\tau \cdot   \left(    \nabla J   \right)
+
\frac{2}{b}  \,J    \left(    \nabla \cdot \tau \right)
+
\frac{2}{b^2}  \,   \left(    \nabla \cdot \tau \right)^2 \\ \nonumber
&-
\frac{d-6}{b^2} \tau^\mu \nabla_\mu \left( \nabla \cdot \tau \right)
+
\frac{(d-6)(d-2)}{b^3}  \tau^\mu \tau^\nu  \left(   \nabla_\mu \tau_\nu       \right)
+
\frac{10-3d}{b^3}     \tau \cdot \tau      \left(    \nabla \cdot \tau     \right)
+
\frac{(d-4)(d-2)}{b^4}    \left(     \tau \cdot \tau    \right)^2\\  \nonumber
&-
\frac{2(d-4)}{b^2}  J  \left(     \tau \cdot \tau    \right) \, ,
\end{align}
As anticipated, for general $d$ not all the terms combine in boundary terms, but it is easy to check that for $d=4$ this expression becomes a total derivative. Moreover, $\sqrt{g}\tilde{\Box} \tilde{J}$ is Weyl invariant in $d=4$ by construction.

As we shall see, all these combinations are especially helpful in the cohomological analysis since it allows to consider the $1$-cochains built with Weyl invariants
which are free to appear in the anomaly without any constraint on their coefficients.

\section{Weyl comohology in $d=2$ for invariant and affinely transforming torsion}\label{sect_2d_cohomological_analisys}

\subsection{Cohomological analysis for the torsion transforming affinely in $d=2$ and integration of the anomaly}\label{sect_2d_cohomological_analisys_affine_torsion}

Now concentrate on the affine transformation $\delta_\sigma \tau_\mu = b \partial_\mu \sigma$ for $b\neq 0$.
The basis in $d=2$ for the possible $1$-cochains $\omega_i[\sigma; g]$ can be conveniently written by using the modified Schouten trace introduced above
\begin{align}\label{eq:1-cochains-curvatures_2d}
\omega_1 =\int {\rm d}^2x \sqrt{g} \sigma \tilde{J} 
\, ,
\qquad
\omega_2 =\int {\rm d}^2x \sqrt{g} \sigma  \left(\nabla \cdot \tau\right) 
\, ,
\qquad
\omega_3=\int {\rm d}^2x \sqrt{g} \sigma  \left(\tau \cdot \tau  \right) \, ,
\end{align}
as done also in \cite{Zanusso:2023vkn}. The consistency condition allows the first a term to enter the anomaly without any constraint on its coefficient since
\begin{equation}
\delta_\sigma \int {\rm d}^2x \sqrt{g} \sigma \tilde{J}
=
0 \, ,
\end{equation}
where $\sigma$ is Grassmannian and $\delta_\sigma$ is the operator \eqref{eq:coboundary_operator}.
Consequently, we have
\begin{align}
\delta_\sigma \sum_{i=1}^{3} c_i \omega_i[\sigma; g, \tau] 
=
-2 b c_3  \int {\rm d}^2x \sqrt{g} \sigma  \left(\nabla_\mu \sigma \right)  \tau^\mu
=
0 
\quad
\Rightarrow 
\quad
c_3
=
0 \, ,
\end{align}
in agreement with the result of Ref.~\cite{Zanusso:2023vkn} at renormalization group fixed points.
Therefore, in $d=2$ the consistent anomaly in the presence of torsion transforming affinely is 
\begin{align}\label{eq:2d_anomaly_affine_torsion}
A_\sigma
=
\int \sqrt{g}  \sigma \Big\{
c_1 \tilde{J} + c_2 \left( \nabla \cdot \tau \right) \Big\} \, .
\end{align}
In other words, we have two $1$-cocycles while the integrand of $\omega_3$ is prohibited in the anomaly (at renormalization group fixed points) unless we take torsion to be untouched by Weyl transformations, i.e., $b=0$, which is the case that we consider in the next subsection.

Let us now consider the trivial anomalies. It is easy to obtain the following basis for the $0$-cochains
\begin{align}\label{eq:0-cochains-tau_2d}
\xi_1=\int {\rm d}^2x \sqrt{g} \tilde{J} 
\, ,
\qquad
\xi_2 =\int {\rm d}^2x \sqrt{g}   \left(\nabla \cdot \tau\right) 
\, ,
\qquad
\xi_3=\int {\rm d}^2x \sqrt{g}   \left(\tau \cdot \tau  \right) \, .
\end{align}
From which obtain
\begin{align}
\delta_\sigma \xi_1 =0
\, ,
\qquad\qquad
\delta_\sigma \xi_2 = 0 
\, ,
\qquad\qquad
\delta_\sigma  \xi_3= - 2 b\int {\rm d}^2x \sqrt{g}  \sigma \left(\nabla \cdot \tau\right)  \, .
\end{align}
Accordingly, $\omega_2$ is a trivial anomaly since $\omega_2 = - \frac{1}{2 b} \delta_\sigma  \xi_3$. It clear that the result does not depende on the chosen bases, because we could have used ${J}$ as well instead of $\omega_{1}$ and, in this way, the anomaly would have been parametrized by a combination of $J$ and $\nabla\cdot\tau$ leadin to the same result.

Let us now move to the problem of obtaining an effective action reproducing the anomaly. We consider \eqref{eq:2d_anomaly_affine_torsion}, thus the nonintegrated anomaly is
\begin{align}\label{eq:2d_emt_affine_torsion}
\mathcal{A}
=
-\langle T^{\mu}{}_{\mu}  \rangle 
-
b \nabla_\mu  \langle \mathcal{D}^{\mu} \rangle
=
c_1 \tilde{J} 
+
c_2 \left( \nabla \cdot \tau \right) 
\, .
\end{align}
The finite conformal transformations of the involved tensors are 
\begin{align}
\sqrt{g} \tilde{J} 
=
\sqrt{g'} \tilde{J'}
\, , \qquad \qquad
\sqrt{g} \left( \nabla \cdot \tau \right)
=
\sqrt{g'} \left( \nabla' \cdot \tau' +  b \Delta_2' \sigma \right)   \, ,
\end{align}
where we recall that $g'$ is the fiducial metric appearing in $g_{\mu\nu}=e^{2\sigma}g'_{\mu\nu}$, while $\tau_\mu=\tau'_\mu + b \partial_\mu \sigma$, and $\Delta'_2=\Box'$ is the 2d Yamabe's operator \cite{Juhl_book} computed using $g'$.

The conformal invariant $\sqrt{g}\tilde{J}$ borrows topological information like typical $a$-anomalies. This is because the inclusion of total derivatives does not change the Euler characteristic, as long as the manifold $M$ has no boundary, $\partial M=\emptyset$ (recall that $\tilde{J} \sim E_2 +\nabla \tau$). In the presence of a boundary the anomaly would acquire new structures akin to the Gibbons-Hawking-York term \cite{Solodukhin:2015eca}, but we do not consider them here. Therefore, if  $\partial M=\emptyset$ as we will we assume, we could call $\tilde{J}$ a mixed $(a + b)$-anomaly. However, this distinction is irrelevant for both the construction and the integration of the anomaly as we see later when integrating it. For this reason we stick with the definition that $a$- and $b$-anomaly are discernible from each other only based on their Weyl transformations. With this definition we cannot have something like an $(a + b)$-anomaly (we will have other mixed anomalies in the following, however).

Let us turn to the term $\left( \nabla \cdot \tau \right)$. This term has some properties of an $a$-anomaly, as its transformation law suggests, but it is also a trivial anomaly since it can be generated by variation of a local action (see above, it is also manifestly a total derivative). We call such a contribution an $(a + a')$-anomaly. It is important to realize such a ``double nature'' since this fact dictates how to integrate the equation of the anomaly. Under this light, we get that the most general action producing the anomaly upon variation is
\begin{align}\label{eq:2d_Gamma_ind_sigma_affine_torsion}
\Gamma_{WZ}[\sigma; g',\tau]
&=
 \int {\rm d}^2x \sqrt{g'}  \sigma \Big\{ 
c_1   \tilde{J'} + \mathfrak{C}_2  \left( \nabla' \cdot \tau'\right) + b \frac{\mathfrak{C}_2}{2}  \Delta'_2 \sigma
\Big\} 
-
\frac{\mathfrak{C}_3}{2 b} \int {\rm d}^2x \sqrt{g}   \left( \tau \cdot \tau \right)  \, ,  \\ \nonumber
& \text{with} \quad \mathfrak{C}_2  + \mathfrak{C}_3 = c_2    \, ,
\end{align}
where $\left( \nabla \cdot \tau \right)$ has been integrated simultaneously as both an $a$- and an $a'$-anomaly, though the two contributions are constrained such that their coefficients match the original $c_2$ charge in the anomaly. 

Several comments are in order. From the cohomological point of view, one might object that an $(a + a')$-anomaly should be intergrated only as a trivial anomaly. This is because, any exact $1$-form is cohomologous to the zero form. This way of proceeding is actually \emph{incorrect} in our opinion, as we now argue. The point is that the structure of the de Rham cohomology constructed with $\delta_\sigma$ is richer that its analogous built with just the exterior derivative $d$ on forms. Indeed, in the former case we have the natural additional notion of the transformation of $p$-forms $p_i$ under a group $G$. In this case it is the Weyl group, but it could be any other. In fact, let us consider the relation $p_1 \sim p_2$, which stands for  ``$p_1$ transforms like $p_2$'' under Weyl transformations. It is simple to show that $\sim$ is an equivalence relation being the following trivially satisfied 
\begin{align}
p_1 \sim p_1 
\, ,
\qquad
\text{if \quad} p_1 \sim p_2 \wedge p_2 \sim p_3 \,\, \Rightarrow \,\, p_1 \sim p_3
\, ,
\qquad
\text{if \quad}  p_1 \sim p_2  \,\, \Rightarrow \,\, p_2 \sim p_1
\, ,
\end{align}
for $\forall$ $p_{1,2,3}$ belonging to the exterior algebra.
Therefore, we can further divide the de Rham cohomology into equivalence sub-classes of objects transforming in the same way, which is a relevant information for the integration of the anomaly. We will call these classes ``Weyl sub-classes.''

For example, consider a trivial anomaly $a' \in [0]_{dR}$, where $ [0]_{dR}$ is the equivalence class of exact $1$-forms in the standard de Rham cohomology. Although cohomologous from the de Rham view point, different representatives of this class can belong to different Weyl sub classes and must be integrated differently. In a similar way, $\tilde{J}$ and $J$ belong to the same de Rham equivalence class but to different Weyl sub-classes and are integrated accordingly. In the following, we will encounter other instances of this fact on which we will comment in due time.

We also have to stress that a ``double'' nature of the kind of $\left( \nabla \cdot \tau \right)$ anomaly leads to an ambiguity that can pose stringent limits to the physical usefulness of the action \eqref{eq:2d_Gamma_ind_sigma_affine_torsion}. Indeed, it forces us to introduce an arbitrary constant, say $\mathfrak{C}_2$ by solving the constraint for $\mathfrak{C}_3$, on which physical observables may depend. However, this constant cannot have any physical meaning since it just corresponds to the arbitrary splitting of a number in the sum of two, $c_2 = \mathfrak{C}_2  + \mathfrak{C}_3 $. We study this situation in more detail in the next subsection by computing as an example the (nonolocal) emt and the Wald's entropy of a $2d$ black hole. While the former is indeed dependent on $\mathfrak{C}_2$, we will show that, surprisingly enough, the Wald's entropy turns out to be independent of this constant as should be expected from a physical observable.

In the remainder of this section we fix $b=-1$ for convenience, although it is not strictly necessary, and the general formulas can be obtained by rescaling $\tau_\mu$.
By following the procedure outlined in subsect.~\ref{subsect_general_formalism}, we arrive at the nonlocal version of this equation
\begin{align}\label{eq:2d_Gamma_ind_nonlocal_affine_torsion}
\Gamma_{NL}[g,\tau]
=
-\int {\rm d}^2x \sqrt{g}   \Bigg\{ 
&\left[ c_1 \tilde{J} +  \frac{\mathfrak{C}_2}{2}  \left(\nabla\cdot\tau\right)   \right]    \frac{1}{\Delta_2}   \left(\nabla\cdot\tau\right)  
\Bigg\}
-
\frac{\mathfrak{C}_2 - c_2}{2} \int {\rm d}^2x \sqrt{g}   \left( \tau \cdot \tau \right)  \\ \nonumber 
&-
\frac{C_4}{2} \int {\rm d}^2x \sqrt{g}   \Bigg\{ 
\tilde{J}   \frac{1}{\Delta_2}   \tilde{J}    
\Bigg\}
 \, ,
\end{align}
which correctly reproduce the anomaly upon variantion (in Appendix \ref{app:non_local_actions} the reader can find the variations computed without fixing $b=-1$). Notice that we have added a conformally invariant nonlocal last term with an arbitrary coefficient $C_4$, representing the ``ambiguity'' of adding an arbitrary conformal action to the anomalous action (the anomaly determines the effective action up to conformally invariant terms). Notice that such ambiguity \emph{vanishes} for a geometry such that $\tilde{J}=0$ and, to some extent, for black holes thermodynamics, as we discussed in the introduction. One might be worried by the absence of the Polyakov action which instead, as might be expected, appears if we parametrize the anomaly using $J$ and $ \left( \nabla \cdot \tau \right)$. However, it is simple to show that the $\Gamma_{NL}$s obtained with these two parametrizations simply differ by a conformally invariant action precisely of the form $ \int {\rm d}^2x \sqrt{g}\tilde{J}(1/\Delta_2 )\, \tilde{J} $, see Appendix \ref{app:non_local_emt_Diff}. We added this contribution also for this reason, i.e., we can use it to interpolate with the standard Polyakov's result.

We conclude this subsection by computing the anomaly in a concrete example. We consider a very simple conformally invariant theory and compute the anomaly covariantly using the Seleey-DeWitt coefficient $\hat{a}_1(x)$ \cite{Barvinsky:1985an}. The simplest example we can think of is that of a scalar field $\phi$ coupled to $\tau$. In $d=2$ we have
\begin{align}
\tau_\mu \rightarrow \tau_\mu + b \partial_\mu \sigma 
\, ,
\qquad
\phi   \rightarrow  \phi
\, ,
\end{align}
because $\phi$ weighs zero.
Then, the action
\begin{align}
S
=
\frac{1}{2} \int {\rm d}^4x \sqrt{g}  \left( 
\nabla^\mu \phi \nabla_\mu\phi
+
\xi \tilde{J}\phi^2
\right)
\end{align}
is invariant. Notice that a term like $\tau_\mu^2\phi^2$ is forbidden. Therefore, the operator of interest is $F(\nabla)=-\Box+\xi \tilde{J}$ and the anomaly is
\begin{align}
 4\pi \, \mathcal{A}
=
\Tr a_1
=
\left(  \frac{1}{3} - \xi  \right) \tilde{J} + \frac{1}{3} \nabla\cdot\tau
\, ,
\end{align}
in agreement with the cohomological analysis and the coefficients $c_i$ are determined by comparing the above formula with \eqref{eq:2d_emt_affine_torsion}.

\subsubsection{Localized action, energy momentum tensor and Wald's entropy}\label{sect_BH_thermo_and_2d_emt_affine_torsion}

In order to compute the renormalized (vev of the) emt, virial current and other physically relevant quantities, it is convenient to write \eqref{eq:2d_Gamma_ind_nonlocal_affine_torsion} in yet another form, namely to use auxiliary scalar fields to make it a local action \cite{Riegert:1984kt}. For this purpose, it is helpful to rewrite \eqref{eq:2d_Gamma_ind_nonlocal_affine_torsion} in the symmetric form by ``completing the square'' in the first term
\begin{align}\label{eq:2d_Gamma_ind_symm_form_affine_torsion}
\Gamma_{NL}[g,\tau]
=
-\frac{\mathfrak{C}_2}{2} \int {\rm d}^2x \sqrt{g}   \Bigg\{ 
&\left[  \left(\nabla\cdot\tau\right)+ \frac{c_1}{\mathfrak{C}_2} \tilde{J}     \right]    \frac{1}{\Delta_2}  \left[  \left(\nabla\cdot\tau\right)+ \frac{c_1}{\mathfrak{C}_2} \tilde{J}     \right]   
\Bigg\}  
+
\frac{c^2_1-C_4\mathfrak{C}_2}{2\mathfrak{C}_2} \int {\rm d}^2x \sqrt{g}   \Bigg\{ 
\tilde{J}   \frac{1}{\Delta_2}   \tilde{J}    
\Bigg\}  \\ \nonumber
&
- 
\frac{ \mathfrak{C}_2-c_2 }{2}\int {\rm d}^2x \sqrt{g}   \left( \tau \cdot \tau \right)
 \, .
\end{align}
This expression is then easily recasted in a local form by introducing two auxiliary scalar fields $\varphi$ and $\psi$ (we comment more on the number auxiliary fields in Subect.~\ref{sect_2d_cohomological_analysis_invariant_torsion}, Sect.~\ref{sect:Conclusions} and Appendix~\ref{app:local_actions_auxiliary_fields}) as
\begin{align}\label{eq:2d_Gamma_ind_auxiliary_fields_affine_torsion}
\Gamma_{NL}[g,\varphi, \psi]
=
 \int {\rm d}^2x \sqrt{g}  \Bigg\{ 
&\frac{1}{2} \varphi \Delta_2 \varphi + \alpha \varphi \tilde{J} + \beta \varphi  \left(\nabla\cdot\tau\right) - \frac{1}{2} \psi \Delta_2 \psi - \gamma \psi \tilde{J} 
\Bigg\} \\ \nonumber
&
-
\frac{ \mathfrak{C}_2-c_2 }{2} \int {\rm d}^2x \sqrt{g}   \left( \tau \cdot \tau \right)
 \, ,
\end{align}
The scalars satisfy the following equations of motion
\begin{align}\label{eq:eom_scalars_2d_affine_torsion}
\Delta_2 \varphi 
=
-
\alpha \tilde{J} 
 -
\beta \left(\nabla\cdot\tau\right) 
\, ,
\qquad 
\Delta_2 \psi
=
-
\gamma \tilde{J}
\, ,
\end{align}
which can be used in \eqref{eq:2d_Gamma_ind_auxiliary_fields_affine_torsion} to obtain the nonlocal action
\begin{align}
\Gamma_{int}[g,\varphi, \psi]
=
- \frac{1}{2}  \int {\rm d}^2x \sqrt{g}  \Bigg\{ 
 \left(  \beta    \left(\nabla\cdot\tau\right) +  \alpha \tilde{J}  \right)   \frac{1}{\Delta_2}   \left( \beta   \left(\nabla\cdot\tau\right) +  \alpha \tilde{J}  \right) 
-
 \left(  \gamma   \tilde{J}      \right)   \frac{1}{\Delta_2}   \left(  \gamma   \tilde{J}     \right) 
\Bigg\} 
 \, ,
\end{align}
that on-shell reproduces \eqref{eq:2d_Gamma_ind_symm_form_affine_torsion} by choosing the parameters
\begin{align}\label{eq:mapping_constants_auxiliary_fields_affine_torsion}
\beta=\sqrt{\mathfrak{C}_2}
\,,  
\qquad
\alpha=\frac{c_1}{\sqrt{\mathfrak{C}_2}}
\,,
\qquad
\gamma^2=\frac{c_1^2}{\mathfrak{C}_2}-C_4
\,.
\end{align}
Notice that these parameters could become imaginary. Typically, the imaginary part of the effective action is associated with the creation of particles by an external field \cite{Schwinger:1951nm}. However, in this paper, we prefer to view the auxiliary fields as formal placeholders for representing nonlocalities which enable us to avoid cumbersome calculations with Green functions, e.g., Eq.~\eqref{eq:2d_Gamma_ind_nonlocal_affine_torsion}. In fact, we will always assume that these fields are on-shell. From this perspective, it seems reasonable to require that the anomalous action acquires an imaginary part only if the anomaly itself has a complex coefficient $c$. Specifically, we reject the possibility of obtaining such an imaginary contribution from the ``localization'' process. Consequently, we take the unphysical parameter to satisfy $\mathfrak{C}_2 > 0$, and the undetermined parameter $C_4 < c_1^2/\mathfrak{C}_2$.

Using the localized action, we can compute the emt from \eqref{eq:2d_Gamma_ind_auxiliary_fields_affine_torsion} by the functional derivative
\begin{align}\label{eq:emt_auxiliary_fields_affine_torsion}
-\frac{2}{\sqrt{g}} \frac{\delta}{\delta g_{\mu\nu}}    \Gamma_{NL}[g,\varphi, \psi]
=&
\langle T^{\mu\nu}  \rangle_{ren} 
=
g^{\mu\nu}\left[ (\beta-\alpha)\tau\cdot \nabla \varphi + \gamma\tau\cdot  \nabla  \psi   + \alpha\Box\varphi - \gamma\Box\psi + \frac{1}{2} \nabla\varphi\cdot\nabla\varphi    -     \frac{1}{2} \nabla\psi\cdot\nabla\psi  \right] \nonumber \\ \nonumber
&+2(\alpha-\beta)\tau^{(\mu}\nabla^{\nu)} \varphi- 2\gamma \tau^{(\mu}\nabla^{\nu)} \psi - \nabla^\mu \varphi  \nabla^\nu \varphi  +     \nabla^\mu \psi  \nabla^\nu \psi - \alpha \nabla^\mu\nabla^\nu \varphi + \gamma \nabla^\mu\nabla^\nu \psi \\ 
&-\frac{\mathfrak{C}_2-c_2}{2}\left(g^{\mu\nu}\tau\cdot\tau   -  2\tau^\mu\tau^\nu  \right)
\, ,
\end{align}
where it is understood that $\varphi$ and $\psi$ satisfy Eq.~\eqref{eq:eom_scalars_2d_affine_torsion}, and the constants are determined by Eq.~\eqref{eq:mapping_constants_auxiliary_fields_affine_torsion} with some appropriate boundary conditions. We also used the fact that in $2$ dimensions the relation $R_{\mu\nu}=R/2 g_{\mu\nu}$ holds locally.
This emt displays an ``unphysical'' dependence on $\mathfrak{C}_2$, in the sense that $\mathfrak{C}_2$ is not determined, jeopardizing in principle the usefulness of the anomalous action.

Nevertheless, for consistency we now verify that  \eqref{eq:2d_Gamma_ind_auxiliary_fields_affine_torsion} reproduce the anomaly.
For convenience, in the following we will drop the suffix ``$ren$.''
The trace of Eq.~\eqref{eq:emt_auxiliary_fields_affine_torsion} gives
\begin{align}
-\langle T^{\mu}{}_{\mu}  \rangle 
=
C_4 \tilde{J}
+
c_1 \left(\nabla\cdot\tau\right) 
\, ,
\end{align}
where we used also the equation of motion of the scalar fields. For what concern the virial current, we get 
\begin{align}\label{eq:2d_div_virial_current_affine_torsion}
\langle \mathcal{D}^{\mu}  \rangle  &= \frac{1}{\sqrt{g}} \frac{\delta}{\delta \tau_{\mu}}   \Gamma_{NL}[g,\varphi, \psi]
=
(c_2-\mathfrak{C}_2)\tau^\mu
+
(\alpha-\beta) \nabla^\mu \varphi
-
\gamma   \nabla^\mu \psi
\, , \\ \nonumber
\nabla_\mu\langle \mathcal{D}^\mu \rangle
&=
\left( c_2 - c_1 \right) \left(\nabla\cdot\tau\right) 
+
\left( c_1 - C_4 \right) \tilde{J}    
\, ,
\end{align}
having used also the equations of motion of $\phi$.
From these results, we recover the anomaly in its original form, $ \mathcal{A} =-\langle T^{\mu}{}_{\mu}  \rangle + \nabla_\mu  \langle \mathcal{D}^{\mu} \rangle =c_1 \tilde{J} +c_2 \left( \nabla \cdot \tau \right)$. Moreover, by taking a covariant derivative of \eqref{eq:emt_auxiliary_fields_affine_torsion} (see Appendix~\ref{app:non_local_emt_Diff}), we verify that
\begin{align}
\nabla_{\mu}  \langle T^{\mu}{}_{\nu}\rangle
=
\langle\mathcal{D}^{\mu}\rangle     \Omega_{\mu\nu}
+
\tau_\nu \nabla_{\mu} \langle \mathcal{D}^{\mu}\rangle
\, ,
\end{align}
implying that Diff are not broken by our anomalous action.

Let us now turn to the computation of the general, i.e., without fixing any particular theory, contribution to the black hole entropy coming from \eqref{eq:2d_Gamma_ind_auxiliary_fields_affine_torsion}. To do this, we can use the Wald's entropy formula. In his original work \cite{Wald:1993nt}, Wald showed that also for higher derivative theories of gravity in $d \geq 4$ and admitting stationary black hole solutions, it is possible to construct a black hole entropy satisfying the second law of thermodynamics. In this approach, the entropy is given by $2\pi$ times an integral of the N\"other charge of Diff arising from the Killing vector that generates the horizon. Afterwards, in the paper \cite{Myers:1994sg} by Myers, the Wald's formalism has been shown to hold also for nonlocal actions. In particular, Myers showed that in $d=2$ we can safely use a localized action of the type that we have developed above in order to compute the entropy. In this case, the Wald's entropy can be written as  \cite{Myers:1994sg}
\begin{align}\label{eq:2d_Wald:entropy}
S
=
-2\pi \left. \left(
\frac{\partial L}{\partial R_{\alpha\beta\mu\nu}} 
\right)
 \epsilon_{\alpha\beta} \epsilon_{\mu\nu} \right|_{x=x_h}
\,
\end{align}
where $x_h$ is the horizon location, $L=L(\varphi, \partial \varphi, \phi, \partial \psi, g, \partial g)$ is the Lagrangian, and $\epsilon$ is an antisymmetric tensor normalized as $\epsilon^{\mu\nu}\epsilon_{\mu\nu}=2$. Accordingly, we obtain contributions to the entropy only from the curvature dependent terms $ \alpha/2 \, \varphi R \subset \alpha \varphi \tilde{J} $ and $\gamma/2 \,  \psi R \subset \gamma \psi \tilde{J} $. A simple computation shows that
\begin{align}
S_{\varphi \tilde{J}}
=
\left.
- 2 \pi \alpha \varphi \right|_{x=x_h}
\, ,
\qquad 
S_{\psi \tilde{J}}
=
\left.
 2 \pi \gamma \psi \right|_{x=x_h}
\, ,
\end{align}
where $\varphi$ and $\psi$ are defined as in \eqref{eq:eom_scalars_2d_affine_torsion}. Notice the different signs in the two contributions. Collecting these results and using the definition of $\tilde{J}$, we obtain
\begin{align}\label{eq:entropy_affine_torsion_2d_with_C4}
\frac{1}{2\pi}S_{\varphi \tilde{J}+\psi \tilde{J}}
=
\left.
C_4 \frac{1}{\Delta_2}  J  
\, \right|_{x=x_h} 
-
\left.
 \left(C_4  - c_1 \right) \frac{1}{\Delta_2} \left(\nabla\cdot\tau\right)   
\, \right|_{x=x_h}
\, .
\end{align}
The second term is an entirely new contribution to the entropy coming only from torsion. On the one hand, it is amusing that the Wald entropy is independent from $\mathfrak{C}_2$, which should not parametrize any observable, but, on the other hand, it depends on the coupling constant $C_4$ of the conformal action in
\eqref{eq:2d_Gamma_ind_symm_form_affine_torsion}, which represents the part of the effective action that is not determined by the integration of the anomaly.  Notice that the cancellation of $\mathfrak{C}_2$ in Eq.~\eqref{eq:entropy_affine_torsion_2d_with_C4} is only possible if we retain $\psi$, i.e., if the Weyl invariant action in Eq.~\eqref{eq:2d_Gamma_ind_symm_form_affine_torsion} is included. By setting $C_4=0$ and, accordingly, $\gamma=\alpha$, the entropy simplifies considerably  
\begin{align}\label{eq:entropy_affine_torsion_2d}
\frac{1}{2\pi}S_{\varphi \tilde{J}+\psi \tilde{J}}
=
\left.
  c_1  \frac{1}{\Delta_2} \left(\nabla\cdot\tau\right)   
\, \right|_{x=x_h}
=
  c_1   \int {\rm d}^2y \sqrt{g} G(x_h,y) \left(\nabla\cdot\tau\right)(y)
\, .
\end{align}
Although the simple computations leading to the previous formula are enough to ensure independence of $S_{\varphi \tilde{J}+\psi \tilde{J}}$ on $\mathfrak{C}_2$, it is rather difficult to draw any physical conclusion from \eqref{eq:entropy_affine_torsion_2d} as it stands. On the one hand we still have the famous ambiguity of adding $\Gamma_{conf}$, as parametrized by the dependence on $C_4$ in \eqref{eq:entropy_affine_torsion_2d_with_C4}. On the other hand, it is well known that the Green's function depends on the boundary conditions and therefore \eqref{eq:entropy_affine_torsion_2d} on the quantum state. Nevertheless, $c_1$ is state independent and determined only by the field content of the model. We give additional comments on this problem in Sect.~\ref{sect:Conclusions} and Subsect.~\ref{sect_BH_thermo_and_2d_emt_invariant_torsion}. In the latter, we compute the Wald's entropy for Rindler space.

\subsection{Cohomological analysis for invariant torsion in $d=2$ and integration of the anomaly}\label{sect_2d_cohomological_analysis_invariant_torsion}

We now concentrate on the case in which $\delta_\sigma \tau_\mu=0$.
In the case of conformally invariant torsion, the coboundary operator simplifies to
\begin{equation}
\delta_\sigma
=
2\int {\rm d}^2x \, \sigma \, g_{\mu\nu} \frac{\delta}{\delta g_{\mu\nu}} \, ,
\end{equation}
 which could be seen as the $b\to 0$ limit of the affinely transforming case of the previous subsection.
Therefore, we now have
\begin{align}\label{eq:Asigma_invariant_tau}
\delta_\sigma \Gamma
=
- \int {\rm d}^dx \sqrt{g} 
\sigma \langle T^{\mu}{}_{\mu} \rangle 
 =
 A_\sigma
 \, .
\end{align}
For invariant torsion the modified curvature tensors are not particularly useful. Therefore, as a basis for the $1$-cochains we use
\begin{align}\label{eq:basis_1cochain_inv_torsion}
\omega_1= \int {\rm d}^2x \sqrt{g} \sigma {J} \, ,
\qquad
\omega_2= \int {\rm d}^2x \sqrt{g} \sigma    \left(\nabla \cdot \tau  \right) \, ,
\qquad
\omega_3= \int {\rm d}^2x \sqrt{g} \sigma    \left(\tau \cdot \tau  \right)
\, .
\end{align}
A simple calculation shows that in $d=2$ 
\begin{align}
\delta_\sigma  \omega_i[\sigma; g, \tau] 
=
0 \,, \quad i=1,2,3  \, ,
\end{align}
and, accordingly, in the anomaly we have three $1$-cocycles
\begin{align}\label{eq:2d_anomaly_invariant_torsion}
A_\sigma
=
\int \sqrt{g}  \sigma \Big\{
c_1 J 
+
c_2 \left( \nabla \cdot \tau \right) 
+
 c_3 \left( \tau \cdot \tau \right) \Big\} 
 \, ,
\end{align}
where the coefficients can be computed with covariant techniques for given specific models. We stress that the heat kernel constraints $c_1$ to the positive number $\Tr \hat{1}$ (up to a multiplicative positive constant), where $\hat{1}$ is the identity in field space.

The analysis of trivial anomalies proceeds in a completely analogous way as in the previous subsection. It is simple to check that for invariant torsion 
\begin{align}
\delta \xi_i=0, \qquad \text{for \,} i=1,2,3 \, ,
\end{align}
where the $\xi_i$ are obtained by removing $\sigma$ from the $1$-cochains given in \eqref{eq:basis_1cochain_inv_torsion}. Therefore, there are now only nontrivial anomalies. In particular, even though $\nabla\cdot\tau$ is still a boundary term, it cannot be interpreted as an $a'$-anomaly since now it is not exact.
As a double check, we can consider the following conformal action
\begin{align}
S=\frac{1}{2} \int {\rm d}^2x \sqrt{g}  \left( \nabla^\mu \phi \nabla_\mu\phi+\xi_1 \left( \nabla \cdot \tau \right) \phi^2+\xi_2  \left( \tau \cdot \tau \right) \phi^2\right)
\, , 
\end{align}
from which, using the results \cite{Barvinsky:1985an}, it is simple to see that $\mathcal{A} \propto \Tr a_1 $ has precisely the form \eqref{eq:2d_anomaly_invariant_torsion}.

Let us now tackle the problem of integrating the anomaly. We express it as
\begin{align}\label{eq:2d_emt_invariant_torsion}
\mathcal{A}
=
\langle T^{\mu}{}_{\mu} \rangle 
=
c_1 J 
+
c_2 \left( \nabla \cdot \tau \right) 
+
 c_3 \left( \tau \cdot \tau \right) 
  \, ,
\end{align}
which can be easily integrated by using the finite conformal transformations
\begin{align}
\sqrt{g} \left( \nabla \cdot \tau \right)
=
\sqrt{g'} \left( \nabla' \cdot \tau \right)
\, , \qquad 
\sqrt{g} \left( \tau \cdot \tau \right)
=
\sqrt{g'} \left( \tau \cdot \tau \right)
\, , \qquad 
\sqrt{g}J=\sqrt{g'}J' -  \Delta'_2 \sigma 
\, .
\end{align}
Differently from the case of the previous subsection, we do not have anomalies of mixed nature. Indeed, these equations correctly suggest to identify $ \left( \nabla \cdot \tau \right) $ and $\left( \tau \cdot \tau \right)$ with $b$-anomalies, while $J$ is the standard two-dimensional $a$-anomaly. In particular, we have already discussed that $\left( \nabla \cdot \tau \right)$ cannot be thought of as an $ a'$-anomaly. Cohomologically, we have that $\left( \nabla \cdot \tau \right)$, $ \left( \tau \cdot \tau \right)$ and $J$ belong to three different de Rham equivalence classes and so the first two fit in different Weyl sub-classes even though they transform in the same way. 
Thus, we have that 
\begin{align}\label{eq:2d_Gamma_ind_invariant_torsion}
\Gamma_{WZ}[\sigma; g',\tau]
=
 \int {\rm d}^2x \sqrt{g'}  \sigma \Big\{
c_1  J' 
-
\frac{1}{2}c_1 \Delta'_2 \sigma
+
 c_2 \left( \nabla' \cdot \tau \right)
+
c_3 \left( \tau \cdot \tau \right)  \Big\}\, 
\end{align}
is such that $\frac{\delta}{\sqrt{g}\delta\sigma}\Gamma_{WZ}=\langle T^{\mu}{}_{\mu} \rangle $, where the right hand side is given in Eq.~\eqref{eq:2d_emt_invariant_torsion}. Notice that by choosing a basis containing some linear combination of $J$ with the invariants $(\nabla \cdot \tau)$ or $(\tau \cdot \tau)$ in \eqref{eq:2d_emt_invariant_torsion} we would change the de Rham equivalence class but not the nature of the Weyl subclasses. For example, a term like $J + (\tau\cdot\tau) \notin [J]_{dR}$ but  $J \sim J + (\tau\cdot\tau) $ and so they are integrated analogously. Of course, ambiguities of this kind are not problematic since they just correspond to different parametrizations of the anomaly. 

The nonlocal action is easy to obtain and it reads (see also Appendix \ref{app:non_local_actions})
\begin{align}\label{eq:2d_Gamma_nonlocal_invariant_torsion}
\Gamma_{NL}[g,\tau]
=
- \int {\rm d}^2x \sqrt{g}   \Bigg\{ 
\Bigg(   \frac{c_1}{2}  J 
+
  c_2 \left(  \nabla \cdot \tau  \right)
+
  c_3 \left(  \tau \cdot \tau  \right)
 \Bigg) \frac{1}{\Delta_2} J
\Bigg\} 
+
\Gamma_{c}[g,\tau]
 \, .
\end{align}
We get the Polyakov action plus some torsion dependent terms. Notice that, as usual, we have an undetermined conformal integration constant. Therefore, even if we do not encounter the ambiguity we have seen earlier, the integration of the anomaly in the presence of torsion is less powerful than the purely metric case.

\subsubsection{Localized action, energy momentum tensor and Wald's entropy}\label{sect_BH_thermo_and_2d_emt_invariant_torsion}
In order to write the effective action in terms of auxiliary scalar fields, we start by writing \eqref{eq:2d_Gamma_nonlocal_invariant_torsion} in the symmetric form 
 \begin{align}\label{eq:2d_Gamma_ind_symm_form_invariant_torsion}
\Gamma_{NL}[g,\tau]
=
&-\frac{c_1}{4} \int {\rm d}^2x \sqrt{g}   \Bigg\{ 
\left[ J + \frac{2c_2}{c_1} \left(\nabla\cdot\tau\right) \right]    \frac{1}{\Delta_2}  \left[   J + \frac{2c_2}{c_1} \left(\nabla\cdot\tau\right)   \right]   
+
\left[ J + \frac{2c_3}{c_1} \left(\tau\cdot\tau\right) \right]
\frac{1}{\Delta_2}  \left[   J + \frac{2c_3}{c_1} \left(\tau\cdot\tau\right)   \right]  
\Bigg\}   \nonumber \\ 
&
+ \frac{1}{c_1} \int {\rm d}^2x \sqrt{g}   \Bigg\{ 
c^2_2 \left(\nabla\cdot\tau\right)    \frac{1}{\Delta_2} \left(\nabla\cdot\tau\right)   
+ c^2_3    \left(\tau\cdot\tau\right)      \frac{1}{\Delta_2}    \left(\tau\cdot\tau\right) 
\Bigg\}
 \, .
\end{align}
Therefore, for the purpose of localizing the action, we now need four scalar fields $\varphi_1$, $\psi_1$, and  $\varphi_2$, $\psi_2$. The localized action is
\begin{align}\label{eq:2d_Gamma_ind_auxiliary_fields_invariant_torsion}
\Gamma_{int}[g,\varphi, \psi]
=
\sum_{i=1}^2 \int {\rm d}^2x \sqrt{g}  &\Bigg\{ 
 \frac{1}{2} \varphi_i \Delta_2 \varphi_i 
+
\alpha_i \varphi_i {J} 
+
\beta_i \varphi_i \mathcal{T}_i 
-
\frac{1}{2} \psi_i \Delta_2 \psi_i 
-
\beta_i \psi_i \mathcal{T}_i 
\Bigg\}
 \, , \\ \nonumber
& 
\mathcal{T}_1=\nabla\cdot\tau \, , 
\qquad
\mathcal{T}_2=   \tau\cdot\tau
 \, ,
\end{align}
and the scalars satisfy the following equations of motion
\begin{align}\label{eq:eom_scalars_2d_invariant_torsion}
\Delta_2 \varphi_i 
=
-
\alpha_i {J} 
-
 \beta_i    \mathcal{T}_i
\, ,
\qquad 
\Delta_2 \psi_i
=
-
\beta_i  \mathcal{T}_i
\, ,
\qquad i=1,2
\, ,
\end{align}
where the abbreviations $ \mathcal{T}_i$ are defined in \eqref{eq:2d_Gamma_ind_auxiliary_fields_invariant_torsion}. Going on-shell in \eqref{eq:2d_Gamma_ind_auxiliary_fields_invariant_torsion} gives the action
\begin{align}
\Gamma_{int}[g,\varphi, \psi]
=
-\sum_{i=1}^2  \frac{1}{2}  \int {\rm d}^2x \sqrt{g}  \Bigg\{ 
 \left( \alpha_i {J} + \beta_i    \mathcal{T}_i \right)   \frac{1}{\Delta_2}   \left( \alpha_i {J} + \beta_i    \mathcal{T}_i \right) 
-
 \left(   \beta_i    \mathcal{T}_i      \right)   \frac{1}{\Delta_2}   \left(    \beta_i    \mathcal{T}_i    \right) 
\Bigg\} 
 \, ,
\end{align}
which reproduces the symmetric form \eqref{eq:2d_Gamma_ind_symm_form_invariant_torsion} by choosing 
\begin{align}
\alpha_1
=
\alpha_2
=\sqrt{\frac{c_1}{2}}
\, ,
\qquad
\beta_1=\sqrt{\frac{2}{c_1}} \, c_2 
\, ,
\qquad
\beta_2=\sqrt{\frac{2}{c_1}} \, c_3 
\, ,
\end{align}
which are real numbers.

As outlined at the end of Subsect.~\ref{sect_BH_thermo_and_2d_emt_affine_torsion}, we adopt an ``agnostic'' perspective on the auxiliary fields in this paper, intentionally refraining from assigning them any physical interpretation. Instead, we regard them as purely auxiliary quantities that aid intermediate calculations and ultimately disappear from the final results via their equations of motion, i.e., by going on-shell. We adopt this interpretation because, to the best of our knowledge, there is no general consensus in the literature regarding both the physical meaning and the precise number of necessary extra fields needed to localize the action. In this analysis, we have included the \emph{maximum} number $n$ of auxiliary fields, given by $n=2(w-1)$, where $w$ denotes the number of nontrivial Weyl cocycles. 
Although this ``maximal'' approach may introduce some redundancy, it allows to avoid the overlooking of any potentially significant physical information. Alternative, more minimal, choices that still reproduce the nonlocal actions are explored in detail in Appendix~\ref{app:local_actions_auxiliary_fields}.

From Eq.~\eqref{eq:2d_Gamma_ind_auxiliary_fields_invariant_torsion_1_auxiliary_field}, we obtain the following renormalized emt  and virial current by functional differentiating w.r.t.\ the metric and the torsion respectively
\begin{align}
\langle T^{\mu\nu}  \rangle
 &=
g^{\mu\nu}\Big[  \alpha_1\Box\varphi_1  + \frac{1}{2} \nabla\varphi_1\cdot\nabla\varphi_1    -     \frac{1}{2} \nabla\psi_1\cdot\nabla\psi_1  \Big] - \nabla^\mu \varphi_1  \nabla^\nu \varphi_1  +     \nabla^\mu \psi_1  \nabla^\nu \psi_1+ \alpha_1 \nabla^\mu\nabla^\nu \varphi_1  \\ 
&+\left(\alpha_1 \rightarrow \alpha_2 \, , \varphi_1   \rightarrow \varphi_2 \, , \psi_1 \rightarrow \psi_2 \right) - 2\beta_1\left( \tau^{(\mu}\nabla^{\nu)} \varphi_1 - \tau^{(\mu}\nabla^{\nu)} \psi_1 \right) + \beta_1 g^{\mu\nu}\left[ \tau \cdot \nabla \varphi_1 - \tau \cdot \nabla \psi_1     \right] \nonumber \\
&+\beta_2    \left( 2\tau^\mu\tau^\nu \varphi_2  - 2\tau^\mu\tau^\nu \psi_2  - g^{\mu\nu}   \left( \tau \cdot  \tau \right) \varphi_2  + g^{\mu\nu}  \left( \tau \cdot  \tau \right) \psi_2    \right)
\, , \nonumber \\ 
%
\langle \mathcal{D}^{\mu}  \rangle 
&=
2\beta_2  \tau^\mu  \left(   \varphi_2  +  \psi_2    \right)
-
\beta_1 \nabla^\mu \left(  \varphi_1  +  \psi_1 \right)
\, .
\end{align}
In these expressions the scalar fields should be taken to satisfy the equations of motion \eqref{eq:eom_scalars_2d_invariant_torsion} solved on a particular background. We come back on this point in the $d=4$ case.
It is important to stress that now the coefficients in the emt depend only on those of the anomaly.
It is just a matter of algebra to verify that the trace of this emt gives the anomaly \eqref{eq:2d_emt_invariant_torsion} and the Diff are not broken at the quantum level (see Appendix \ref{app:non_local_emt_Diff}).

Let us now briefly turn our attention to the black hole entropy. Again, only the curvature dependent terms contribute to the Wald entropy. Accordingly, we get the quantum contributions
\begin{align}
S_{\varphi_i {J}}
=
\left.
- 2 \pi \alpha_i \varphi_i \right|_{x=x_h}
\, ,
\qquad i=1,2
\, .
\end{align}
We recall that $x_h$ is the horizon location. Putting everything together, we get
\begin{align}\label{eq:entropy_corrections_2d_inv_torsion}
\frac{1}{2\pi}S_{\varphi_{1,2} {J}}
=
-\sum_{i=1}^2 \left.\alpha_i \frac{1}{\Delta_2}  \left( - \alpha_i J - \beta_i \mathcal{T}_i \right)\right|_{x=x_g}
=
\left.\Bigg\{
c_1\frac{1}{\Delta_2}  J  
+
c_2 \frac{1}{\Delta_2}  \nabla\cdot\tau
+
c_3 \frac{1}{\Delta_2}  \tau\cdot\tau  
\,
\Bigg\} \right|_{x=x_g}
\, .
\end{align}
The first term is the contributions coming from the Polyakov action, which has been evaluated on a torsionless background in \cite{Myers:1994sg}.

For concreteness, let us consider the example of a $2d$ Euclidean Rindler space. This can be interesting because such a space provides an approximation of the near horizon geometry of a black hole. Moreover, being Rindler space flat, we could expect the corrections \eqref{eq:entropy_corrections_2d_inv_torsion} to the black hole entropy to be zero. 
The Euclidean Rindler metric reads $ds^2=dr^2+ \left(\frac{1}{\beta_H}\right)^2 r^2d\eta^2$, in which $\eta$ is the Euclidean time and $\beta_H$ is the inverse Hawking temperature, which is defined by the requirement that there is no conical singularity at $r=0$. We now take $\eta$ to be periodic with period $2\pi \beta$, for $\beta \neq \beta_H$ \cite{Solodukhin:1994yz}. By performing the change of variables $\theta=\beta^{-1} \eta$, the Rindler metric becomes
\begin{align}\label{eq:rindler_metric_off_shell}
ds^2=dr^2+ \alpha^2 r^2d\theta^2 
\, ,
\qquad \alpha=\frac{\beta}{\beta_H} 
\, .
\end{align}
Of course, the previous equation displays the well-known conical singularity at the origin $r=0$, which correctly disappears for $\beta=\beta_H$  \cite{Solodukhin:1994yz}. By using an auxiliary metric, it is possible to see that the scalar curvature has the form $R=\frac{2(\alpha-1)}{\alpha}\delta(r)$. In Ref.~\cite{Solodukhin:1994yz} it is also shown that the Green's function of the Laplacian on a Rindler geometry with metric \eqref{eq:rindler_metric_off_shell} is
\begin{align}
G(\vec{r},\vec{r}_1\,)
=
\frac{2(\alpha-1)}{\alpha} \ln \abs{ \vec{r}-\vec{r}_1  }
\, ,
\end{align}
up to harmonic functions. In the previous equation $\abs{ \vec{r}-\vec{r}_1} = r^2 + r_1^2 - 2 r r_1  \cos \theta$. On the Rindler space, let us also consider a spherically symmetric radial vector field $\tau$ vanishing at infinity with the power law
\begin{align}
\tau_r
\sim
\frac{(\alpha-1)}{r^{1+\epsilon}}
\, .
\end{align}
We have inserted the factor $(\alpha-1)$ because, for $\alpha=1$, the background geometry then becomes that of a Wick rotated $2d$ Minkowski space for which we choose the torsion to vanish.
Thus, the contributions to the entropy from the Wald's formula are
\begin{align}\label{eq:entropy_correction_1}
 \left. \frac{1}{\Delta_2}  \tau\cdot\tau  \right|_{r_1=0}
=
\frac{2(\alpha-1)^3}{\alpha}  \int_{0}^{2\pi} \int_{a}^{\infty}  r \alpha  {\rm d}\theta{\rm d}r \,\frac{\ln r}{r^{2(1+\epsilon)}}
=
(\alpha-1)^3 \, \frac{\pi\epsilon^{-2}(1 + 2\epsilon \ln a)}{a^{2\epsilon}} 
\, ,  
\end{align}
and, by using $\nabla \cdot \tau = g^{-1/2}\partial_r \left(g^{1/2}  \tau^r  \right)    = (r \alpha)^{-1}\partial_r\left(r \alpha  \frac{\alpha-1}{r^{1+\epsilon}}  \right)$, we get
\begin{align}\label{eq:entropy_correction_2}
\left. \frac{1}{\Delta_2}  \nabla\cdot\tau  \right|_{r_1=0}
=
-\frac{2(\alpha-1)^2(1 + \epsilon)}{\alpha}  \int_{0}^{2\pi} \int_{a}^{\infty}  r \alpha  {\rm d}\theta{\rm d}r \,\frac{\ln r}{r^{2(1+\epsilon)}}
=
-(\alpha-1)^2 \, \frac{4\pi\epsilon^{-2}(1 + \epsilon)(1 + \epsilon \ln a)}{a^{\epsilon}} 
\, .
\end{align}
Some comments are in order. To begin with, one might wonder which is the quantum state selected by this procedure. Physically, to choose $\beta=\beta_H$ as periodicity, means that an observer at $r=1/a$ measures the temperature $T=\frac{a}{2\pi}$. Therefore, with the identification $\alpha=1$, we are implicitly choosing to be in the Rindler thermal state, which can be shown to be the Minkowski vacuum \cite{Hartman_book}. Moreover, in Eqs.~\eqref{eq:entropy_correction_1} and \eqref{eq:entropy_correction_2} we have a logarithmic divergence as we approach the tip of the cone, i.e., $a \to 0$. However, when we take $\alpha=1$ to remove the conical singularity, these expressions are actually zero. As anticipated, we obtain that, on the Rindler geometry, the contribution to the entropy coming from the anomalous action is zero. Of course this does not mean that the total entropy is zero since the leading contribution is still there.

It would be of interest to consider a more realistic set up. A more correct way to proceed (and generalizing the results of \cite{Myers:1994sg}), would be to start with a model of $2d$ gravity with dynamical torsion, presumably extending dilaton gravity to the presence of torsion.  It would then be necessary to obtain a static, spherically symmetric solution on which to evaluate the Green's function of $\Delta_2$ with appropriate boundary conditions, an endeavor we plan to undertake in the future.

\section{Weyl cohomology in $d=4$ for invariant and affinely transforming torsion}\label{sect_4d_cohomological_analisys}

%

For convenience, we start by recalling the standard result for the purely metric case (i.e., without torsion), and then we include the torsion vector with the two possible transformations in the next subsections.
We can construct the following basis in $d=4$ for the possible $1$-cochains $\omega_i[\sigma; g]$ involving only curvatures of the metric
\begin{align}\label{eq:1-cochains-curvatures}
&\omega_1 =\int {\rm d}^4x \sqrt{g} \sigma W^{\alpha\beta\rho\gamma} W_{\alpha\beta\rho\gamma} \, ,
&& \omega_2 =\int {\rm d}^4x \sqrt{g} \sigma K^2_{\mu\nu}   \, ,
\nonumber\\
&  \omega_3 =\int {\rm d}^4x \sqrt{g} \sigma J^2    \, ,
&& \omega_4 = \int {\rm d}^4x \sqrt{g} \sigma   \Box J \, .
\end{align}
Therefore, we can parametrize $\omega=\sum_{i=1}^4 c_i \omega_i[\sigma; g]$ and thus Eq.~\eqref{eq:1-cocycle} becomes
\begin{align}
\delta_\sigma \sum_{i} c_i \omega_i[\sigma, g] 
=
0 \, .
\end{align}
Computing the variations $\delta_\sigma$ (where $\delta_\sigma$ can equivalently be \eqref{eq:coboundary_operator} or \eqref{eq:coboundary_operator_inv_torsion} since \eqref{eq:1-cochains-curvatures} do not depend on $\tau_\mu$) and using the anticommuting nature of $\sigma$ along with integration by parts, it is easy to see that the nonzero contributions are
\begin{align*}
\delta_\sigma \omega_2[\sigma, g]
&=
- 2 \int {\rm d}^4x \sqrt{g}  \sigma \nabla_\mu \sigma (\nabla^\mu J )
=
- 2 \int {\rm d}^4x \sqrt{g}\sigma \nabla_\mu \sigma \left( (\nabla^\mu \tilde{J} ) - \nabla^\mu \nabla \cdot \tau  + \tau^\nu \nabla^\mu\tau_\nu   \right)
\, ,
 \\ \nonumber
\delta_\sigma \omega_3[\sigma, g]
&
=
- 2 \int {\rm d}^4x \sqrt{g}  \sigma \nabla_\mu \sigma (\nabla^\mu J )
=
- 2 \int {\rm d}^4x \sqrt{g}\sigma \nabla_\mu \sigma \left( (\nabla^\mu \tilde{J} ) - \nabla^\mu \nabla \cdot \tau  + \tau^\nu \nabla^\mu\tau_\nu   \right)
\, .
\end{align*}
Consequently, the consistency condition imposes $c_2=-c_3$, so that the two variations cancel each other. Recalling the form of the Euler density in terms of the Shouten tensor and its trace, $E_4=8 (J^2 - K^2_{\mu\nu}) +W^2_{\alpha\beta\rho\gamma} $, the consistency conditions dictates the celebrated fact that curvature terms must appear in the anomaly only as $W^2$, $E_4$ and $\Box J$ for the Weyl group \cite{Bonora:1985cq}. It is also easy to see that the only trivial anomaly is the total derivative term. In particular, we have
\begin{align*}
\frac{2g_{\mu\nu}  }{\sqrt{g} } \frac{\delta}{\delta g_{\mu\nu}} \int d^4 x \sqrt{g} J^2 
=
 -2 \square J
\, .
\end{align*}
Including trivial anomalies and renaming in a more traditional way the coefficients so to highlight the type of anomaly, we obtain the following purely metric contribution to the anomaly
\begin{align}\label{eq:4d_curvature_dependendent_anomaly}
A^g_\sigma
=
\int\sqrt{g}  \sigma \left\{
b W^2_{\alpha\beta\rho\gamma}
+
a  E_4 
+
a' \Box R
\right\}
\, .
\end{align}
As a final remark, we notice that for purely metric case the easiest basis to work with would contain precisely $(W^2, E_4,\Box J)$. However, because of the basis we will choose for the $\tau$-dependent 1-cochains, we take that of Eq.~\eqref{eq:1-cochains-curvatures}.

\subsection{Cohomological analysis for the torsion transforming affinely in $d=4$}\label{sect_4d_cohomological_analisys_affine_torsion}

We now consider $1$-cochains $\omega_i[\sigma; g, \tau]$ depending on the torsion vector too. It is convenient to use a basis involving the modified conformal tensors of the subsect.~\ref{subsect_modified_conf_tensors}. For example, we immediately have 
\begin{align}
\delta_\sigma \int {\rm d}^4x \sqrt{g}  \sigma \Omega^2_{\mu\nu} 
=
\delta_\sigma \int {\rm d}^4x \sqrt{g}  \sigma \tilde{J}^2
=
\delta_\sigma \int {\rm d}^4x \sqrt{g}  \sigma  \tilde{K}^2_{\mu\nu} 
=
\delta_\sigma \int {\rm d}^4x \sqrt{g}  \sigma \tilde{\Box} \tilde{J}
=
0
\, ,
\end{align}
so that they are allowed in the anomaly without any constraint on their coefficients. We can choose the following minimal basis for the $1$-cochains (see Appendix \ref{app:basis_cochains_and_variations})
\begin{align}\label{eq:1-cochains-bulk-tau}
&\omega_5 =\int {\rm d}^4x \sqrt{g} \sigma \left(\tau \cdot \tau \right)^2 \, , 
&&\omega_6 =\int {\rm d}^4x \sqrt{g} \sigma \tilde{J}^2  \, ,
&& \omega_7 =\int {\rm d}^4x \sqrt{g}    \sigma \tilde{K}^2_{\mu\nu}   \, ,
\nonumber\\
&    \omega_8 =\int {\rm d}^4x \sqrt{g}    \sigma \Omega^2_{\mu\nu}         \, ,
&&  \omega_9 =\int {\rm d}^4x \sqrt{g}    \sigma \tilde{J} \left(\tau \cdot \tau \right)   \, ,
&&    \omega_{10} =\int {\rm d}^4x \sqrt{g} \sigma  \left( \nabla\tilde{J} \right) \cdot \tau   \, ,
\nonumber\\
&  \omega_{11} = \int {\rm d}^4x \sqrt{g} \sigma  \tilde{K}^{\mu\nu} \tau_\mu \tau_\nu  \, ,
&& \omega_{12} = \int {\rm d}^4x \sqrt{g} \sigma \,  \tau \cdot \tau  \left(\nabla \cdot \tau \right) \,,
&&   \omega_{13} = \int {\rm d}^4x \sqrt{g} \sigma \left(\nabla \cdot \tau \right)^2     \, ,
\nonumber\\
&    \omega_{14} = \int {\rm d}^4x \sqrt{g} \sigma  \left(\tau^\mu \tau^\nu \right)   \nabla_\mu  \tau_\nu   \, ,
&&     \omega_{15} = \int {\rm d}^4x \sqrt{g} \sigma \,     \tau_\mu \Box \tau^\mu \, ,
&& \omega_{16} = \int {\rm d}^4x \sqrt{g} \sigma \,   \left( \nabla_\mu \tau_\nu   \right)^2  \, ,
\nonumber\\
& \omega_{17} = \int {\rm d}^4x \sqrt{g} \sigma \,   \tau^\mu \nabla_\mu \nabla\cdot \tau         \, , 
&& \omega_{18}=\int {\rm d}^4x \sqrt{g}    \sigma   \tilde{\Box} \tilde{J}  \, \,,
\end{align}
where $ \left( \nabla\tilde{J} \right) \cdot \tau = \left(\nabla_\mu \tilde{J}\right) \tau^\mu$. The label of the new terms runs from $5$ to $18$ as they should be added to the original purely metric-dependent basis.
By including all terms, we get the consistency condition 
\begin{align}
\delta_\sigma \left(  
 \sum_{i}^4 c_i \omega_i[\sigma, g]
 +\sum_{i=1}^{14} f_i \omega_{i+4}[\sigma; g, \tau]
 \right)
 =0 \, .
\end{align}
We can exploit the basis \eqref{eq:1_basis_2_cochains} to get
\begin{align}\label{eq:c_conditions4d}
\sum_{j,n}
\int {\rm d}^4x \sqrt{g}  \Bigg\{ h_j(f_i,c_k) \left(\nabla_\mu \sigma \Box\sigma\right)  w_j^{\mu}
+
l_n(f_i,c_k)  \left(\sigma \nabla_\mu \sigma \right) v_n^{\mu}
\Bigg\}
=
0 
\quad
\Longrightarrow
\quad
h_j= 0 \, ,\quad  l_n= 0 \, ,
\end{align}
and therefore the relations (see also Appendix \ref{app:basis_cochains_and_variations})
\begin{align}
& f_1 =\frac{f_6}{2b^3}\, ,
&&  f_5 = \frac{f_6}{b} \, ,
&&   f_7=0  \, ,
&&  f_8 =   \frac{f_{10} }{2} - \frac{f_6}{b^2} \, ,
\nonumber\\
&   f_9= \frac{f_6}{2b} - f_{11}    \, ,
&& f_{12} = f_{11}  \, ,
&& f_{13} = - f_{11}   \, ,
&&    c_2=  - \frac{bf_6}{2} - c_1  \, ,
\nonumber
\end{align}
while the subset of constants $ f_2 \, , f_3 \, , f_4 \, , f_{14} $ is obviously left arbitrary, i.e., they are not fixed by the consistency requirement. Up to already known purely-metric contributions $A^g_\sigma$ of \eqref{eq:4d_curvature_dependendent_anomaly}, the resulting torsion-dependent anomaly $A^{\tau}_\sigma$ becomes 
\begin{align}\label{eq:anomaly_affine_torsion}
A^{\tau}_\sigma
=
\int\sqrt{g}  \sigma \Bigg\{&
\left(f_2 - \frac{bf_6}{2}  \right) \tilde{J}^2
+
f_3  \tilde{K}^2_{\mu\nu} 
+
f_4 \Omega^2_{\mu\nu}
+
f_{14}  \tilde{\Box} \tilde{J} 
+
f_6 \nabla_\mu  \left(    \tilde{J}\tau^\mu \right)
+
\frac{f_{10}}{2}  \nabla_\mu \left(  \tau^\mu \,\tau \cdot \tau     \right)
  \\ \nonumber
&+
f_{11} \left(  \frac{1}{2} \Box    \tau^2     - \nabla_\mu \left( \tau^\mu \,\nabla \cdot \tau \right)    \right)
\Bigg\}
\, .
\end{align}
In the following, we redefine $\left(f_2 - \frac{bf_6}{2}  \right) \equiv f_2$ to simplify the notation.

We now turn to the study of trivial anomalies, which begins by constructing a basis for the $0$-cochains. The procedure is carried out in Appendix \ref{app:basis_cochains_and_variations} and simply amounts to ``remove'' $\sigma$ and perform integration by parts, as done in the $d=2$ case. The result is that we can choose the following basis
\begin{align}
&\xi_1 =\int {\rm d}^4x \sqrt{g}  \left(\tau \cdot \tau \right)^2 \, ,
&&  \xi_2 =\int {\rm d}^4x \sqrt{g}    \left(\nabla \cdot \tilde{J}  \right)\tau \, ,
&& \xi_3 =\int {\rm d}^4x \sqrt{g}     \tilde{J} \left(\tau \cdot \tau \right)  \, ,
\nonumber\\
&   \xi_4 = \int {\rm d}^4x \sqrt{g}   {\tilde{K}}^{\mu\nu} \tau_\mu \tau_\nu  \, ,
&& \xi_5 = \int {\rm d}^4x \sqrt{g}  \,  \tau \cdot \tau  \left(\nabla \cdot \tau \right) \, ,
&& \xi_6 = \int {\rm d}^4x \sqrt{g}  \tau^\mu \nabla_\mu \nabla\cdot \tau \, ,
\nonumber\\
&    \xi_{7} = \int {\rm d}^4x \sqrt{g}  \,   \left( \nabla_\mu \tau_\nu   \right)^2\, ,
\nonumber
\end{align}
the elements of which satisfy the right properties listed above. When computing the variation $\delta_\sigma$, we integrate by parts as much as possible, so that no derivatives act on $\sigma$. We report here the result in terms of functional derivatives 
\begin{align}\label{eq:4d_trivial_anomalies_affine_torsion}
&\frac{1}{\sqrt{g}} \left\{ 2 g_{\mu\nu} \frac{ \delta}{\delta g_{\mu\nu}} - b \nabla_\mu   \frac{ \delta}{\delta \tau_{\mu}}   \right\} 
 \xi_3 
=
-2b\nabla_\mu  \left(    \tilde{J}\tau^\mu \right)
 \, , \\ \nonumber     
& \frac{1}{\sqrt{g}}  \left\{ 2 g_{\mu\nu} \frac{ \delta}{\delta g_{\mu\nu}} - b \nabla_\mu   \frac{ \delta}{\delta \tau_{\mu}}   \right\} \left(  \frac{1}{2b}  \xi_5    -\frac{1}{4 b^2}  \xi_1 \right)  
=
  \left(  \frac{1}{2} \Box    \tau^2     - \nabla_\mu \left( \tau^\mu \,\nabla \cdot \tau \right)    \right) 
 \, , \\ \nonumber     
& \frac{1}{\sqrt{g}}  \left\{ 2 g_{\mu\nu} \frac{ \delta}{\delta g_{\mu\nu}} - b \nabla_\mu   \frac{ \delta}{\delta \tau_{\mu}}   \right\} 
 \xi_1
=
-4b \nabla_\mu  \left( \tau^\mu  \tau\cdot\tau\right) 
 \, , \\ \nonumber     
& \frac{1}{\sqrt{g}}    \left\{ 2 g_{\mu\nu} \frac{ \delta}{\delta g_{\mu\nu}} - b \nabla_\mu   \frac{ \delta}{\delta \tau_{\mu}}   \right\} 
\left( \frac{1}{b}  \xi_2 + \frac{2}{ b^2}  \xi_3 \right) 
=
-\tilde{\Box} \tilde{J}
 \, .
\end{align}
These equations show that the first three terms are indeed trivial anomalies, i.e., they can be generated from the the variation of a local action functional, or, more geometrically, they are exact $1$-forms. However, also the Weyl invariant $\tilde{\Box} \tilde{J}$ is a trivial anomaly. Therefore, it is a potentially problematic $\left( b + a'  \right)$-anomaly.

\subsubsection{A simple example}\label{sect_4d_examples_anomaly_affine_torsion}

As a check, in this subsection we compute the anomaly with the heat kernel in an explicit example. We consider a scalar QED-like theory, in which $\tau_\mu$ plays a role analog to that of the gauge potential, and compute the resulting second Seleey-Dewitt coefficient $\hat a_2$. We define the covariant derivative, $D_\mu\phi=\partial_\mu\phi + \frac{1}{b}\tau_\mu\phi$, which ensures that $D_\mu\phi$ transforms as $\phi$ itself under Weyl transformations. Thus, we have the following transformation laws
\begin{align}
\tau_\mu \rightarrow \tau_\mu + b \partial_\mu \sigma 
\, ,
\qquad
\phi   \rightarrow   e^{-\sigma} \phi
\, ,
\qquad
D_\mu\phi \rightarrow e^{-\sigma}D_\mu\phi
\, ,
\end{align}
and it becomes straightforward to write an action
\begin{align}
S
=
\int {\rm d}^4x \sqrt{g}  \left( 
\frac{1}{2}D^\mu \phi D_\mu\phi
+
\frac{\xi}{2} \tilde{J}\phi^2
+
\frac{1}{4} \Omega^2_{\mu\nu}
\right)
\end{align}
that is conformally invariant. Notice, however, that the derivative $D_\mu$ cannot be straightforwardly integrated by parts. A way out would be to use the Weyl potential $S_\mu$ in place of $\tau_\mu$ and a Weyl covariant derivative which attributes the correct gauge-covariant derivative to $g_{\mu\nu}$ as in \cite{Sauro:2022hoh} (which requires a disformation term to compensate the transformation of the Christoffel symbols). Consequently, given that $\phi D^\mu\phi$ has Weyl weight $-4$, we could safely integrate by parts.

However, for our purpose it is simpler and more straightforward to just expand $D_\mu$ and integrate by parts the regular covariant derivatives. Proceeding in this way, the Hessian reads
\begin{align}
H
=
-\Box
+
\frac{1}{b^2}\tau\cdot\tau
-
\frac{1}{b}\nabla\cdot\tau
+
\xi\tilde{J}
\, .
\end{align}
After substituting the modified curvatures and some lengthy algebra, the trace of SeeleyDeWitt coefficient $a_2$ gives the anomaly
\begin{align}\label{eq:anomaly_4d_example}
(4\pi)^2 \mathcal{A}  =\Tr a_2
&=
\frac{1}{120}W^2
-
\frac{1}{360}E_4
+
\frac{1}{30}\Box J
+
\frac{1}{2}\left(\xi-1\right)^2 \tilde{J}^2
-
\frac{1}{6}\left(\xi-1\right) \tilde{\Box}\tilde{J} 
+
\frac{1}{3b}\left(\xi-1\right)  \nabla_\mu \left(   \tilde{J}   \tau^\mu   \right) 
\, ,
\end{align}
which is in agreement with our general analysis for the special set of parameters
\begin{align}
f_2=\frac{1}{2}\left(\xi-1\right)^2
\,,
\quad
f_3=f_4=f_{11}=0
\, ,
\quad
f_{10}=\frac{1}{3b}\left(\xi-1\right)
\, , 
\quad
f_{14}=-\frac{1}{6}\left(\xi-1\right)
\, .
\end{align}
Notice that for the choice $\xi=1$, only the first three contributions in \eqref{eq:anomaly_4d_example} survives and there are no charges associated to torsion-dependent terms. This is of course to be expected since for this value of $\xi$ the Hessian reduces to the standard $4d$ Yamabe operator $H \equiv \Delta_2 = -\Box + J = -\Box+ \frac{R}{6}$.

A more complicated example would involve the coupling of $\tau_\mu$ to a non-gauge vector $V_\mu$, so that an interaction like $\tilde{K}^{\mu\nu}V_\mu V_\nu$ would be allowed. However, the heat kernel analysis becomes more involved and we do not delve further into it.

\subsection{Integration of the anomaly for torsion transforming affinely in $d=4$}\label{sect_4d_integration_anomaly_affine_torsion}

We recall the tensors that are invariant under transformations in $d=4$
\begin{align}\label{eq:invariants_4d}
&\sqrt{g} \tilde{J}^2 
=
\sqrt{g'} \tilde{J}'^2
\, , \qquad 
\sqrt{g} \tilde{K}^2_{\mu\nu}
=
\sqrt{g'}  \tilde{K}'^2_{\mu\nu}
\, , \qquad 
\sqrt{g} \Omega^2_{\mu\nu}
=
\sqrt{g'}\Omega'^2_{\mu\nu}
\, , \qquad 
\sqrt{g}  \tilde{\Box} \tilde{J}
=
\sqrt{g'} \tilde{\Box}' \tilde{J}'
\, .
\end{align}
The first three invariants in \eqref{eq:invariants_4d} can be interpreted as ``pure'' $b$-anomalies, but the total derivative (the fourth term in $d=4$), $\sqrt{g} \tilde{\Box} \tilde{J}$, is in fact a $\left( b + a'  \right)$-anomaly, as we have already noticed earlier.
On the other hand, the finite transformations of the other total derivative terms are complicated expressions involving up to three $\sigma$. In particular, they (or their general linear combinations) do not transform linearly with $\sigma$. Therefore, these can be interpreted as ``pure'' $a'$-anomalies. Cohomologically, the tensors $ \tilde{J}^2 $, $\tilde{K}^2_{\mu\nu}$ and $ \Omega^2_{\mu\nu}$ belong to different nontrivial de Rham classes and $\tilde{\Box} \tilde{J}$ is an exact form that belongs to the Weyl sub-class of invariants.

Let us now integrate the anomaly. The effect of integration of the purely metric structure alone is well-known. 
In particular, the contribution to the emt trace can be written in the convenient form
\begin{align}\label{eq:4d_emt_metric}
\mathcal{A}_g
=
b_1 W^2
+
a_1 Q_4
+
a_1'  \Box J
\, .
\end{align}
In the previous equation, we introduced the fourth $Q$-curvature \cite{BransonQ}
\begin{align}\label{eq:Q4-def+transf}
Q_4=\frac{1}{4}\left( E_4   - 4 \Box J \right) -  W^2 \, ,
\qquad
\sqrt{g}  Q_4=\sqrt{g'}  ( {Q'}_4 + \Delta'_4 \sigma)
\, ,
\end{align}
where ${\Delta}_4 $ is the Paneitz operator \cite{Paneitz:2008afy}, also known as the Fradkin-Tseytlin-Riegert operator \cite{Fradkin:1983tg,Riegert:1984kt}, which is self-adjoint and gives a fourth-order conformal action.
The total derivative is a trivial anomaly since it can be produced by the variation of the local action $\frac{2g_{\mu\nu}  }{\sqrt{g} } \frac{\delta}{\delta g_{\mu\nu}} \int d^4 x \sqrt{g} J^2 = -2 \square J$. Notice that $Q_4$ is technically in a different de Rham class than $E_4$, but this is unimportant because our use the $Q$-curvature amounts just to a reparametrization of the anomaly that is convenient for the transformation property given in \eqref{eq:Q4-def+transf}. Therefore, the action reproducing the metric part of the anomaly is
\begin{align}\label{eq:WZ_metric_4d}
\Gamma_{WZ}[\sigma;g ]
&=
\Gamma_{conf}[g]
+
\int {\rm d}^4x  \sqrt{g'} \sigma \left(
b_1  {W}'^2  
+
a_1 {Q'}_4   + \frac{a_1}{2}      {\Delta'}_4 \sigma 
\right)
-
\frac{1}{2}a_1'  \int {\rm d}^4x \sqrt{g} \, 
J^2  \\ \nonumber
&=
\Gamma_{conf}[g]
+
\Gamma_{NL}[\sigma; g']
+
\Gamma_{L}[g]
\, .
\end{align}
Going through the usual procedure, it can be checked that this action reproduces the Riegert's one \cite{Riegert:1984kt}.
The Wess-Zumino action for the torsionful part is instead
\begin{align}
\Gamma_{WZ}[\sigma;g,\tau]
&=
\Gamma_{conf}[g,\tau]
+
\int {\rm d}^4x  \sqrt{g'} \sigma \left(
f_2  \tilde{J}^2
+
f_3  \tilde{K}^2_{\mu\nu} 
+
f_4 \Omega^2_{\mu\nu}
+
\mathfrak{c}_{14}  \tilde{\Box} \tilde{J} 
\right)
+
\mathfrak{C}_{14}\int {\rm d}^4x \sqrt{g} \mathcal{X}
+
\Gamma_{loc}
\\    \nonumber
&=
\Gamma_{conf}[g,\tau]
+
\Gamma_{NL}[\sigma; g',\tau']
+
\Gamma_{L}[g,\tau]
\, , 
 \qquad\qquad \text{with} \,  \,  \,     \mathfrak{c}_{14}+ \mathfrak{C}_{14}=f_{14}
\, ,
\end{align}
where $\int {\rm d}^4x \sqrt{g} \mathcal{X}$ is short-hand for the linear combination in the fourth line of Eq.~\eqref{eq:4d_trivial_anomalies_affine_torsion} while in $\Gamma_{loc}=\Gamma_{loc}[g,\tau]$ are contained all the other combinations appearing in Eq.~\eqref{eq:4d_trivial_anomalies_affine_torsion}. The local anction $\Gamma_L[g,\tau]$ is then the sum $\int {\rm d}^4x \sqrt{g} \mathcal{X}+\Gamma_{loc}[g,\tau]$, and the full Wess-Zumino action is the sum of the two contributions, $\Gamma_{WZ}=\Gamma_{WZ}[\sigma;g]+\Gamma_{WZ}[\sigma;g,\tau]$.

From the previous considerations and including everything, the nonlocal action we get is
\begin{align}\label{eq:4d_Gamma_nonlocal_affine_torsion}
\Gamma_{NL}[g,\tau]
&=
\int {\rm d}^4x \sqrt{g}   \Bigg\{ 
\Bigg(  
b_1  {W}^2  
+
f_2  \tilde{J}^2
+
f_3  \tilde{K}^2_{\mu\nu} 
+
f_4 \Omega^2_{\mu\nu}
+
\mathfrak{c}_{14}  \tilde{\Box} \tilde{J} 
+
\frac{a_1}{2}  Q_4 
 \Bigg) 
 \frac{1}{\Delta_4} Q_4
\Bigg\}  \\ \nonumber
&
-
\frac{a_1'}{2}\int {\rm d}^4x \sqrt{g}   
J^2
+
\Gamma_{loc}[g,\tau]
+
\mathfrak{C}_{14}\int {\rm d}^4x \sqrt{g} \mathcal{X}
+
\Gamma_{c}[g,\tau]
\, ,  \\ \nonumber
 &{\rm for}\qquad  \mathfrak{c}_{14}+ \mathfrak{C}_{14}=f_{14}
\, .
\end{align}
See also Appendix \ref{app:non_local_actions} for more steps of the computation.
Evidently, we have an ambiguity similar to that observed in $d=2$, which now arises from a $\left(b + a'\right)$- rather than $\left(a + a'\right)$-anomaly. Notice, the sign difference with respect the $2d$ case that can be traced back to the transformation law of $Q_4$.

\subsubsection{Localized action and energy momentum tensor for torsion transforming affinely}\label{sect_4d_emt_affine_torsion}

Also in four dimensions we can obtain the localized action, which now will contain $10$ scalars fields.
To see this, we write the nonlocal part of the action \eqref{eq:4d_Gamma_nonlocal_affine_torsion} as
 \begin{align}\label{eq:4d_Gamma_ind_symm_form_affine_torsion}
&\Gamma_{NL}[g,\tau]
=
\frac{a_1}{10} \int {\rm d}^4x \sqrt{g}   \Bigg\{ 
\left[ Q_4 + 5 \frac{b_1}{a_1}  {W}^2       \right]    \frac{1}{ {\Delta}_4  }  \left[   Q_4  + 5 \frac{b_1}{a_1}    {W}^2    \right]   
+
 \left[ Q_4 + 5 \frac{f_2}{a_1}     \tilde{J}^2  \right]    \frac{1}{ {\Delta}_4  }  \left[   Q_4+ 5 \frac{f_2}{a_1}  \tilde{J}^2   \right]  
  \\ \nonumber
&+
 \left[ Q_4 + 5 \frac{f_3}{a_1}   \tilde{K}^2_{\mu\nu}   \right]    \frac{1}{ {\Delta}_4  }  \left[   Q_4 + 5 \frac{f_3}{a_1}    \tilde{K}^2_{\mu\nu}   \right]  
 +
 \left[ Q_4 +  5 \frac{f_4}{a_1}    \Omega^2_{\mu\nu}  \right]    \frac{1}{ {\Delta}_4  }  \left[   Q_4 +  5 \frac{f_4}{a_1}   \Omega^2_{\mu\nu}   \right] 
 +
 \left[ Q_4 + 5 \frac{\mathfrak{c}_{14}}{a_1}    \tilde{\Box} \tilde{J}   \right]    \frac{1}{ {\Delta}_4  }  \left[   Q_4 +5 \frac{\mathfrak{c}_{14}}{a_1}    \tilde{\Box} \tilde{J}     \right] \Bigg\}   \\ \nonumber
&
- \frac{1}{2} \frac{5}{a_1} \int {\rm d}^4x \sqrt{g}   \Bigg\{ 
b_1^2   {W}^2    \frac{1}{{\Delta}_4  }  {W}^2
+ f^2_2    \tilde{J}^2     \frac{1}{{\Delta}_4  }  \tilde{J}^2  +f^2_3 \tilde{K}^2_{\mu\nu}       \frac{1}{{\Delta}_4  }\tilde{K}^2_{\mu\nu} 
+f^2_4     \Omega^2_{\mu\nu}       \frac{1}{{\Delta}_4  }    \Omega^2_{\mu\nu} 
+\mathfrak{c}_{14}^2     \tilde{\Box} \tilde{J}       \frac{1}{{\Delta}_4  }   \tilde{\Box} \tilde{J} 
\Bigg\}
 \, .
\end{align}
So, we now need five couples of scalar fields $(\varphi_i$, $\psi_i)$, with  $i=1,\cdots,5$. Before going on, it is important to notice that the coefficient $a_1$ may be negative, at least for free fields of spin $s=0,1/2,1$ with a second order Hessian. Indeed, from the definition of $Q_4$ it is easy to see that $a_1=4 a$, where $a$ is the coefficient of $E_4$ in the heat kernel coefficient $\Tr \hat{a}_2$, which is negative. An explicit computation gives 
\begin{align}
(4\pi)^2\,a&=-\frac{1}{360}  N_\phi  - \frac{11}{360} N_\uppsi  - \frac{31}{180} N_A
\, ,
\end{align}
where $N_\phi, N_\uppsi $ and $ N_A$ are the number of fields with $s=0,1/2,1$, respectively. Therefore, the requirement to have real coefficients in the localized action fixes its overall sign as follows
\begin{align}\label{eq:4d_Gamma_ind_auxiliary_fields_affine_torsion}
\Gamma_{int}[g,\varphi_i, \psi_i]
=
\sum_{i=1}^5& \int {\rm d}^4x \sqrt{g}  \Bigg\{ 
 \frac{1}{2} \varphi_i {\Delta}_4  \varphi_i 
-
\alpha_i \varphi_i  Q_4 
-
\beta_i \varphi_i \mathcal{T}_i 
-
\frac{1}{2} \psi_i {\Delta}_4 \psi_i 
-
\beta_i \psi_i \mathcal{T}_i 
\Bigg\} \\ \nonumber
& 
\mathcal{T}_1= {W}^2 \, , 
\qquad
\mathcal{T}_2=   \tilde{J}^2\, , 
\qquad
\mathcal{T}_3= \tilde{K}^2_{\mu\nu} 
\qquad
\mathcal{T}_4=   \Omega^2_{\mu\nu} \, , 
\qquad
\mathcal{T}_5=    \tilde{\Box} \tilde{J} 
 \, .
\end{align}
From Eq.~\eqref{eq:4d_Gamma_ind_auxiliary_fields_affine_torsion} we get the following equations of motion
\begin{align}\label{eq:eom_scalars_4d_affine_torsion}
{\Delta}_4  \varphi_i 
=
\alpha_i  Q_4 + \beta_i    \mathcal{T}_i
\, ,
\qquad 
{\Delta}_4  \psi_i
=
-\beta_i  \mathcal{T}_i
\, ,
\qquad i=1,2,3,4,5
\, ,
\end{align}
where $ \mathcal{T}_i$ are defined as in \eqref{eq:4d_Gamma_ind_auxiliary_fields_affine_torsion}. Finally, going on-shell in \eqref{eq:4d_Gamma_ind_auxiliary_fields_affine_torsion}, we obtain
\begin{align}
\Gamma_{int}[g,\varphi, \psi]
=
-\sum_{i=1}^5  \frac{1}{2}  \int {\rm d}^4x \sqrt{g}  \Bigg\{ 
 \left( \alpha_i Q_4  + \beta_i    \mathcal{T}_i \right)   \frac{1}{\Delta_4}   \left( \alpha_i Q_4  + \beta_i    \mathcal{T}_i \right) 
-
 \left(   \beta_i    \mathcal{T}_i      \right)   \frac{1}{\Delta_4}   \left(    \beta_i    \mathcal{T}_i    \right) 
\Bigg\} 
 \, ,
\end{align}
which reproduces \eqref{eq:4d_Gamma_ind_symm_form_affine_torsion} by choosing 
\begin{align}\label{eq:4d_affine_torsion_coefficientes_local_action}
& \alpha_{i=1,...,5}=\sqrt{-\frac{a_1}{5}} 
\, ,
\qquad
\beta_1=-b_1 \sqrt{-\frac{5}{a_1}}
\, ,
\qquad
\beta_2=-f_2  \sqrt{-\frac{5}{a_1}}
\, ,
\qquad
\beta_3=-f_3  \sqrt{-\frac{5}{a_1}}
\, , \\ \nonumber
\qquad
&\beta_4=-f_4  \sqrt{-\frac{5}{a_1}} 
\, ,
\qquad\quad
\beta_5=-\mathfrak{c}_{14}   \sqrt{-\frac{5}{a_1}}
\, .
\end{align}
Our choice of the overall sign is important because by changing the right-hand side of Eq.~\eqref{eq:4d_Gamma_ind_auxiliary_fields_affine_torsion} by minus sign, we would have gotten purely imaginary $\alpha_i$ and $\beta_i$ coefficients.

From Eq.~\eqref{eq:4d_Gamma_ind_auxiliary_fields_affine_torsion} it is possible to obtain the trace of the emt and the divergence of the virial current which display a dependence on the parameter $\mathfrak{c}_{14}$
that is not fixed by the integration. 
Naturally, we could compute the emt and virial current as in the previous sections. However, the equations are more complicated, besides depending on $\mathfrak{c}_{14}$. Therefore, we will report them only later for the case of invariant torsion. Still we have computed them and, after tedious algebra, we confirm that the anomaly is correctly reproduced. Indeed, when expressing the anomaly in terms of the auxiliary fields, we have
\begin{align}
\mathcal{A}
=
\frac{1}{\sqrt{g}}
\left(2g_{\mu\nu}\frac{\delta}{\delta g_{\mu\nu}}
-
b \nabla_\mu \frac{\delta}{\delta \tau_{\mu}} \right)
\Gamma_{int}[g,\varphi_i, \psi_i]
&=
- \sum_{i=1}^5 \alpha_i {\Delta}_4  \varphi_i 
\nonumber\\
&=  
b_1  {W}^2  
+
f_2  \tilde{J}^2
+
f_3  \tilde{K}^2_{\mu\nu} 
+
f_4 \Omega^2_{\mu\nu}  
+
\mathfrak{c}_{14}  \tilde{\Box} \tilde{J} 
+
{a_1}  Q_4 
\end{align}
where the equations of motion and the relations \eqref{eq:4d_affine_torsion_coefficientes_local_action} have been used. 

For a more minimal choice of auxiliary fields, which results in the same anomaly, but at the cost of modifying the conformal actions, see the discussion in Sect.~\ref{sect:Conclusions} and Appendix~\ref{app:local_actions_auxiliary_fields}. Although the conformal actions are often overlooked, as we saw in Subsect.~\ref{sect_2d_cohomological_analisys_affine_torsion}, they may still play a significant role.

\subsection{Cohomological analysis for conformally invariant torsion in $d=4$}\label{sect_4d_cohomological_analisys_invariant_torsion}

For invariant torsion, the nihilpotent coboundary operator simplifies to 
\begin{align}
\delta_\sigma
=
\delta^{g}_\sigma
=
2\int {\rm d}^dx \, \sigma \, g_{\mu\nu} \frac{\delta}{\delta g_{\mu\nu}} \,.
\end{align}
Clearly, the analysis of the $1$-cochains involving only curvatures is the same as the one given at the very beginning of this section.
However, in the case of invariant torsion, we have a new conformal invariant allowed in the anomaly given by $\left(\tau \cdot \tau \right)^2$. The corresponding $1$-cochain is
\begin{align}
\omega_5[\sigma; g, \tau]
=
 \int {\rm d}^4x \sqrt{g} \sigma \left(\tau \cdot \tau \right)^2 \, \, . 
\end{align}
It is simple to see that $\delta_\sigma \omega_1[\sigma; g, \tau]=0$ because of the Grassmannian nature of $\sigma$. In this case, the modified curvature tensors are not very useful, since they are no longer conformal invariants. Therefore, we choose a \emph{new} basis for the torsion-dependent $1$-cochains
\begin{align}
&\omega_5 =\int {\rm d}^4x \sqrt{g} \sigma \left(\tau \cdot \tau \right)^2 \, ,
&&\omega_6 =\int {\rm d}^4x \sqrt{g} \sigma {J} \left(\nabla \cdot \tau \right)  \, ,
&&\omega_7 =\int {\rm d}^4x \sqrt{g}    \sigma {K}^{\mu\nu}  \left( \nabla_\mu \tau_\nu   \right)  \, ,
\nonumber\\
& \omega_8 =\int {\rm d}^4x \sqrt{g}    \sigma \Omega^2_{\mu\nu}   \, ,
&& \omega_9 =\int {\rm d}^4x \sqrt{g}    \sigma {J} \left(\tau \cdot \tau \right)   \, ,
&& \omega_{10} =\int {\rm d}^4x \sqrt{g} \sigma  \left( \nabla{J} \right) \cdot \tau   \, ,
\nonumber\\
&    \omega_{11} = \int {\rm d}^4x \sqrt{g} \sigma  {K}^{\mu\nu} \tau_\mu \tau_\nu \, ,
&& \omega_{12} = \int {\rm d}^4x \sqrt{g} \sigma \,  \tau \cdot \tau  \left(\nabla \cdot \tau \right) \,,
&&  \omega_{13} = \int {\rm d}^4x \sqrt{g} \sigma \left(\nabla \cdot \tau \right)^2  \, ,
\nonumber\\
& \omega_{14} = \int {\rm d}^4x \sqrt{g} \sigma  \left(\tau^\mu \tau^\nu \right)   \nabla_\mu  \tau_\nu\,
&&  \omega_{15} = \int {\rm d}^4x \sqrt{g} \sigma \,     \tau_\mu \Box \tau^\mu \, , \, ,
&&    \omega_{16} = \int {\rm d}^4x \sqrt{g} \sigma \,   \left( \nabla_\mu \tau_\nu   \right)^2    ,
\nonumber\\
&  \omega_{17} = \int {\rm d}^4x \sqrt{g} \sigma \,   \tau^\mu \nabla_\mu \nabla\cdot \tau \, ,
&&  \omega_{18}=\int {\rm d}^4x \sqrt{g}    \sigma  \Box \nabla \cdot \tau  \,.
\nonumber
\end{align}

The consistency conditions look the same as Eq.~\eqref{eq:c_conditions4d}. From these, we get the following constraints the coefficients of the anomaly have to satisfy (see Appendix \ref{app:basis_cochains_and_variations})
\begin{align}
& f_3=0 \, ,
&&  f_{5} =f_{12}-f_{11}  \, ,
&&  f_6 = f_{2} \, ,
&& f_{7}=2(f_{12}-f_{11}) \, ,
\nonumber\\
& f_{8}=\frac{f_{10}}{2}     \, ,
&&  f_{9} =f_{11}-f_{12}+f_{13}    \, ,    
&& c_{2}=-c_1    \, ,
&& f_{18}=\frac{f_7}{2}   \, ,
\nonumber
\end{align}
with the parameters $ f_1 \, , f_4  $ that are left arbitrary. Up to the usual metric contributions and some manipulations given in Appendix \ref{app:basis_cochains_and_variations}, the consistent anomaly is  
\begin{align}\label{eq:anomaly_inv_torsion}
A^{\tau}_\sigma
&=
\int \sqrt{g} \sigma \Bigg\{
f_1  \left(\tau \cdot \tau \right)^2
+
f_4 \Omega^2_{\mu\nu}
+
(f_{12}-f_{11}) \left(   \tau^\mu \,\nabla_\nu \nabla_\mu \tau^\nu  + \left( \nabla_\mu \tau_\nu   \right)^2   \right) 
+
\frac{f_{11}}{2} \Box \tau^2_\mu
+
 \frac{f_{10}}{2}   \nabla_\mu \left( \tau^\mu \,\tau \cdot \tau \right) 
 \\ \nonumber
&
 +
(f_{11}-f_{12}+f_{13})\nabla_\mu \left( \tau^\mu \,\nabla \cdot \tau \right) 
 +
 f_{2}   \left(  \frac{1}{2}  \Box \nabla \cdot \tau        +     \nabla_\mu \left( \tau^\mu \,J \right)                \right)
  \Bigg\}
\, .
\end{align}
In the following we are going to redefine $f_{12}-f_{11} \equiv f_5$ and $f_{11}-f_{12}+f_{13} \equiv f_9$ for notational simplicity.

The trivial anomalies can be obtained without particular difficulties as in the previous section. In particular, by a direct calculation, we have that 
\begin{align}
&
-2\frac{ g_{\mu\nu}}{\sqrt{g}} \frac{ \delta}{\delta g_{\mu\nu}}
\int {\rm d}^4x \sqrt{g}   \left( \nabla  J\right)   \cdot    \tau 
=
-2 \left( \nabla_\mu \left( \tau^\mu \,J \right)           +    \frac{1}{2}  \Box \nabla \cdot \tau             \right)  \, , \nonumber
\qquad
-2\frac{ g_{\mu\nu}}{\sqrt{g}} \frac{ \delta}{\delta g_{\mu\nu}}
\int {\rm d}^4x \sqrt{g} J  \left( \tau \cdot \tau \right)
=
 \Box \tau^2_\mu \, ,  
\\   \nonumber
&-2\frac{ g_{\mu\nu}}{\sqrt{g}} \frac{ \delta}{\delta g_{\mu\nu}}
\int {\rm d}^4x \sqrt{g}   \tau^\mu \nabla_\mu  \nabla \cdot \tau 
=
-4 \nabla_\mu \left( \tau^\mu    \nabla \cdot \tau           \right) \, , 
 \qquad
-2\frac{ g_{\mu\nu}}{\sqrt{g}} \frac{ \delta}{\delta g_{\mu\nu}}
\int {\rm d}^4x \sqrt{g}   \tau \cdot \tau \left( \nabla \cdot \tau \right)
=
2\nabla_\mu \left( \tau^\mu    \tau \cdot \tau           \right) \, . 
\end{align}
Also in this case the total derivative terms are akin to trivial anomalies.

We close this subsection with a simple example. Consider the conformal action
\begin{align}
S
=
 \frac{1}{2}\int {\rm d}^4x \sqrt{g} \Big\{ 
 \phi \Delta_4 \phi
+
\xi_1 \left(\tau \cdot \tau \right)^2 \phi^2
+
\xi_2 \Omega^2_{\mu\nu} \phi^2
+
\xi_3 ( \Box \tau^2_\mu )\phi^2
+
\xi_4 \left(   \tau^\mu \,\nabla_\nu \nabla_\mu \tau^\nu  + \left( \nabla_\mu \tau_\nu   \right)^2   \right)  \phi^2
 \Big\} \, .
\end{align}
Of course, we could have included all the structures appearing in the anomaly, but we have chosen to leave out the trivial anomalies. Therefore, the Hessian reads 
\begin{align}
H=\Delta_4 + P=\Delta_4 + \xi_1 \left(\tau \cdot \tau \right)^2+\xi_2 \Omega^2_{\mu\nu} +\xi_3 ( \Box \tau^2_\mu )+\xi_4 \left(   \tau^\mu \,\nabla_\nu \nabla_\mu \tau^\nu  + \left( \nabla_\mu \tau_\nu   \right)^2   \right) 
\, .
\end{align}
From the results of \cite{Barvinsky:1985an}, excluding the well known contributions coming the Paneitz operator, we have that the anomaly is indeed proportional to $P$ as defined in the previous equation.

\subsection{Integration of the anomaly for conformally invariant torsion in $d=4$ and the $\Psi$-anomaly}\label{sect_4d_integration_anomaly_invariant_torsion}

The trace of the emt can be expressed as
\begin{align}\label{eq:4d_emt_invariant_torsion}
-\langle T^{\mu}{}_{\mu} \rangle_{\tau}
&=
f_1  \left(\tau \cdot \tau \right)^2
+
f_4 \Omega^2_{\mu\nu}
+
f_{5} \left(   \tau^\mu \,\nabla_\nu \nabla_\mu \tau^\nu  + \left( \nabla_\mu \tau_\nu   \right)^2   \right) 
+
 \frac{f_{10}}{2}    \nabla_\mu \left( \tau^\mu \,\tau \cdot \tau \right) 
 +
 f_9\nabla_\mu \left( \tau^\mu \,\nabla \cdot \tau \right) 
  \\  \nonumber
&+
\frac{f_{11}}{2} \Box \tau^2_\mu
+
 f_{2}   \left(  \frac{1}{2}  \Box \nabla \cdot \tau        +     \nabla_\mu \left( \tau^\mu \,J \right)                \right)
  +
  \langle T^{\mu}{}_{\mu} \rangle_{g} \, ,
\end{align}
where $ \langle T^{\mu}{}_{\mu} \rangle_{g}$ comes from the purely metric contributions \eqref{eq:4d_emt_metric}. To integrate this equation we notice that 
\begin{align}
&\sqrt{g} \left( \tau \cdot \tau \right)^2
=
\sqrt{g'} \left( \tau \cdot \tau \right)^2
\, , \qquad 
\sqrt{g} \Omega^2_{\mu\nu}
=
\sqrt{g'}\Omega^2_{\mu\nu}
\, , \\ 
&\sqrt{g}\left(   \tau^\mu \,\nabla_\nu \nabla_\mu \tau^\nu  + \left( \nabla_\mu \tau_\nu   \right)^2   \right) =\sqrt{g'}\left(  \tau^\mu \,\nabla'_\nu \nabla'_\mu \tau^\nu  + \left( \nabla'_\mu \tau_\nu   \right)^2 - \nabla'_\mu \left( \tau\cdot\tau  \, \nabla'^\mu\sigma \right)\right)  \, .
\end{align}
Let us remark that both curvature and torsion vector are invariant: $\Omega^2_{\mu\nu}=\Omega'^2_{\mu\nu}$ and $\tau_\mu=\tau'_\mu$.

An important point comes about: we notice that the first line gives analogues of $b$-anomalies, while the second line is a \emph{new} kind of anomaly, since is neither a topological term, nor a conformal invariant, nor a trivial anomaly. We will refer to this new type of anomaly as a $\Psi$-anomaly. It is somewhat similar to an $a$ anomaly because, it transform with a total derivative $\nabla_\mu \left( \tau\cdot\tau  \, \nabla^\mu\sigma \right)=\left( \tau\cdot\tau\right) \Box \sigma  + 2 (\tau^\mu\nabla_\nu\tau_\mu)  \nabla^\nu\sigma$.
The transformation, however, includes a new operator which is not $\Delta_4$ of the GJMS type \cite{Juhl_book}.

Nevertheless, this operator satisfies two important properties
\begin{align}
&\sqrt{g'}\sigma \left( \left( \tau\cdot\tau\right) \Box'   + 2 (\tau^\mu\nabla'_\nu\tau_\mu)  \nabla'^\nu \right)\sigma
=
\sqrt{g}\sigma \left(\left( \tau\cdot\tau\right) \Box   + 2 \tau^\mu (\nabla_\nu\tau_\mu)  \nabla^\nu \right)\sigma \, , \\ \nonumber
&  \int {\rm d}^4x \sqrt{g} \Big\{ 
\sigma_2 \left( \tau\cdot\tau\right) \Box \sigma_1  + 2 \sigma_2 \tau^\mu (\nabla_\nu\tau_\mu)  \nabla^\nu \sigma_1
\Big\}
=
 \int {\rm d}^4x \sqrt{g} \Big\{ 
(\Box\sigma_2) \left( \tau\cdot\tau\right)  \sigma_1  + 2 (\nabla^\nu\sigma_2) \tau^\mu (\nabla_\nu\tau_\mu)   \sigma_1 
\Big\}
\, .
\end{align}
That is to say, it is a conformally covariant operator and formally self-adjoint much like $\Delta_4$.
In the following we denote the differential operator with $O$
\begin{align}
O(g,\tau) 
\equiv
\left( \tau\cdot\tau\right) \Box   
+
2 \tau^\mu (\nabla_\nu\tau_\mu)  \nabla^\nu
\,,
\end{align}
and the tensor $\Psi$
\begin{align}
\Psi(g',\tau) 
\equiv
\tau^\mu \,\nabla'_\nu \nabla'_\mu \tau^\nu  
+
\left( \nabla'_\mu \tau_\nu   \right)^2 
 \,.
\end{align}
From the cohmological viewpoint, $\Psi$ belongs to a different de Rham class than $\Omega^2_{\mu\nu}$ and $ \left( \tau \cdot \tau \right)^2$, and also to a new kind of Weyl sub-class transforming differently from, e.g., $Q_4$. However, the operator $O(g,\tau)$ has all the properties that are formally needed to integrate the anomaly in the same way of a regular $a$-anomaly, for which the discussion of subsect.~\ref{subsect_general_formalism} trivially generalizes. For example, any Weyl invariant $F$ can be nonlocally integrated in two different ways
\begin{align}
 \int {\rm d}^4x\sqrt{g}  F(g,\tau) \frac{1}{O}\Psi \, ,
 \qquad
 \int {\rm d}^4x\sqrt{g}  F(g,\tau) \frac{1}{\Delta_4} Q_4 \, . 
\end{align}
At first sight, it might look that we are facing a new ambiguity in the integration procedure, similar to what we saw for anomalies mixed with trivial $1$-cocycles. However, there is now a simple workaround the problem: we can integrate $Q_4$ and $\Psi$ simultaneously, as we can do with the definition \eqref{eq:def_modif_Q_and_delta} below. This is different for partially $a'$-anomalies, whose ambiguity cannot be resolved in such a way since it comes from just one term. Another difference is that the ambiguity of $\Psi$-like anomalies does not really show up at the level of the Wess-Zumino action, see \eqref{eq:4d_Gamma_ind_nonlocal_invariant_torsion}, but it concerns only the nonlocal action. The extension to more contributions analogous to the $\Psi$-anomaly is straightforward. 

Thus, we get (up to integration constants) the Wess-Zumino action
\begin{align}\label{eq:4d_Gamma_ind_nonlocal_invariant_torsion}
\Gamma_{WZ}
=
\Gamma_{WZ}[\sigma; g',\tau]
+
\Gamma_{WZ}[\sigma; g']
=
 &\int {\rm d}^4x \sqrt{g'}  \sigma \Bigg\{ f_1   \left( \tau \cdot \tau \right)^2
+
f_4 \Omega^2_{\mu\nu}
+
f_5  \Psi(g',\tau) - \frac{f_5 }{2} O(g',\tau)\sigma   
 \Bigg\} \\ \nonumber
 &
 +
\Gamma_{WZ}[\sigma; g']
 \, ,
\end{align}
reproducing the bulk terms when functionally differentiating with respect to $\sigma$. Naturally, $\Gamma_{WZ}[\sigma; g']$ is the purely metric contribution that was previously introduced in Eq.~\eqref{eq:WZ_metric_4d}. The local part is
\begin{align}\label{eq:4d_Gamma_ind_local_invariant torsion}
\Gamma_{L}[g,\tau]
=
- \int {\rm d}^4x \sqrt{g}   \Bigg\{
\frac{f_{11}}{2} J  \left( \tau \cdot \tau \right)
+
\frac{f_{10}}{4}   \tau \cdot \tau \left( \nabla \cdot \tau \right)
-
\frac{f_{9}}{4}  \tau^\mu \nabla_\mu  \nabla \cdot \tau 
+
\frac{f_2}{2} J  \left( \nabla \cdot \tau \right)
 \Bigg\}\, ,
\end{align}
which reproduces the boundary terms in the anomaly upon application of $-2\frac{ g_{\mu\nu}}{\sqrt{g}} \frac{ \delta}{\delta g_{\mu\nu}}$.
Combing these results together, the effective action reproducing the anomaly is
\begin{align}\label{eq:4d_Gamma_tot_nonlocal_invariant torsion}
\Gamma^{tot}_{WZ}
=
\Gamma_{conf}
+
\Gamma_{WZ}
+
\Gamma_{L}
 \, ,
\end{align}
where $\Gamma_{L}=\Gamma_{L}[g]+\Gamma_{L}[g,\tau]$, and we defined the conformal action $\Gamma_{conf} = \Gamma_{conf}[g]+\Gamma_{conf}[g,\tau]$, which acts as a kind of integration constant. The nonlocal version of $\Gamma^{tot}_{WZ}$ can be written as
\begin{align}\label{eq:Wess_Zumino_action_invariant_torsion}
\Gamma_{NL}
=
\Gamma_{conf}
+
 \int {\rm d}^4x \sqrt{g'}  \Bigg\{ \left(
f_1  \left( \tau \cdot \tau \right)^2          
+
f_4  \Omega^2_{\mu\nu} 
+
b_1 {W}^2 
+
\frac{1}{2} Q_4^\tau \right)\frac{1}{{\Delta}^\tau_4} Q_4^\tau
 \Bigg\} 
+
\Gamma_{L}
 \, ,
 \end{align}
which includes the contributions coming from the purely metric structure and the torsion. In the previous equation, we introduced the quantity $Q_4^\tau$ which is such that
\begin{align}\label{eq:def_modif_Q_and_delta}
Q_4^\tau
=
a_1 Q_4
+
f_5 \Psi(g,\tau)
\, ,
\qquad
\sqrt{g} Q_4^\tau
=
\left(
\sqrt{g'} {Q'}_4^\tau + {{\Delta}'}^\tau_4 
\right)
\, ,
\end{align}
where we defined the differential operator ${\Delta}^\tau_4$ as
\begin{align} 
{\Delta}^\tau_4  
&=
a_1\Box^2 
+
 a_1R^{\mu\nu}\nabla_\mu \nabla_\nu 
-
\left(  f_5   \tau \cdot \tau +\frac{2 }{3} a_1R \right) \Box
+
\left( \frac{a_1}{2} \nabla^\mu R -  2 f_5  (\tau^\nu \nabla^\mu     \tau_\nu) \right) \nabla_\mu \\ \nonumber
&\equiv
a_1 {\Delta}_4  
-
 f_5 O(g,\tau)
\, ,
\end{align}
so that ${\Delta}^\tau_4$ is self adjoint and $\int {\rm d}^4x \sqrt{g} \sigma{\Delta}^\tau_4 \sigma$ is a conformal action by construction.

\subsubsection{Localized action and energy momentum tensor for conformal invariant torsion}\label{sect_4d_emt_invariant_torsion}
It is easy to write  \eqref{eq:Wess_Zumino_action_invariant_torsion}  in the symmetric form (cf.\ following the discussion of Appendix~\ref{app:local_actions_auxiliary_fields})
 \begin{align}\label{eq:4d_Gamma_ind_symm_form_invariant_torsion}
\Gamma_{NL}[g,\tau]
&=
\frac{1}{6} \int {\rm d}^4x \sqrt{g}   \Bigg\{
\left[ Q_4^\tau + 3 b_1  {W}^2       \right]    \frac{1}{ {\Delta}^\tau_4  }  \left[   Q_4^\tau  + 3 b_1  {W}^2    \right]   
+
 \left[ Q_4^\tau + 3 f_4   \Omega^2_{\mu\nu}    \right]    \frac{1}{ {\Delta}^\tau_4  }  \left[   Q_4^\tau + 3 f_4   \Omega^2_{\mu\nu}    \right]  
  \nonumber \\ \nonumber
 & \qquad \qquad \qquad \qquad
 +\left[ Q_4^\tau + 3 f_1   \left( \tau \cdot \tau \right)^2   \right]    \frac{1}{ {\Delta}^\tau_4  }  \left[   Q_4^\tau + 3 f_1   \left( \tau \cdot \tau \right)^2    \right]  
\Bigg\} 
\\&
- \frac{3}{2} \int {\rm d}^4x \sqrt{g}   \Bigg\{ 
b_1^2   {W}^2    \frac{1}{{\Delta}^\tau_4  }  {W}^2
+ f^2_4    \Omega^2_{\mu\nu}      \frac{1}{{\Delta}^\tau_4  }   \Omega^2_{\mu\nu}
+f^2_1   \left( \tau \cdot \tau \right)^2       \frac{1}{{\Delta}^\tau_4  }  \left( \tau \cdot \tau \right)^2 
\Bigg\}
 \, .
\end{align}
Accordingly, we use three pairs of scalar fields $(\varphi_i$, $\psi_i)$, with  $i=1,2,3$. The localized action reads
\begin{align}\label{eq:4d_Gamma_ind_auxiliary_fields_invariant_torsion}
\Gamma_{int}[g,\varphi, \psi]
=
\sum_{i=1}^3 \int {\rm d}^4x \sqrt{g}  &\Bigg\{ 
- \frac{1}{2} \varphi_i {\Delta}^\tau_4  \varphi_i 
+
\alpha_i \varphi_i  Q_4^\tau 
+
\beta_i \varphi_i \mathcal{T}_i 
+
\frac{1}{2} \psi_i {\Delta}^\tau_4 \psi_i 
+
\beta_i \psi_i \mathcal{T}_i 
\Bigg\} \\ \nonumber
& 
\mathcal{T}_1=\left( \tau \cdot \tau \right)^2 \, , 
\qquad
\mathcal{T}_2=   \Omega^2_{\mu\nu} \, , 
\qquad
\mathcal{T}_3=    {W}^2
 \, ,
\end{align}
and the equations of motion for auxiliary fields are
\begin{align}\label{eq:eom_scalars_4d_invariant_torsion}
{\Delta}^\tau_4  \varphi_i 
=
\alpha_i  Q_4^\tau + \beta_i    \mathcal{T}_i
\, ,
\qquad 
{\Delta}^\tau_4  \psi_i
=
-\beta_i  \mathcal{T}_i
\, ,
\qquad i=1,2,3
\, ,
\end{align}
where the $ \mathcal{T}_i$ are defined as in \eqref{eq:4d_Gamma_ind_auxiliary_fields_invariant_torsion}. Thus, by using \eqref{eq:eom_scalars_4d_invariant_torsion} in \eqref{eq:4d_Gamma_ind_auxiliary_fields_invariant_torsion} we obtain the action \eqref{eq:4d_Gamma_ind_symm_form_invariant_torsion} for the choice
\begin{align}\label{eq:4d_inv_torsion_coefficientes_local_action}
\alpha_{1,2,3}= \frac{1}{\sqrt{3}}
\, ,
\qquad
\beta_1=\sqrt{3} \,  f_1
\, ,
\qquad
\beta_2=\sqrt{3} \, f_4
\, ,
\qquad
\beta_3=\sqrt{3} \, b_1
\, .
\end{align}
We can now derive the renormalized energy momentum tensor from the localized action by functional differentiation. For convenience, we split the emt in different contributions. From the fourth order kinetic terms we have
\begin{align}
-\frac{2}{\sqrt{g}}\frac{\delta}{\delta g_{\mu\nu}}
&
  \int {\rm d}^4x \sqrt{g}  \left\{ - \frac{1}{2}
\varphi_i {\Delta}^\tau_4  \varphi_i 
\right\}
=
\frac{f_7}{2}   g^{\mu\nu}    \tau  \cdot \tau      \left(  \nabla \varphi_i \cdot  \nabla \varphi_i \right)
- 
f_7 \tau^\mu \tau^\nu \left(  \nabla \varphi_i \cdot  \nabla \varphi_i \right)
-
f_7   \tau  \cdot \tau \nabla^\mu \varphi_i  \nabla^\nu \varphi_i 
\\ \nonumber
&
+
 \frac{a_1}{2} g^{\mu\nu} \Box \varphi_i   \Box \varphi_i
 -
2a_1  \Box \varphi_i \nabla^\mu \nabla^\nu \varphi_i
+
\frac{4a_1}{3} \nabla_\alpha \nabla^\mu  \varphi_i   \nabla^\alpha \nabla^\nu \varphi_i
-
 \frac{a_1}{3}   g^{\mu\nu}     \nabla_\alpha  \nabla_\beta \varphi_i   \nabla^\alpha \nabla^\beta \varphi_i
 -
 \frac{a_1}{3} g^{\mu\nu} \nabla_\alpha \Box \varphi_i   \nabla^\alpha \varphi_i
\\ \nonumber
& +
 \frac{a_1}{2} \nabla^{(\mu} \Box \varphi_i \nabla^{\nu)}  \varphi_i
-
\frac{2a_1}{3} \nabla_\alpha \nabla^\mu \nabla^\nu \varphi_i   \nabla^\alpha \varphi_i
-
\frac{4a_1}{3} g^{\mu\nu}  K_{\alpha\beta} \nabla^\alpha  \varphi_i   \nabla^\beta  \varphi_i
+
\frac{4a_1}{3} {W^{\mu}{}_{\alpha}}^{\nu}{}_{\beta} \nabla^\alpha  \varphi_i   \nabla^\beta  \varphi_i
 \\ \nonumber
&
+
\frac{16a_1}{3}  K^{(\mu}{}_{\alpha} \nabla^\alpha  \varphi_i   \nabla^{\nu)}  \varphi_i
\equiv
T^{\mu\nu}_{K, \, \varphi_i}
\, ,
\end{align}
while from the $Q_4^\tau$-dependent term we get
\begin{align}
-\frac{2}{\sqrt{g}}\frac{\delta}{\delta g_{\mu\nu}}
&
  \int {\rm d}^4x \sqrt{g}   \varphi_i   Q_4^\tau
=
f_7 \tau^\mu \tau^\nu \left(  \nabla \varphi_i \cdot  \nabla \varphi_i \right)
-
2 f_7   \tau^{(\mu}    \Omega^{\nu)\alpha}   \nabla_\alpha  \varphi_i 
-
2 f_7   \tau^{\alpha}    \nabla_\alpha  \tau^{(\mu}  \nabla^{\nu)}  \varphi_i 
+
2 f_7    \Omega^{\mu\alpha}  \Omega^{\nu}{}_{\alpha} 
\\ \nonumber
&
-
\frac{f_7}{3} g^{\mu\nu} \Omega^2_{\mu\nu}\varphi_i
+
 f_7   g^{\mu\nu} \tau^{\alpha}   \nabla_\alpha \tau^{\beta} \nabla_\beta \varphi_i
 +
a_1 g^{\mu\nu} \left( \frac{5}{3}   J \Box \varphi_i   + 2 K_{\alpha\beta} \nabla^\alpha \nabla^\beta  \varphi_i   - \frac{1}{3} \Box^2 \varphi_i  -  \nabla J \cdot \nabla\varphi_i  \right)
\\ \nonumber
&
-
2a_1 B^{\mu\nu} \varphi_i
-
\frac{8a_1}{3}  K^{\mu\nu} \Box \varphi_i
+
4 a_1 K^{\alpha(\mu} \nabla^{\nu)} \nabla_\alpha \varphi_i
+
4 a_1 \nabla^{(\mu}K^{\nu)}{}_{\alpha} \nabla^\alpha \varphi_i
-
4 a_1 (\nabla_\alpha  K^{\mu\nu})   \nabla^\alpha \varphi_i
\\ \nonumber
&
+ 
2a_1  \nabla^{(\mu} J  \nabla^{\nu)}\varphi_i 
-
2 a_1 J    \nabla^{\mu}   \nabla^{\nu}\varphi_i 
+
\frac{a_1}{3}   \nabla^{\mu}   \nabla^{\nu}\Box\varphi_i 
\equiv
T^{\mu\nu}_{Q, \, \varphi_i}
\, ,
\end{align}
where $B^{\mu\nu} $ is the Bach tensor.
We are left only with the terms depending on the $T_i$, which contribute to the emt as
\begin{align}
&-\frac{2}{\sqrt{g}}\frac{\delta}{\delta g_{\mu\nu}}
  \int {\rm d}^4x \sqrt{g}  \varphi_i \left( \tau \cdot \tau \right)^2
=
4 \varphi_i \tau^\mu \tau^\nu \left( \tau \cdot \tau \right)
-
g^{\mu\nu} \varphi_i \left( \tau \cdot \tau \right)^2
\equiv
T^{\mu\nu}_{\mathcal{T}_1, \, \varphi_i}
\, ,
\\ \nonumber
&
-\frac{2}{\sqrt{g}}\frac{\delta}{\delta g_{\mu\nu}}
 \int {\rm d}^4x \sqrt{g}  \varphi_i\Omega_{\mu\nu}^2
=
4 \varphi_i  \Omega^{\mu\alpha}  \Omega^{\nu}{}_{\alpha} 
-
g^{\mu\nu} \varphi_i \Omega_{\mu\nu}^2
\equiv
T^{\mu\nu}_{\mathcal{T}_2, \, \varphi_i}
\, ,
\\ \nonumber
&
-\frac{2}{\sqrt{g}}\frac{\delta}{\delta g_{\mu\nu}}
 \int {\rm d}^4x \sqrt{g}  \varphi_i W^2
=
8\varphi_iB^{\mu\nu}
-
g^{\mu\nu} \varphi_i  W^2
+
4 \varphi_i W^{\mu\alpha\beta\rho}W^{\nu}{}_{\alpha\beta\rho}
+
16 C^{(\mu\nu)}{}_{\alpha} \nabla^\alpha \varphi_i
+
8\varphi_i {W^{\mu}{}_{\alpha}}^{\nu}{}_{\beta} \nabla^\alpha  \varphi_i   \nabla^\beta  
\\ \nonumber
&\hspace{4.5cm}\equiv T^{\mu\nu}_{\mathcal{T}_3, \, \varphi_i}
\, .
\end{align}
In the previous equation, we also introduced the Cotton tensor $C^{\mu\nu\alpha}$. See, for example, \cite{Osborn:2015rna} for the definitions of the Bach and Cotton tensors.
With this notation, and up to local terms, the renormalized emt can be written as
\begin{align}\label{eq:emt_4d_invariant_torsion}
\langle T^{\mu\nu}  \rangle
=
 \sum_{i=1}^3
 \left\{
 T^{\mu\nu}_{K, \, \varphi_i}
 +
\alpha_i T^{\mu\nu}_{Q, \, \varphi_i}
+
\beta_i T^{\mu\nu}_{\mathcal{T}_i, \, \varphi_i}
-
 T^{\mu\nu}_{K, \, \psi_i}
 +
\beta_i T^{\mu\nu}_{\mathcal{T}_i, \, \psi_i}
 \right\}
 \, .
\end{align}
In this expression the only contribution to the trace comes from $T^{\mu\nu}_{Q, \, \varphi_i} $. Indeed, we have
\begin{align}
\langle T^\mu{}_\mu  \rangle
=
-\frac{2g_{\mu\nu}}{\sqrt{g}}\frac{\delta}{\delta g_{\mu\nu}}
\Gamma_{int}[g,\varphi, \psi]
=
- \sum_{i=1}^3 \alpha_i {\Delta}^\tau_4  \varphi_i
=
-
a_1 Q_4
-
f_7 \Psi 
-
f_1 \left( \tau \cdot \tau \right)^2 
-
f_4  \Omega^2_{\mu\nu}
-
b_1 W^2 \, ,
\end{align}
where the equations of motion and the relations \eqref{eq:4d_inv_torsion_coefficientes_local_action} has been used. Therefore, \eqref{eq:emt_4d_invariant_torsion} correctly reproduces the anomaly \eqref{eq:4d_emt_invariant_torsion}.

The relevance of this emt resides in the fact that it would appear on the right-hand side of a semiclassical gravitational field equations. Although \eqref{eq:emt_4d_invariant_torsion} is certainly an approximation, being the anomalous action built up to a conformal action $\Gamma_c$, it can be useful in those situations where such an action is irrelevant, e.g., on conformally flat spaces, or not crucial as for black hole thermodynamics. An advantage is that, in this way we can avoid to go through complicated renormalization procedures which are in general necessary in a direct computation of the emt in a field theoretical framework. Moreover, this expression does not crucially depend on the spin of the fields integrated away. On the other hand, this equation displays a certain degree of state dependency. Indeed, to proceed further we should determine the auxiliary fields by using their equations of motion \eqref{eq:eom_scalars_4d_invariant_torsion} and the choice of the boundary conditions also select the quantum state w.r.t.\ we are computing the vev $\langle T^{\mu\nu}  \rangle$, see Refs.\ \cite{Balbinot:1999ri,Anderson:2007eu}. Not surprisingly, we are again faced with the problem of choosing an explicit background on which to solve the equation of motion of the auxiliary scalar fields. We discuss some possible black hole solutions on which it could be of interest to evaluate \eqref{eq:emt_4d_invariant_torsion} in the concluding section.

\section{Conclusions and outlook}\label{sect:Conclusions}

The primary objective of this work has been to determine the most general conformal anomalies for the Weyl group in the presence of vectorial torsion in $d=2,4$ for different conformal transformations, and the associated anomalous actions. 
The cohomological analysis reveals interesting new insights on these issues. Particularly noteworthy is the presence of the new $\Psi$-anomaly and partially $a'$-anomalies. The former can be integrated unambiguously by redefining the anomaly, while the latter introduces a dependence on an undetermined parameter in the anomalous action, potentially limiting its applicability. However, we have observed with a specific example that some physical quantities might be not sensitive to this issue, such as the Wald entropy for black hole thermodynamics in $d=2$. We believe it would be valuable to further clarify this matter.

We also should make a remark on the number of auxiliary fields that we have used to localize the anomalous actions. When expressing the $\Gamma_{NL}[g,\tau]$ in the symmetric form, we always kept the subtraction terms, which are conformal invariants. It is important to notice that their coefficients are determined by those of the anomaly. However, in principle, the additional conformal invariant terms could be excluded, cutting down the number of auxiliary fields by half. Specifically, the $\psi_i$ fields used in the main text could be avoided in this way.

In fact, as we discuss in detail in Appendix~\ref{app:local_actions_auxiliary_fields}, it is also possible to work with a single scalar field.
However, in this paper, we have chosen to exploit the auxiliary scalar fields only as computational tools, which drop out from the final result once we go on-shell. Thus, we deliberately avoid assigning any specific physical meaning to these fields, even if a single scalar field could naturally be interpreted as a dilaton, i.e., the Goldstone boson arising in the low energy spontaneously broken phase of a conformal field theory \cite{Schwimmer:2010za}. An important result highlighted in Ref.~\cite{Schwimmer:2010za} is the presence of problematic double poles due to the propagator of the fourth-order Paneitz operator $\Delta_4$, which are absent in the so-called Fradkin-Vilkovisky (FV) effective action \cite{Fradkin:1978yw,Barvinsky:2023exr}. 
An analogous result was noted also in Ref.~\cite{Coriano:2022jkn} at the level of four-point functions, where it was suggested that the conformal (nonlocal) actions could play a role in resolving these issues. For example, although an action such as $\int  \sqrt{g}  W^2 \frac{1}{\Delta_4} W^2$ appears to be Weyl invariant, the computation of the Green function for $\Delta_4$ requires the choice of boundary conditions, which could break conformal invariance and contribute nontrivially. 

Given these considerations, and recalling the discussion of Subsect.~\ref{sect_2d_cohomological_analisys_affine_torsion} of the well-definedness of the Wald entropy, we have chosen not to discard these actions, as is often done in the literature. Instead, we opted to work with the maximum number of scalar fields, determined by the number of nontrivial Weyl cocycles as discussed in the subsection~\ref{sect_BH_thermo_and_2d_emt_invariant_torsion}, not to lose any physical effects.

As an additional remark, we note that the use of the FV action does not circumvent the ambiguities associated with mixed anomalies. This is best seen in the formalism of \cite{Barvinsky:2023exr}, where the FV action is derived via a gauge-fixing procedure from the Wess-Zumino (WZ) action. Specifically, the FV gauge corresponds to the choice $R[g'_{\mu\nu}]=R[e^{-2\sigma}g_{\mu\nu}]=0$, which can be solved for $\sigma$ and substituted back in the WZ action. However, since the ambiguity related to mixed anomalies arises at the level of the WZ action's coefficient, employing the FV action does not resolve this kind of ambiguity, even if it alleviates the problem of double poles as discussed in Refs.~\cite{Schwimmer:2010za,Coriano:2022jkn}.

Let us provide some clarification on the generality of our results. In this paper, we have focused on the anomaly of the ``full'' Weyl group, which means that $\sigma$ does not need to satisfy any differential constraint. However, it is possible to consider substructures of the Weyl group, which are usually defined by the requirement of leaving some geometrical tensor invariant. In such cases, as recently discussed in Ref.\ \cite{Martini:2024tie}, additional terms in the anomaly may emerge if compared to our current analysis. Additionally, we have tacitly assumed that the anomaly arises only from free field theories (through our examples). However, in an interacting theory, background matter fields and renormalization group beta functions would indeed contribute to the anomaly. To generalize our results, we could include elements in the basis of $1$-cochains cochains that depend on the desired matter field and possess the necessary properties. While investigating such scenarios could be intriguing due to their model independence, it would be considerably more complicated. 
For completeness, let us examine a simple example for conformally invariant torsion: an interacting scalar field with a two-derivatives kinetic term in $d=4$. Of course, in this setting the scalar field has mass dimension one. Consequently, the basis of $1$-cochains should include the following additional terms
\begin{align}
&\omega_{19} =\int {\rm d}^4x \sqrt{g} \sigma \left(\tau \cdot \tau \right)  \varphi^2  \, ,
&&  \omega_{20} =\int {\rm d}^4x \sqrt{g}\sigma \left(\nabla \cdot \tau \right)  \varphi^2  \, ,
&&  \omega_{21} =\int {\rm d}^4x \sqrt{g} \sigma \tau^\mu \varphi \nabla_\mu \varphi   \, ,  
\nonumber\\
& \omega_{22} =\int {\rm d}^4x \sqrt{g} \sigma \varphi^4 \,,
&&  \omega_{23} =\int {\rm d}^4x \sqrt{g} \sigma {J} \varphi^2 \,,  
&& \omega_{24}=\int {\rm d}^4x \sqrt{g} \sigma \Box \varphi^2 \,,
\nonumber\\
& \omega_{24}=\int {\rm d}^4x \sqrt{g} \sigma   \varphi  \Box \varphi \, .
\nonumber
\end{align}
Indeed, these contributions meet the required criteria of being linear in $\sigma$, having mass dimension four, being parity-even and invariant under diffeomorphisms. It is straightforward to see that for strong conformal transformations, we can replace ${J}$ with the modified Schouten's trace $\tilde{J}$ that we have used throughout this paper.

Another contribution that we discarded from the beginning is the Pontryagin density, even though it is a conformal invariant and is not inherently forbidden in the anomaly. Of course, the fact that it is allowed by the cohomology does not necessarily mean it is realized in a specific model, it simply indicates that it is not intrinsically inconsistent and the literature on the topic is not yet settled on its realization in similar models, see, e.g., \cite{Bonora:2014qla} and \cite{Bastianelli:2016nuf,Larue:2023tmu}. We hope to come back to these questions in a future publication since this term would have important and far reaching consequences \cite{Mauro:2014eda}.

Another significant future direction for this work involves applying the anomalous actions that we have obtained to relevant physical contexts. We have already discussed a potential investigation in $d=2$ concerning black hole physics. In $d=4$, there are other interesting possibilities to explore. Let us begin by considering the case in which the torsion vector transforms affinely. If we require a nonvanishing (semiclassical) torsion in vacuum and field equations with up to two derivatives, the only possible gravitational classical action is
\begin{align}\label{eq:Weyl_carghed_BH_action}
S_{cl}[g,\tau]
=
\int {\rm d}^4x \sqrt{g} \left\{
m^2_p R
+
\frac{1}{4} \Omega^2_{\mu\nu}
\right\}
\, ,
\end{align}
where $m_p$ is the Plank mass and $R$ is the Riemannian scalar curvature. In this way, we would get a torsionful black hole solution for free, because, as a matter of fact, the field equations derived from this action are formally equivalent to the Einstein-Maxwell system that yields a Reissner-Nordstrom-like solution. However, the resulting black hole is not electrically charged, but, instead, it would carry a ``Weyl charge.'' In this situation, we could consider the action 
\begin{align}
\Gamma[g,\tau]
=
\int {\rm d}^4x \sqrt{g} \left\{
m^2_p R
+
\frac{1}{4} \Omega^2_{\mu\nu}
\right\}
+
\Gamma_{int}[g,\varphi_i, \psi_i]
\, ,
\end{align}
where $e^{-\Gamma_{int}}=\int [d\phi] e^{-S_{cl}[g,\tau,\phi]}$ is the anomalous action of Eq.~\eqref{eq:4d_Gamma_ind_auxiliary_fields_affine_torsion} that is obtained by functionally integrating some conformally coupled free matter field $\phi$ with action $S_{cl}[g,\tau,\phi]$. In a path-integral approach to thermodynamics, the free energy can be approximated as
\begin{align}\label{eq:partition_function_torsion_affine}
\ln \mathcal{Z}(\beta)
\approx
-
S_{cl}[g,\tau]
-
\Gamma_{int}[g,\varphi_i, \psi_i]
\, .
\end{align}
Notice that, unfortunately, by proceeding in this way we would encounter the same unphysical ambiguity we already discussed, not to mention the technical difficulty of explicitly evaluating the anomalous action. From what we obtained in $d=2$, a possible way out could be to use Wald's formula in four dimension to compute the contributions to the entropy coming from $\Gamma_{int}[g,\varphi_i, \psi_i]$. Although this is not an easy task, it could be interesting to see whether or not the unphysical dependence on $\mathfrak{c}_{14}$ survives. However, this seems unlikely simply because the ambiguity arises from a $\left( b + a' \right)$- rather than an $\left( a + a' \right)$-anomaly, cf.\ Eqs.~\eqref{eq:mapping_constants_auxiliary_fields_affine_torsion} and~\eqref{eq:4d_affine_torsion_coefficientes_local_action}. The point is that, whereas in the $2d$ case the $\mathfrak{C}_2$ coefficient quantifying the ambiguity appears in different points, so it is possible for it to cancel out by going on-shell, in the $4d$ case this does not happen for $\mathfrak{c}_{14}$. 

When instead torsion is invariant under Weyl, another interesting case of study would be to add a mass term and/or torsion self-interactions to the action \eqref{eq:Weyl_carghed_BH_action}. In this way, we would obtain a solution similar to that of Proca stars \cite{Brito:2015pxa}. Its thermodynamics could be studied with a similar logic to that leading to Eq.~\eqref{eq:partition_function_torsion_affine}. Also in this case, it is probably easier to resort to the computation of Wald entropy. However, even if we do not have the ambiguity related to mixed anomalies, the technical difficulties are still present. In particular, to explicitly evaluate the entropy, it is necessary to express the Green function of ${\Delta}^\tau_4$ on a black-hole background. We hope to come back to this matters in a future work.

\smallskip

\paragraph*{Acknowledgments.} The authors are grateful to D.~Sauro for numerous helpful comments during the development of this work and for reading the draft.

\appendix

\section{Obtaining the nonlocal actions for invariant and affinely transforming torsion}\label{app:non_local_actions}
In this appendix we show how to compute the variations of the nonlocal action with respect to $\sigma$ to obtain the anomaly by using their conformal properties. 

\subsection{$2d$ case for invariant and affinely transforming torsion}
We start by considering $\int {\rm d}^2x \sqrt{g}  \tilde{J}  \frac{1}{\Delta_2}   \left(\nabla\cdot\tau\right)$, which appears for the torsion vector transforming affinely. We have
\begin{align}
&\frac{\delta}{\delta\sigma} \int {\rm d}^2x  {\rm d}^2y (\sqrt{g}  \tilde{J})_x G(x,y)  (\sqrt{g} \left(\nabla\cdot\tau\right))_y
=
\frac{\delta}{\delta\sigma} \int {\rm d}^2x  {\rm d}^2y (\sqrt{g'}  \tilde{J}')_x G'(x,y) \left(   \sqrt{g'} \left(\nabla'\cdot\tau'    + b \Delta'_2 \sigma \right)   \right)_y \\ \nonumber
&=
\frac{\delta}{\delta\sigma} \int {\rm d}^2x  {\rm d}^2y \left[ (\sqrt{g'}  \tilde{J}')_x G'(x,y)  (\sqrt{g'} \left(\nabla' \cdot\tau' \right))_y +b   \int {\rm d}^2x  {\rm d}^2y (\sqrt{g'}  \tilde{J}')_x \left(  \sqrt{g'}_y  \Delta'_{2,y}G'(x,y)      \right)  \sigma(y) \right] \\ \nonumber
&=
b \sqrt{g}  \tilde{J} \, ,
\end{align}
where the conformal invariance of the Green's function G of $\Delta_2$, along with its defintion as a Green's function, have been used. Similarly, we find for  $\int {\rm d}^2x \sqrt{g}   \left(\nabla\cdot\tau\right) \frac{1}{\Delta_2}   \left(\nabla\cdot\tau\right)$ that
\begin{align}
\frac{1}{2}\frac{\delta}{\delta\sigma} \int {\rm d}^2x  {\rm d}^2y (\sqrt{g}   \left(\nabla\cdot\tau\right))_x G(x,y)  (\sqrt{g} \left(\nabla\cdot\tau\right))_y
=
b \sqrt{g}  \left(\nabla\cdot\tau\right)
\, ,
\end{align}
where we used again the conformal properties of the Green's function and of $\sqrt{g}\left(\nabla\cdot\tau\right)$. To check that the action \eqref{eq:2d_Gamma_ind_nonlocal_affine_torsion} correctly reproduces the anomaly written in the main text, we can simply set $b=-1$ in the previous formulae. For the contributions $\int {\rm d}^2x \sqrt{g} J \frac{1}{\Delta_2}   {J}$, and $\int {\rm d}^2x \sqrt{g}  \left(\nabla\cdot\tau\right)\frac{1}{\Delta_2}   {J}$, which appear by parametrizing the anomaly with $J$ and $\left(\nabla\cdot\tau\right)$, we have
\begin{align}
&\frac{1}{2}\frac{\delta}{\delta\sigma} \int {\rm d}^2x  {\rm d}^2y (\sqrt{g}  J )_x G(x,y)  (\sqrt{g}J)_y
=
- \sqrt{g}  J\, , \\ \nonumber
& \frac{\delta}{\delta\sigma} \int {\rm d}^2x  {\rm d}^2y (\sqrt{g}  \left(\nabla\cdot\tau\right)  )_x G(x,y)  (\sqrt{g}J)_y
=
b \sqrt{g} \left( J - \frac{1}{b}  \left(\nabla\cdot\tau\right)   \right) 
\, .
\end{align}
Combining these results we can immediately verify the results reported in the main text.

For what concerns the case of invariant torsion, the variation of $\int {\rm d}^2x \sqrt{g} J \frac{1}{\Delta_2}   {J}$ is obvioulsy the same as above. Instead, the variations of $\int {\rm d}^2x \sqrt{g} \left(\tau\cdot\tau\right) \frac{1}{\Delta_2}   {J}$ and  $\int {\rm d}^2x \sqrt{g}  \left(\nabla\cdot\tau\right)\frac{1}{\Delta_2}   {J}$ are now
\begin{align}
&\frac{\delta}{\delta\sigma} \int {\rm d}^2x  {\rm d}^2y (\sqrt{g}  \left(\tau\cdot\tau\right) )_x G(x,y)  (\sqrt{g}J )_y
=
- \sqrt{g}   \left(\tau\cdot\tau\right)  \, ,\\ \nonumber
&\frac{\delta}{\delta\sigma} \int {\rm d}^2x  {\rm d}^2y (\sqrt{g} \left(\nabla\cdot\tau\right) )_x G(x,y)  (\sqrt{g} J )_y
=
- \sqrt{g}  \left(\nabla\cdot\tau\right)  \, ,
\end{align}
being $\sqrt{g}  \left(\tau\cdot\tau\right)$ and $\sqrt{g} \left(\nabla\cdot\tau\right)$ invariants.

\subsection{$4d$ case for invariant and affinely transforming torsion}
We start with the $4d$ case for transforming torsion. Take any conformal invariant denoted $\sqrt{g}F(g,\tau)$. Then, the variation of the functional $ \int {\rm d}^4x\sqrt{g}  F(g,\tau) \frac{1}{\Delta_4} Q_4$ is
\begin{align}
\frac{\delta}{\delta\sigma} \int {\rm d}^4x  {\rm d}^4y  (\sqrt{g} F(g,\tau))_x G(x,y)  (\sqrt{g}Q_4)_y
=
 \sqrt{g} F(g,\tau) \, ,
\end{align}
where $\sqrt{g}F(g,\tau)=\sqrt{g}\left\{ \tilde{J}^2, \tilde{K}^2_{\mu\nu}, {\Omega}^2_{\mu\nu}, \tilde{\Box}\tilde{J}    \right\}$ and where now $ G(x,y)$ is the Green's function of $\Delta_4$. Notice the different sign w.r.t.\ the previous case due to the transformation of $Q_4$. As we have noticed in the main text, $\tilde{\Box}\tilde{J} $ can be also obtained from the variation of a local action.

Let us now turn the case of invariant torsion. We start considering 
\begin{align*}
\Gamma_{NL}[\sigma; g', \tau]
&=
- \int {\rm d}^4x \sqrt{g'}  \sigma \Bigg\{ 
f_1   \left( \tau \cdot \tau \right)^2
+
f_4 \Omega^2_{\mu\nu}
+
f_5  \Psi(g',\tau)  - \frac{f_5 }{2} O(g',\tau)\sigma  +\\ \nonumber
&+ 
b_1 \, {W'}^2  
+
a_1 {Q'}_4   + \frac{a_1}{2}      {\Delta'}_4 \sigma  
 \Bigg\} \, ,
 \end{align*}
where $\Psi(g',\tau) = \tau^\mu \,\nabla'_\nu \nabla'_\mu \tau^\nu  + \left( \nabla'_\mu \tau_\nu   \right)^2$, as defined in the main text. Moreover, both $O$ and $\Delta_4$ enjoy the crucial properties of being conformally covariant operators and self adjoints. In particular, we have that the $b$-anomalies can also be obtained from the variation of  $ \int {\rm d}^4x\sqrt{g}  F(g,\tau) \frac{1}{O}\Psi$. In fact,
\begin{align}
\frac{\delta}{\delta\sigma} \int {\rm d}^4x  {\rm d}^4y (\sqrt{g}  F(g,\tau))_x P(x,y) (\sqrt{g}\Psi)_y
=
- \sqrt{g} F(g,\tau) \, ,
\end{align}
where $P(x,y)$ is the Green's function of $O$. Therefore we can use them separately to get the following nonlocal action 
\begin{align*}
\Gamma_{NL}[ g, \tau]
&=
 \int {\rm d}^4x \sqrt{g}  \Bigg\{ 
  \left( \tau \cdot \tau \right)^2 \left(\mathfrak{F}_1    \frac{1}{O(g,\tau)}  \Psi(g,\tau)  + \mathcal{F}_1    \frac{1}{\Delta}_4 {Q}_4             \right)
+
 \Omega^2_{\mu\nu}    \left(\mathfrak{F}_4    \frac{1}{O(g,\tau)}  \Psi(g,\tau)  + \mathcal{F}_4    \frac{1}{\Delta}_4 {Q}_4             \right) \\ \nonumber
&+
 {W}^2  \left(\mathfrak{B}_1  \frac{1}{O(g,\tau)}  \Psi(g,\tau)  + \mathcal{B}_1   \frac{1}{\Delta}_4 {Q}_4             \right)
 - 
\frac{f_5}{2}   \Psi(g,\tau)  \frac{1}{O(g,\tau)}  \Psi(g,\tau)  
+
 \frac{a_1}{2}  {Q}_4  \frac{1}{\Delta}_4 {Q}_4  
 \Bigg\} 
  \\ \nonumber
&\text{with} \qquad  \mathcal{F}_1 - \mathfrak{F}_1 = f_1 \, , \quad  \mathcal{F}_4 - \mathfrak{F}_4 = f_4 \, ,  \quad \mathcal{B}_1 - \mathfrak{B}_1 = b_1 \, ,
 \end{align*}
which correctly reproduces the anomaly, although in this way we meet a similar kind of ambiguity that we have seen in the case of transforming torsion. However, in this case there is a way out, which consists in defining the quantities
\begin{align*}
Q_4^\tau
=
a_1 Q_4
+
f_5 \Psi(g,\tau)
\, ,
\qquad
\sqrt{g} Q_4^\tau
=
\left(
\sqrt{g'} {Q'}_4^\tau + {{\Delta}'}^\tau_4 
\right)
\, ,
\end{align*}
where
\begin{align*} 
{\Delta}^\tau_4  
&=
a_1\Box^2 
+
 a_1R^{\mu\nu}\nabla_\mu \nabla_\nu 
-
\left(  f_5   \tau \cdot \tau +\frac{2 }{3} a_1R \right) \Box
+
\left( \frac{a_1}{2} \nabla^\mu R -  2 f_5  (\tau^\nu \nabla^\mu     \tau_\nu) \right) \nabla_\mu \\ \nonumber
&\equiv
a_1 {\Delta}_4  
-
 f_5 O(g,\tau)
\, ,
\end{align*}
Therefore, ${\Delta}^\tau_4$ is self-adjoint and $\int {\rm d}^4x \sqrt{g} \sigma{\Delta}^\tau_4 \sigma$ is a conformal action by construction. Accordingly, we could write the nonlocal effective action in the following form
\begin{align*}
\Gamma_{NL}
=
 \int {\rm d}^4x \sqrt{g'}  \Bigg\{ \left(
f_1  \left( \tau \cdot \tau \right)^2          
+
f_4  \Omega^2_{\mu\nu} 
+
b_1 {W}^2 
+
\frac{1}{2} Q_4^\tau \right)\frac{1}{{\Delta}^\tau_4} Q_4^\tau
 \Bigg\} 
  \, ,
 \end{align*}
where we integrated the terms $Q_4$ and $\Psi$ simultaneously.

\section{Nonlocal actions, energy-momentum tensor and diffeomorphism invariance in $d=2$}\label{app:non_local_emt_Diff} 
 
In this appendix, we report the computations leading the ``nonlocal'' energy momentum tensors and virial currents, together with the checks that diffeomorphism invariance is not broken at the quantum level for invariant and transforming torsion.
Moreover, we check that, as expected, the nonlocal action obtained with different parametrizations of the anomaly differs from that of the main text by a conformal term $\Gamma_c$. 

\subsection{The case of affinely transforming torsion}

We start by showing the different parametrizations of the anomaly just amount to add a conformal action to the one used in the main text. To see this, we repeat the calculation leading to the nonlocal action starting from the basis with $J$ rather than $\tilde{J}$. We have
\begin{align}\label{eq:2d_emt_affine_torsion_with_J}
\langle T^{\mu}{}_{\mu}  \rangle 
+
b \nabla_\mu  \langle \mathcal{D}^{\mu} \rangle
=
c_1 J
+
r_2 \left( \nabla \cdot \tau \right) 
\, ,
\end{align}
where $r_2=c_2-c_1$. The terms transform as
\begin{align}
\sqrt{g} J
=
\sqrt{g'} \left( J' -  \Delta_2' \sigma \right)   
\, , \qquad 
\sqrt{g} \left( \nabla \cdot \tau \right)
=
\sqrt{g'} \left( \nabla' \cdot \tau' - \Delta_2' \sigma \right)   \, ,
\end{align}
where we have fixed $b=-1$.
In this case, $J$ is the standard $2d$ $a$-anomaly, while $ \left( \nabla \cdot \tau \right)$ is again an $(a + a')$-anomaly. Accordingly, the local anomalous action is
\begin{align}\label{eq:2d_Gamma_ind_sigma_affine_torsion_with_J}
&\Gamma_{WZ}[\sigma; g',\tau]
=
 \int {\rm d}^2x \sqrt{g'}  \sigma \Big\{ 
c_1  J'+ R_2  \left( \nabla' \cdot \tau'\right) -\left(   \frac{R_2}{2} +\frac{c_1}{2}\right)  \Delta'_2 \sigma
\Big\} 
+
\frac{R_3}{2} \int {\rm d}^2x \sqrt{g}   \left( \tau \cdot \tau \right) 
 \, ,  \\ \nonumber
& \text{with} \quad R_2  + R_3 = r_2    \, ,
\end{align}
where, from \eqref{eq:2d_Gamma_ind_sigma_affine_torsion}, it is clear that $R_3=\mathfrak{C}_3$, and $R_2=\mathfrak{C}_2-c_1$. For the nonlocal action we obtain 
\begin{align}\label{eq:2d_Gamma_ind_nonlocal_affine_torsion_with_R}
\Gamma_{NL}[g,\tau]
=
-\frac{1}{2} &\left(c_1 + R_4  \right) \int {\rm d}^2x \sqrt{g}   \Bigg\{ 
J   \frac{1}{\Delta_2}  J
\Bigg\}
-
\frac{1}{2}\left(R_2 + R_4   \right) \int {\rm d}^2x \sqrt{g}   \Bigg\{ 
 \left(\nabla\cdot\tau\right)    \frac{1}{\Delta_2}    \left(\nabla\cdot\tau\right)  
 \Bigg\} \\ \nonumber 
&
+
R_4 \int {\rm d}^2x \sqrt{g}   \Bigg\{ 
 \left(\nabla\cdot\tau\right)    \frac{1}{\Delta_2}    J \Bigg\} 
 -
\frac{R_2-r_2}{2} \int {\rm d}^2x \sqrt{g}  \Bigg\{ \left( \tau \cdot \tau \right) 
\Bigg\}
 \, ,
\end{align}
since $\frac{\delta}{\delta\sigma} \int {\rm d}^2x \sqrt{g}   \Big\{  \left(\nabla\cdot\tau\right)    \frac{1}{\Delta_2}    J \Big\}= -\sqrt{g}(J + \nabla\cdot\tau)$. Notice that the nonlocal Polyakov action explicitly appears as first term, and that $R_4$ is an arbitrary parameter. At this point, it is simple to see that the difference between \eqref{eq:2d_Gamma_ind_nonlocal_affine_torsion} and \eqref{eq:2d_Gamma_ind_nonlocal_affine_torsion_with_R} is the conformally invariant action
\begin{align}
\frac{C_4+c_1+R_4}{2}\int {\rm d}^2x \sqrt{g}   \Bigg\{ 
\tilde{J}   \frac{1}{\Delta_2}   \tilde{J}    
\Bigg\}
 \, .
\end{align}
In particualar, the actions \eqref{eq:2d_Gamma_ind_nonlocal_affine_torsion} and \eqref{eq:2d_Gamma_ind_nonlocal_affine_torsion_with_R} coincide if we choose $C_4=- r_1 - R_4 $.
Obviously, choosing the constants in this way, it is completely equivalent to work with the nonlocal action that we have shown above or with that of the main text.

Let us now show the results that allow to check that the Diff group is not broken by the anomalous action~\eqref{eq:2d_Gamma_ind_auxiliary_fields_affine_torsion}. We start by computing the divergence of the emt \eqref{eq:emt_auxiliary_fields_affine_torsion} and the virial current. For the first we get
\begin{align*}
\nabla_\mu \langle T^{\mu\nu}  \rangle 
&=
 (\alpha-\beta)\left( \tau^\nu\Box  \varphi  + 2\nabla^{[\rho} \tau^{\nu]} \nabla_\rho \varphi  + \nabla \cdot \tau \nabla^\nu \varphi  \right) 
-
\gamma \tau^\nu\Box   \psi   
-
2\gamma \nabla^{[\rho} \tau^{\nu]} \nabla_\rho \psi 
-
 \Box\varphi \nabla^\nu \varphi 
 +
 \Box\psi \nabla^\nu \psi
 \\ \nonumber
&
-
\frac{1}{2} \alpha R \nabla^\nu \varphi 
+
\frac{1}{2} \gamma R \nabla^\nu \psi 
-
\gamma \nabla \cdot \tau \nabla^\nu \psi
+
(\mathfrak{C}_2-c_2)\left(\tau^\nu \nabla \cdot \tau + \tau^\mu \nabla_\mu \tau^\nu  - \tau^\mu \nabla^\nu \tau_\mu   \right)
\, ,
\end{align*}
and for the virial current
\begin{align*}
\frac{1}{\sqrt{g}} \frac{\delta}{\delta \tau_{\mu}}   \Gamma_{NL}[g,\varphi, \psi]
=
\langle \mathcal{D}^{\mu}  \rangle 
=
(\mathfrak{C}_2-c_2)\tau^\mu
+
(\alpha-\beta) \nabla^\mu \varphi
-
\gamma   \nabla^\mu \psi
\, ,
\end{align*}
while its divergence is reported in Eq.~\eqref{eq:2d_div_virial_current_affine_torsion} of the main text. Combining these results, and going on-shell in the auxiliary fields, it is just a matter of (lengthy and tedious) algebra to show that the N\"other identity of Diff is valid also quantum mechanically.

\subsection{The case of conformally invariant torsion}

By functional differentiating \eqref{eq:2d_Gamma_ind_auxiliary_fields_invariant_torsion} wrt $g_{\mu\nu}$, we get the emt that we report also here for readability
\begin{align*}
&\langle T^{\mu\nu}  \rangle 
=
g^{\mu\nu}\left[  \alpha_1\Box\varphi_1  - \frac{1}{2} \nabla\varphi_1\cdot\nabla\varphi_1    +     \frac{1}{2} \nabla\psi_1\cdot\nabla\psi_1  \right] + \nabla^\mu \varphi_1  \nabla^\nu \varphi_1  -     \nabla^\mu \psi_1  \nabla^\nu \psi_1+ \alpha_1 \nabla^\mu\nabla^\nu \varphi_1  \\ \nonumber
&+\left(\alpha_1 \rightarrow \alpha_2 \, , \varphi_1   \rightarrow \varphi_2 \, , \psi_1 \rightarrow \psi_2 \right) - 2\beta_1\left( \tau^{(\mu}\nabla^{\nu)} \varphi_1 + \tau^{(\mu}\nabla^{\nu)} \psi_1 \right) + \beta_1 g^{\mu\nu}\left[ \tau \cdot \nabla \varphi_1 + \tau \cdot \nabla \psi_1     \right] \\ \nonumber
&+\beta_2    \left( 2\tau^\mu\tau^\nu \varphi_2  + 2\tau^\mu\tau^\nu \psi_2  - g^{\mu\nu}   \left( \tau \cdot  \tau \right) \varphi_2  -g^{\mu\nu}  \left( \tau \cdot  \tau \right) \psi_2    \right)
\, .
\end{align*}
By taking the trace of this emt, and going on shell, with some work we can verify that it reproduces the anomaly $\langle T^{\mu}{}_{\mu}  \rangle = c_1 J + c_2  \left( \nabla \cdot  \tau \right) +c_3  \left( \tau \cdot  \tau \right) \, $. Instead, the virial current is
\begin{align*}
\langle \mathcal{D}^{\mu}  \rangle 
=
2\beta_2  \tau^\mu  \left(   \varphi_2  +  \psi_2    \right)
-
\beta_1 \nabla^\mu \left(  \varphi_1  +  \psi_1 \right)
\, ,
\end{align*}
from which it is easy to obtain its divergence. Similarly to the previous cases, from these equations we get that 
\begin{align}
\nabla_{\mu}  \langle T^{\mu}{}_{\nu}\rangle
=
\langle\mathcal{D}^{\mu}\rangle     \Omega_{\mu\nu}
+
\tau_\nu \nabla_{\mu} \langle \mathcal{D}^{\mu}\rangle
\, 
\end{align}
holds, implying that Diff symmetry is preserved.

\section{Local actions and auxiliary fields}\label{app:local_actions_auxiliary_fields}


It is important to emphasize that some freedom exists in the choice of the number of auxiliary fields needed to localize the action. For instance, we could choose to absorb the conformally invariant actions introduced along this work to symmetrize the nonlocal terms into the conformal action, denoted with $\Gamma_c$, which remains undetermined by the integration of the conformal anomaly. This approach effectively halves the number of scalar fields used in the main text. 
Additionally, by disregarding invariant actions entirely, it is possible to proceed with only a single auxiliary scalar field, as we now explore. 
 
In each dimension and for both types of conformal transformations, the resulting nonlocal actions take the schematic form
 \begin{align}
\Gamma_{NL}
=
- \int  \sqrt{g}   \left(
\sum_i b_i F_i[g,\tau] 
+
\frac{a}{2}E[g,\tau]  \right)\frac{1}{O} E[g,\tau] 
 \, ,
 \end{align}
where we follow the notation of Subsect.~\ref{subsect_general_formalism}, with the $F_i[g,\tau]$ representing Weyl invariants and the $E[g,\tau]$ captures the $a$-anomaly (or some modification of it like the $Q$-curavture), which transforms with a conformally covariant operator $O$.
It is straightforward to rewrite this action in a symmetric form via ``square completion''
 \begin{align}
\Gamma_{NL}
=
-\frac{a}{2}&  \int  \sqrt{g}   
\left[ 
E[g,\tau]
+
\sum_i \frac{b_i}{a} F_i[g,\tau]   
\right]
\frac{1}{O}
\left[
E[g,\tau]
+
\sum_i \frac{b_i}{a} F_i[g,\tau]   
\right]
\nonumber \\ 
&
+ 
 \frac{1}{2a} \int  \sqrt{g}   
\left[ 
\sum_i b_i F_i[g,\tau]   
\right]
\frac{1}{O}
\left[
\sum_i b_i F_i[g,\tau]   
\right]
\, .
\end{align}
If we choose to discard the conformal terms in the second line, i.e., absorbing it in the definition of $\Gamma_c$, this action can be localized with only one scalar field as follows
\begin{align}
\Gamma_{int}[g,\varphi]
=
\int {\rm d}^2x \sqrt{g}  &\Bigg\{ 
 \frac{1}{2} \varphi_i O \varphi_i 
+
\alpha \varphi E[g,\tau]
+
\sum_i \beta_i \varphi F_i[g,\tau]   
\Bigg\}
 \, ,
\end{align}
where the constants can be determined by setting the auxiliary fields on-shell through
\begin{align}
O \varphi 
=
-\alpha E[g,\tau] - \sum_i \beta_i F_i[g,\tau]   \, .
\end{align}

To clarify this process, consider the simple example of $d=2$ with invariant torsion.
Here, in place of Eq.~\eqref{eq:2d_Gamma_ind_symm_form_invariant_torsion}, we can express the nonlocal action as
 \begin{align}\label{eq:2d_Gamma_ind_symm_form_invariant_torsion_1_auxiliary_field}
\Gamma_{NL}[g,\tau]
=
-\frac{c_1}{2}& \int {\rm d}^2x \sqrt{g}    
\left[ 
J + \frac{c_2}{c_1} \left(\nabla\cdot\tau\right) + \frac{c_3}{c_1} \left(\tau\cdot\tau\right) 
\right]    
\frac{1}{\Delta_2} 
\left[  
J + \frac{c_2}{c_1} \left(\nabla\cdot\tau\right) + \frac{c_3}{c_1} \left(\tau\cdot\tau\right)  
\right]    + 
 \nonumber \\ 
&
+ \frac{1}{2c_1} \int {\rm d}^2x \sqrt{g}  
 \left[  
c_2 \left(\nabla\cdot\tau\right)   +    c_3  \left(\tau\cdot\tau\right)  
\right]
 \frac{1}{\Delta_2} 
 \left[
c_2 \left(\nabla\cdot\tau\right)   +    c_3  \left(\tau\cdot\tau\right)  
 \right]
 \, ,
\end{align}
which can be localized using one $\varphi$ as
\begin{align}\label{eq:2d_Gamma_ind_auxiliary_fields_invariant_torsion_1_auxiliary_field}
\Gamma_{int}[g,\varphi]
=
\int {\rm d}^2x \sqrt{g}  &\Bigg\{ 
 \frac{1}{2} \varphi \Delta_2 \varphi
+
\alpha \varphi {J} 
+
\beta_1 \varphi \nabla\cdot\tau
+
\beta_2 \varphi  \tau\cdot\tau
\Bigg\}
 \, ,
\end{align}
which reproduces Eq.~\eqref{eq:2d_Gamma_ind_symm_form_invariant_torsion_1_auxiliary_field} up to the conformal action with the choices
\begin{align}
\alpha
=\sqrt{c_1}
\, ,
\qquad
\beta_1=\frac{c_2}{\sqrt{c_1}} \, 
\, ,
\qquad
\beta_2=\frac{c_3}{\sqrt{c_1}} 
\, ,
\end{align}
as can be easily verified by leveraging on the equations of motion $\Delta_2 \varphi =-\alpha {J} - \beta_1 \nabla\cdot\tau - \beta_2 \tau\cdot\tau$.

Notice that, independently from the number of auxiliary fields, these localized actions all yield the same nonlocal action on-shell, but differ by conformally invariant terms. 
However, as discussed throughout the main paper, we decided not to discard these conformal actions, as they can contribute nontrivially. Instead, we intentionally worked with the maximum number of auxiliary fields, which is fixed by the number of nontrivial cocycles.

\section{Basis and variations of $0$- and $1$-cochains with $\tau$ for transforming and invariant torsion in $d=4$}\label{app:basis_cochains_and_variations}

\subsection{Affinely tansforming torsion}

We start with $1$-cochains for transforming torsion. The ``bulk'' terms can be chosen as
\begin{align}\label{eq:1-cochains-bulk-tau}
&\omega_5 =\int {\rm d}^4x \sqrt{g} \sigma \left(\tau \cdot \tau \right)^2 \, ,
&& \omega_6 =\int {\rm d}^4x \sqrt{g} \sigma \tilde{J}^2  \, ,
&& \omega_7 =\int {\rm d}^4x \sqrt{g}    \sigma \tilde{K}^2_{\mu\nu}  \, ,
\nonumber\\
&  \omega_8 =\int {\rm d}^4x \sqrt{g}    \sigma \Omega^2_{\mu\nu}    \, ,
&&   \omega_9 =\int {\rm d}^4x \sqrt{g}    \sigma \tilde{J} \left(\tau \cdot \tau \right)   \, ,
&& \omega_{10} =\int {\rm d}^4x \sqrt{g} \sigma  \left( \nabla\tilde{J} \right) \cdot \tau             \, ,
\nonumber\\
& \omega_{11} = \int {\rm d}^4x \sqrt{g} \sigma  \tilde{K}^{\mu\nu} \tau_\mu \tau_\nu \, ,
&& \omega_{12} = \int {\rm d}^4x \sqrt{g} \sigma \,  \tau \cdot \tau  \left(\nabla \cdot \tau \right) \,,
&&  \omega_{13} = \int {\rm d}^4x \sqrt{g} \sigma \left(\nabla \cdot \tau \right)^2    \, ,
\nonumber\\
&     \omega_{14} = \int {\rm d}^4x \sqrt{g} \sigma  \left(\tau^\mu \tau^\nu \right)   \nabla_\mu  \tau_\nu  \,
&&   \omega_{15} = \int {\rm d}^4x \sqrt{g} \sigma \,     \tau_\mu \Box \tau^\mu\, ,
&&    \omega_{16} = \int {\rm d}^4x \sqrt{g} \sigma \,   \left( \nabla_\mu \tau_\nu   \right)^2    \, 
\nonumber\\
 &   \omega_{17} = \int {\rm d}^4x \sqrt{g} \sigma \,   \tau^\mu \nabla_\mu \nabla\cdot \tau  \, ,
\nonumber
\end{align}
Notice that $\nabla_\mu \tau_\nu \nabla^\nu\tau^\mu$, $\tilde{J}\, \left(\nabla \cdot \tau\right)$ and $\tilde{K}^{\mu\nu} \, \left( \nabla_\mu\ \tau_\nu \right)$ have been replaced by $\frac{1}{2}\Omega^2_{\mu\nu} =\left( \nabla_\mu \tau_\nu   \right)^2-\nabla_\mu \tau_\nu \nabla^\nu\tau^\mu \, $, $\tilde{J}^2$ and $\tilde{K}^2_{\mu\nu}$, respectively. The precise relations can be easily read off from the equation for $\Omega^2_{\mu\nu}$ and Eqs.\ \eqref{eq:TildeJ_Squared} and \eqref{eq:TildeK_Squared}.
A (redundant) basis for the ``total derivatives terms'' is
\begin{align}
&\omega_{18} =\int {\rm d}^4x \sqrt{g} \sigma \nabla_\mu \left( \tau^\mu \,\tau \cdot \tau \right)  \, ,
&& \omega_{19} =\int {\rm d}^4x \sqrt{g} \sigma   \nabla_\mu \left( \tau^\mu \,\nabla \cdot \tau \right) \, ,
&& \omega_{20} =\int {\rm d}^4x \sqrt{g}    \sigma  \nabla_\mu \left( \tau^\alpha \,\nabla^\mu \tau_\alpha \right)  \, ,
\nonumber\\
&   \omega_{21} =\int {\rm d}^4x \sqrt{g}    \sigma     \nabla_\mu \left( \tau^\alpha \,\nabla_\alpha \tau^\mu \right)  \, ,
&&  \omega_{22} =\int {\rm d}^4x \sqrt{g}    \sigma   \Box  \left(    \nabla \cdot \tau    \right)  \, ,
&&    \omega_{23} =\int {\rm d}^4x \sqrt{g} \sigma  \nabla_\mu  \left( \tau^\mu J \right) \, ,  
\nonumber\\
&  \omega_{24} = \int {\rm d}^4x \sqrt{g} \sigma   \nabla_\mu  \left( K^{\mu\nu}\tau_\nu  \right) \, ,   
&& \omega_{25} = \int {\rm d}^4x \sqrt{g} \sigma \Box  \left(    \tau \cdot \tau    \right)\, ,
&&  \omega_{26} = \int {\rm d}^4x \sqrt{g} \sigma     \nabla_\mu  \nabla_\nu \left( \tau^\mu \tau^\nu \right)   \, ,
\nonumber\\
& \omega_{27} = \int {\rm d}^4x \sqrt{g} \sigma   \tilde{\Box}\tilde{J}   \, .
\nonumber
\end{align}
In particular the following properties hold
\begin{itemize}  
\item  $\omega_{18}=2\omega_{14}+\omega_{12}$ ,     \qquad   $\omega_{19}=\omega_{17}+\omega_{13}$ ,   \qquad      $\omega_{20}=\omega_{15}+\omega_{16}=2\omega_{25}$ ,
\item  $\omega_{21}=\omega_{14}+\int {\rm d}^4x \sqrt{g} \sigma\left(  \nabla_\mu \tau_\nu \nabla^\nu\tau^\mu+2{K}^{\mu\nu} \,  \tau_\mu\ \tau_\nu + J \tau \cdot \tau \right)$ , \\
$\omega_{23}=\int {\rm d}^4x \sqrt{g} \sigma \left(    \left(\nabla  J\right)\cdot \tau+  J\, \left(\nabla \cdot \tau\right) \right) \,,$      
\item  $\omega_{24}=\int {\rm d}^4x \sqrt{g} \sigma\left( {K}^{\mu\nu} \, \left( \nabla_\mu\ \tau_\nu \right)+ \left( \nabla{J} \right) \cdot \tau \right)$ , \\
$\omega_{26}= 2\omega_{17}+\omega_{13}+\int {\rm d}^4x \sqrt{g} \sigma\left(  \nabla_\mu \tau_\nu \nabla^\nu\tau^\mu + 2{K}^{\mu\nu} \,  \tau_\mu\ \tau_\nu + J \tau \cdot \tau \right) \,,$
\end{itemize}
which can be straightforwardly reduced to the ``bulk'' terms thanks to the relations \eqref{eq:TildeJ} and \eqref{eq:TildeK}, and also the relations involving $\Omega^2_{\mu\nu}$, i.e., \eqref{eq:TildeJ_Squared} and \eqref{eq:TildeK_Squared}. Therefore only $\omega_{22}$ is actually independent. However, this term can be traded for $\omega_{27} =\int {\rm d}^4x \sqrt{g}    \sigma   \tilde{\Box} \tilde{J}$, see \eqref{eq:BoxJ}. Clearly, the most convenient term to work with is $\omega_{27}$, which is that reported in the main text as $\omega_{18}$. Now it should also be clear which basis we have chosen in the case of conformally invariant torsion.

We now report the variations $\delta_\sigma$ of these $1$-cochains, from which it is possible to obtain the consistent anomaly. We have already noted that $\delta_\sigma \omega_6= \delta_\sigma \omega_7=\delta_\sigma \omega_8=\delta_\sigma \omega_{18} = 0 $ and expressed the variations of purely metric terms using the modified curvature tensors. The other variations are 
\begin{align*}
&\delta_\sigma \omega_5 =-4b\int {\rm d}^4x \sqrt{g}\sigma \nabla_\mu \sigma \, \tau^\mu (\tau \cdot \tau) \, , \\
&\delta_\sigma \omega_9 =  -2 b\int {\rm d}^4x \sqrt{g} \sigma \nabla_\mu \sigma  \left( \tilde{J}\tau^\mu\right) \, , \\ 
&\delta_\sigma \omega_{10} =\int {\rm d}^4x \sqrt{g} \sigma \nabla_\mu \sigma      \left(2 \tilde{J} \tau^\mu     -  b \nabla^\mu \tilde{J}   \right) \, , \\
&\delta_\sigma \omega_{11} = - \int {\rm d}^4x \sqrt{g}  \sigma \nabla_\mu \sigma  \left( 2 b \tilde{K}^\mu{}_{\nu} \tau^\nu    \right)  \, ,\\
&\delta_\sigma \omega_{12} = 2 \int {\rm d}^4x \sqrt{g}  \sigma \nabla_\mu \sigma  \left(  b \tau^\nu (\nabla^\mu  \tau_\nu)  -     \tau^\mu (\tau \cdot \tau)   -  b   \tau^\mu (\nabla \cdot \tau)      \right) \, , \\
&\delta_\sigma \omega_{13} = 2 \int {\rm d}^4x \sqrt{g}    \sigma \nabla_\mu \sigma  \left(  b  \nabla^\mu (\nabla \cdot \tau)    -     2   \tau^\mu (\tau \cdot \tau)               \right)  \, ,     \\
&\delta_\sigma \omega_{14} = \int {\rm d}^4x \sqrt{g}   \sigma \nabla_\mu \sigma  \left(     \tau^\mu (\tau \cdot \tau)   +   \nabla^\mu \nabla \cdot \tau       -      \tau^\nu \nabla^\mu  \tau_\nu                  \right)  \,,   \\
&\delta_\sigma \omega_{15} = \int {\rm d}^4x \sqrt{g} b \nabla_\mu \sigma (\Box \sigma) \tau^\mu    - \delta_\sigma \omega_{12}  \, ,     \\
&\delta_\sigma \omega_{16} = \int {\rm d}^4x \sqrt{g}   \sigma \nabla_\mu \sigma  \left(    b \tilde{J} \tau^\mu         +      2 b \tilde{K}^\mu{}_{\nu} \tau^\nu      2 \tau^\nu \nabla_\nu  \tau^\mu     -     3    \tau^\mu \nabla \cdot \tau      +      b \Box\tau^\mu    +  \tau^\nu \nabla^\mu \tau_\nu   + b\nabla^\mu \nabla \cdot \tau        \right) \, ,      \\
&\delta_\sigma \omega_{17} = \int {\rm d}^4x \sqrt{g}  \left( 2 \sigma \nabla_\mu \sigma  \left(  2 \tau^\mu \nabla \cdot \tau    - \nabla^\mu \nabla \cdot \tau                       \right)  + b \nabla_\mu \sigma (\Box \sigma) \tau^\mu           \right)  \, ,
\end{align*}
for which we have chosen the basis \eqref{eq:1_basis_2_cochains}. Putting everything together, the consistency conditions \eqref{eq:c_conditions4d} give the following. Since we have only one tensor counting as one derivative, i.e., $\tau^\mu$, we have only one $h_{j=1}=f_{11}+f_{13}=0$, while we have the following $l_{j=1,\dots,9}$:
\begin{align*}
&   f_{12}-f_{11} - 2 f_{13} + 2 f_9 + 2 \frac{l_1 + l_2}{b^2}=0    \, ,
&& \quad  f_{12}-f_{11} - 2 f_5 + 2 \frac{ f_6}{b}=0  
\\
&   3 f_{11} - 3 f_{12} + 4 f_{13} + b (f_{10} - 2 f_8) - 4 f_9  = 0     \, , 
&&  \quad   f_{11}-f_{12}+f_{7}=0  \, ,
\\
&  f_{12}-f_{11}  - b (f_{10} - 2 f_8) - 4 \frac{l_1 + l_2}{b^2}  =0   \, ,
&& \quad bf_6+2(c_1+c_2)=0 \, , 
\\
&    f_{12}-f_{11}=0     \, ,
&& \quad    f_{10}-4 b f_1  - 2 f_8=0   \, .   
\end{align*}
We can notice that the first three equation of the first column are not independent. Therefore, we have $12$ variables and $8$ independent equations and we can determine $8$ variables in terms of four. The solution of the system $h_{1}=0 \,, \, l_{j=1,\cdots,9}=0$ is reported in the main text. As a double check on the expression of the anomaly in the main text, let us compute also the variations
\begin{align*}
&\delta_\sigma \omega_{19} = b \int {\rm d}^4x \sqrt{g}  \nabla_\mu \sigma (\Box \sigma) \tau^\mu       \, , 
\qquad
\delta_\sigma \omega_{25} =2  b \int {\rm d}^4x \sqrt{g} \nabla_\mu \sigma (\Box \sigma) \tau^\mu           \, ,  
\qquad
\delta_\sigma  \int {\rm d}^4x \sqrt{g} \nabla_\mu \left(\tilde{J}\tau^\mu\right) =0          \, ,   
\\ \nonumber
&\delta_\sigma \omega_{18}=0 \, .
\end{align*}
Thus, in the anomaly \eqref{eq:anomaly_affine_torsion} we correctly have $ \nabla_\mu \left(\tilde{J}\tau^\mu\right) $ and $ \nabla_\mu \left(\tau^\mu \tau\cdot\tau\right) $ with arbitrary coefficients, while $\omega_{19}$ and $\omega_{25}$ appear with the combination $ \frac{1}{2} \omega_{25}    - \omega_{19} $.

Since we have chosen a basis without derivatives on $\sigma$, the $0$-cochains can be easily constructed. Therefore, we start with
\begin{align}
&\xi_1 =\int {\rm d}^4x \sqrt{g}  \left(\tau \cdot \tau \right)^2 \, ,
&& \xi_2 =\int {\rm d}^4x \sqrt{g}   \left(\nabla \cdot \tilde{J}  \right)\tau \, ,
&& \xi_3 =\int {\rm d}^4x \sqrt{g}     \tilde{K}^{\mu\nu} \nabla_\mu \tau_\nu                     \, ,
\nonumber\\
&  \xi_4 =\int {\rm d}^4x \sqrt{g}   \nabla_\mu \tau_\nu \nabla^\nu\tau^\mu     \, ,
&&       \xi_5 =\int {\rm d}^4x \sqrt{g}     \tilde{{J}} \left(\tau \cdot \tau \right)   \, ,
&&\xi_6 =\int {\rm d}^4x \sqrt{g}   \left( \nabla{\tilde{J}} \right) \cdot \tau                         \, ,
\nonumber\\
&        \xi_7 = \int {\rm d}^4x \sqrt{g}   \tilde{K}^{\mu\nu} \tau_\mu \tau_\nu     \,  ,
&& \xi_8 = \int {\rm d}^4x \sqrt{g}  \,  \tau \cdot \tau  \left(\nabla \cdot \tau \right) \,,
&&      \xi_9 = \int {\rm d}^4x \sqrt{g}  \left(\nabla \cdot \tau \right)^2              \, ,
\nonumber\\
&   \xi_{10} = \int {\rm d}^4x \sqrt{g}   \left(\tau^\mu \tau^\nu \right)   \nabla_\mu  \tau_\nu\,  \, ,
&&       \xi_{11} = \int {\rm d}^4x \sqrt{g}  \,     \tau_\mu \Box \tau^\mu  \, ,
&&    \xi_{12} = \int {\rm d}^4x \sqrt{g}  \,   \left( \nabla_\mu \tau_\nu   \right)^2      \, ,
\nonumber\\
&    \xi_{13} = \int {\rm d}^4x \sqrt{g}  \,   \tau^\mu \nabla_\mu \nabla\cdot \tau \,,
\nonumber
\end{align}
and then, since
\begin{itemize}  
\item  $\int {\rm d}^4x \sqrt{g}   K^{\mu\nu} \nabla_\mu \tau_\nu = \int {\rm d}^4x \sqrt{g}  J      \left(\nabla \cdot \tau \right)\,, $      \quad   $\int {\rm d}^4x \sqrt{g}    \nabla_\mu \tau_\nu \nabla^\nu\tau^\mu=- \int {\rm d}^4x \sqrt{g}  \, \left(  \tau^\mu \nabla_\mu \nabla\cdot \tau + 2 K^{\mu\nu} \tau_\mu \tau_\nu  + J \left(\tau \cdot \tau \right)   \right) \, ,$  
\item  $ \int {\rm d}^4x \sqrt{g}  \,   \left( \nabla_\mu \tau_\nu   \right)^2 = - \int {\rm d}^4x \sqrt{g}  \,     \tau_\nu \Box \tau^\nu$ ,     \qquad    $\int {\rm d}^4x \sqrt{g}   \left(\tau^\mu \tau^\nu \right)   \nabla_\mu  \tau_\nu=\frac{1}{2} \int {\rm d}^4x \sqrt{g}  \,  \tau \cdot \tau  \left(\nabla \cdot \tau \right)$  ,     
\item   $\int {\rm d}^4x \sqrt{g}  \,   \tau^\mu \nabla_\mu \nabla\cdot \tau = - \int {\rm d}^4x \sqrt{g}  \left(\nabla \cdot \tau \right)^2$ ,  \qquad     $\int {\rm d}^4x \sqrt{g}   \left( \nabla{J} \right) \cdot \tau = - \int {\rm d}^4x \sqrt{g}  {J}   \left( \nabla\cdot \tau\right)\,,$
\end{itemize}
we obtain the basis reported in the main text. The variations of the independent $0$-cochains for transforming torsion are 
\begin{align*}
&\delta_\sigma \int {\rm d}^4x \sqrt{g}  \left(\tau \cdot \tau \right)^2 = - 4b \int {\rm d}^4x \sqrt{g}   \sigma \nabla_\mu \left( \tau^\mu \,\tau \cdot \tau \right)  \,, \\
&\delta_\sigma \int {\rm d}^4x \sqrt{g}      \left(\nabla \cdot \tilde{J}  \right)\tau  = - \int {\rm d}^4x \sqrt{g} \sigma  \left(   2 \left(  \nabla \tilde{J}\right) \cdot \tau +  2 \tilde{J} \left(  \nabla \cdot \tau \right)  -b \Box \tilde{J}    \right)      \,  \\
&\delta_\sigma\int {\rm d}^4x \sqrt{g}     \tilde{J} \left(\tau \cdot \tau \right)  = -2b \int {\rm d}^4x \sqrt{g} \sigma \nabla_\mu  \left( \tilde{J} \tau_\mu      \right)   \, ,\\
&\delta_\sigma \ \int {\rm d}^4x \sqrt{g}   \tilde{K}^{\mu\nu} \tau_\mu \tau_\nu = -2b \int {\rm d}^4x \sqrt{g} \sigma   \left( \nabla_{\mu} \tilde{K}^{\mu\nu}   \tau_\nu +    \tilde{K}^{\mu\nu} \nabla_{\mu} \tau_\nu  \right)  \, , \\
&\delta_\sigma \int {\rm d}^4x \sqrt{g}  \,  \tau \cdot \tau  \left(\nabla \cdot \tau \right)  = -2 \int {\rm d}^4x \sqrt{g} \sigma   \left(  b \tau^\mu\nabla_\mu \nabla \cdot \tau  +2 \tau^\mu \tau^\nu \nabla_\nu\tau_\mu + \tau \cdot \tau  \left(\nabla \cdot \tau \right)  \right. \\
&\hspace{4cm}\left. + b\left( \nabla \cdot \tau\right)^2  - b  \tau^\mu \Box \tau_\mu - \left(\nabla_\mu \tau_\nu      \right)^2    \right)  \, ,     \\
&\delta_\sigma \int {\rm d}^4x \sqrt{g}  \,   \tau^\mu \nabla_\mu \nabla\cdot \tau = 2 \int {\rm d}^4x \sqrt{g} \sigma   \left(  2 \tau^\mu \nabla_\mu \nabla\cdot \tau + 2 \left(\nabla \cdot \tau \right)^2 - b \Box \nabla \cdot \tau    \right)  \, ,     \\
&\delta_\sigma \int {\rm d}^4x \sqrt{g}  \,   \left( \nabla_\mu \tau_\nu   \right)^2 = 2 \int {\rm d}^4x \sqrt{g} \sigma   \left(  J\left( \tau\cdot\tau \right) + 2 K_{\mu\nu}\tau^\mu\tau^\nu + 3 b \left(  \nabla J\right) \cdot \tau + b J \left(  \nabla \cdot \tau \right)  -  \left( \nabla \cdot \tau\right)^2   \right. \\  
&\hspace{4cm}\left. +   \tau^\mu \Box \tau_\mu     + b  \Box \nabla \cdot \tau       + 2 b  K_{\mu\nu}\nabla^\mu\tau^\nu  + \nabla_\mu \tau_\nu  \nabla^\nu \tau^\mu   +   \left(\nabla_\mu \tau_\nu      \right)^2     \right)  \, .   \\ 
\end{align*}
From which we deduce, in particular, the variations in Eq.~\eqref{eq:4d_trivial_anomalies_affine_torsion}. The variation $\delta_\sigma \int {\rm d}^4x \sqrt{g}  \,   \left( \nabla_\mu \tau_\nu   \right)^2$ could be written in terms of modified conformal tensors, but we do not actually need it.

\subsection{Conformally invariant torsion}

In the case of invariant torsion the basis is reported in Subsect.~\ref{sect_4d_cohomological_analisys_affine_torsion} and its construction should be clear from the case of transforming torsion analyzed in the previous subsection. The variations of the $1$-cochains are
\begin{align*}
&\delta_\sigma \omega_6 = - \int {\rm d}^4x \sqrt{g} \sigma \nabla_\mu \sigma  \left( 2 J\tau^\mu + \nabla^\mu \nabla \cdot \tau     \right)  \,, 
&&\delta_\sigma \omega_7 = - \int {\rm d}^4x \sqrt{g} \sigma \nabla_\mu \sigma      \left(  \frac{3}{2} J\tau^\mu   + \frac{1}{2} \Box \tau^\mu    + \frac{1}{2} \nabla^\mu \nabla \cdot \tau      \right)          \,,
  \\
&\delta_\sigma \omega_9 = - \int {\rm d}^4x \sqrt{g}  \sigma \nabla_\mu \sigma  \left(  \tau^\nu \nabla^\mu \tau_\nu  \right)  \, ,
&&\delta_\sigma \omega_{10} =  \int {\rm d}^4x \sqrt{g}  \sigma \nabla_\mu \sigma  \left(     \left(    2 J\tau^\mu + \nabla^\mu \nabla \cdot \tau         \right) - \nabla_\mu \sigma (\Box \sigma) \tau^\mu           \right)  \, , 
\\
&\delta_\sigma \omega_{11} = - \int {\rm d}^4x \sqrt{g}    \sigma \nabla_\mu \sigma  \left(   \tau^\nu \nabla_\nu \tau^\mu +     \tau^\mu \nabla \cdot \tau          \right)  \, ,     
&&\delta_\sigma \omega_{12} = - 2 \int {\rm d}^4x \sqrt{g}   \sigma \nabla_\mu \sigma  \left(        \tau^\mu \tau \cdot \tau            \right)  \,,  
 \\
&\delta_\sigma \omega_{13} = -4 \int {\rm d}^4x \sqrt{g} \sigma \nabla_\mu \sigma   \left(    \tau^\mu \nabla \cdot \tau     \right)    \, ,     
&&\delta_\sigma \omega_{14} = \int {\rm d}^4x \sqrt{g}   \sigma \nabla_\mu \sigma  \left(           \tau^\mu \tau \cdot \tau       \right) \, ,     
 \\
&\delta_\sigma \omega_{15} = 2 \int {\rm d}^4x \sqrt{g} \sigma \nabla_\mu \sigma   \left(    \tau^\mu \nabla \cdot \tau     -  \tau^\nu \nabla_\nu \tau^\mu   -      \tau^\nu     \nabla^\mu  \tau_\nu       \right)   \, , 
&&\delta_\sigma \omega_{16} = - 2 \int {\rm d}^4x \sqrt{g} \sigma \nabla_\mu \sigma   \left(    \tau^\mu \nabla \cdot \tau     -  \tau^\nu \nabla_\nu \tau^\mu   -      \tau^\nu     \nabla^\mu  \tau_\nu       \right)   \, ,
 \\
&\delta_\sigma \omega_{17} = 4 \int {\rm d}^4x \sqrt{g}  \sigma \nabla_\mu \sigma  \left(      \tau^\mu \nabla \cdot \tau                  \right)   \, , 
&&\delta_\sigma \omega_{18} = 2 \int {\rm d}^4x \sqrt{g}       \nabla_\mu \sigma (\Box \sigma) \tau^\mu                       \, .
\end{align*}
Combining the previous results, we have one $h_{j=1}=2 f_{18} - f_6=0$ and the  $l_{j=1,\cdots,8}$: 
\begin{align*}
&2 f_{11} - 2 f_{12} + 4 f_{13} - f_7 - 4 f_9=0 \, ,
&&f_{3}=0 \, ,
&& 2f_{12}-2 f_{11}  - f_7=0 \, , 
\\
&f_{10} - 2 f_8=0 \, ,
&& f_6-f_2 - \frac{f_3}{2}=0  \, ,
&&c_1 + c_2=0 \, , 
\\
&f_{11} - f_{12} + f_5=0 \, ,
&& 4  f_6 - 4 f_2 -3  f_3=0\, .
\end{align*}
Notice that the equations in the second column are not linearly independent. We have $8$ equations for $14$ variables. The solution of $h_{1}=0 \,, \, l_{j=1,\cdots,8}=0$ is reported in the main text. By plugging the solution in the ansatz for the anomaly, we get (up to metric contributions)
\begin{align*}
A^{\tau}_\sigma
&=
\int \sqrt{g} \sigma \Bigg\{
f_1  \left(\tau \cdot \tau \right)^2
+
f_4 \Omega^2_{\mu\nu}
+
\frac{f_{11}}{2} \Box \tau^2_\mu
+
 \frac{f_{10}}{2}   \nabla_\mu \left( \tau^\mu \,\tau \cdot \tau \right) 
  +
 f_{2}   \left(  \frac{1}{2}  \Box \nabla \cdot \tau        +     \nabla_\mu \left( \tau^\mu \,J \right)                \right)
 \\ \nonumber
&
+
f_{13}     \nabla_\mu \left( \tau^\mu \,\nabla \cdot \tau \right) 
+
(f_{12}-f_{11}) \left( R_{\mu\nu}  \tau^\mu \tau^\nu +  \left( \nabla_\mu \tau_\nu   \right)^2 -( \nabla \cdot \tau)^2   \right)  \Bigg\}
\, ,
\end{align*}
which can be conveniently rewritten in the basis of \eqref{eq:anomaly_inv_torsion} by using that $( \nabla \cdot \tau)^2  =  \tau^\mu\nabla_\mu  \nabla \cdot \tau - \nabla_\mu \left( \tau^\mu \nabla \cdot \tau \right)$ and that $  \tau^\mu\nabla_\mu  \nabla \cdot \tau  =   \tau^\mu  \nabla _\nu \nabla_\mu \tau^\nu - R_{\mu\nu}  \tau^\mu \tau^\nu$.

As before, as a double check on the expression of the anomaly in the main text, let us report also the variations
\begin{align*}
&\delta_\sigma \int {\rm d}^4x \sqrt{g}  \sigma \nabla_\mu \left( \tau^\mu \,\tau \cdot \tau \right)  =0 \, , 
\quad   
&&\delta_\sigma  \int {\rm d}^4x \sqrt{g}  \sigma  \Box \nabla \cdot \tau  = 2 \int {\rm d}^4x \sqrt{g}  \nabla_\mu \sigma (\Box \sigma) \tau^\mu       \,, 
\\
&\delta_\sigma \int {\rm d}^4x \sqrt{g}  \sigma  \nabla_\mu \left( \tau^\mu \,J \right)      =-  \int {\rm d}^4x \sqrt{g} \nabla_\mu \sigma (\Box \sigma) \tau^\mu           \, ,  
\quad
&&\delta_\sigma \int {\rm d}^4x \sqrt{g} \sigma \,   \tau^\mu \nabla_\nu \nabla_\mu  \tau^\nu = - \delta_\sigma \int {\rm d}^4x \sqrt{g} \sigma \,  \left(  \nabla_\mu  \tau_\nu \right)^2  = -\delta_\sigma \omega_{16}  \, ,
\end{align*}
which, as in the previous case, confirms the structures present in the anomaly \eqref{eq:anomaly_inv_torsion}.

In the case of invariant torsion, the variations for a basis of $0$-cochains in $d=4$ are
\begin{align*}
&\delta_\sigma \int {\rm d}^4x \sqrt{g}  \left(\tau \cdot \tau \right)^2 = 0  \,, \\
&\delta_\sigma \int {\rm d}^4x \sqrt{g}  \left(\nabla J  \right)   \cdot \tau   =  \int {\rm d}^4x \sqrt{g} \sigma  \left(   2 \left(  \nabla J\right) \cdot \tau + \Box \nabla \cdot \tau  +  2 J \left(  \nabla \cdot \tau \right)   \right)      \,  \\
&\delta_\sigma\int {\rm d}^4x \sqrt{g}     {J} \left(\tau \cdot \tau \right)  = -2 \int {\rm d}^4x \sqrt{g} \sigma   \left(  \tau^\mu \Box \tau_\mu +    \left(\nabla_\mu \tau_\nu      \right)^2    \right)   \, ,\\
&\delta_\sigma \ \int {\rm d}^4x \sqrt{g}   {K}^{\mu\nu} \tau_\mu \tau_\nu = - \int {\rm d}^4x \sqrt{g} \sigma   \left( J\left( \tau\cdot\tau \right) + 2K_{\mu\nu}\tau^\mu\tau^\nu + 2 \tau^\mu\nabla_\mu \nabla \cdot \tau  + \left( \nabla \cdot \tau\right)^2  +  \nabla_\mu \tau_\nu  \nabla^\nu \tau^\mu      \right)  \, , \\
&\delta_\sigma \int {\rm d}^4x \sqrt{g}  \,  \tau \cdot \tau  \left(\nabla \cdot \tau \right)  = -2 \int {\rm d}^4x \sqrt{g} \sigma   \left( 2 \tau^\mu \tau^\nu \nabla_\nu\tau_\mu + \tau \cdot \tau  \left(\nabla \cdot \tau \right)     \right)  \, ,     \\
&\delta_\sigma \int {\rm d}^4x \sqrt{g}  \,   \tau^\mu \nabla_\mu \nabla\cdot \tau = 2 \int {\rm d}^4x \sqrt{g} \sigma   \left(  2 \tau^\mu \nabla_\mu \nabla\cdot \tau + 2 \left(\nabla \cdot \tau \right)^2   \right)  \, ,     \\
&\delta_\sigma \int {\rm d}^4x \sqrt{g}  \,   \left( \nabla_\mu \tau_\nu   \right)^2 = 2 \int {\rm d}^4x \sqrt{g} \sigma   \left(  J\left( \tau\cdot\tau \right) + 2 K_{\mu\nu}\tau^\mu\tau^\nu  -  \left( \nabla \cdot \tau\right)^2  +   \tau^\mu \Box \tau_\mu     + \nabla_\mu \tau_\nu  \nabla^\nu \tau^\mu   +   \left(\nabla_\mu \tau_\nu      \right)^2     \right)  \, .    
\end{align*}
From which we get in particular that
\begin{align*}
&\int {\rm d}^4x \sqrt{g}  \sigma \nabla_\mu \left( \tau^\mu \,\tau \cdot \tau \right) = -\frac{1}{4 b} \delta_\sigma \xi_1  \, , \\ \nonumber     
& \int {\rm d}^4x \sqrt{g}  \sigma  \left(  \frac{1}{2} \Box    \tau^2     - \nabla_\mu \left( \tau^\mu \,\nabla \cdot \tau \right)    \right) = -\frac{1}{4 b^2} \delta_\sigma \xi_1 + \frac{1}{2b} \delta_\sigma \xi_5 \, .
\end{align*}


\end{document}